\documentclass[aps,prx,twocolumn,groupedaddress,amsmath,amssymb,superscriptaddress]{revtex4-2}
\usepackage{graphicx}
\usepackage{float}
\usepackage{multirow}
\usepackage[breaklinks=true]{hyperref}
\usepackage[dvipsnames,svgnames]{xcolor}
\usepackage{xcolor}
\usepackage{amssymb}
\usepackage{stackengine,graphicx}
\usepackage{amsmath}
\usepackage{hyperref}
\usepackage{color}
\usepackage{stackengine}
\usepackage{subfigure}
\usepackage{verbatim}
\usepackage{bbold}
\usepackage{stmaryrd} %for short arrows
\usepackage{overpic}
\usepackage[normalem]{ulem}
\usepackage{soul}

\usepackage{pdfpages}
\makeatletter
\AtBeginDocument{\let\LS@rot\@undefined}
\makeatother

\newcommand{\eps}{\varepsilon}

\newcommand{\bas}{{\bf e}_z}

\newcommand{\cd}{\tilde{\chi}}
\newcommand{\vc}[1]{\boldsymbol{\mathrm{#1}}}


\begin{document}

\title{Intrinsic electromagnetic properties
%Time-reversal symmetry breaking invisible in scattering\\
%\red{Hidden axion electrodynamics in reciprocal media}
%\\
%\red{Axion response invisible to electromagnetic scattering}
}

\author{Eduardo Barredo-Alamilla}
\altaffiliation{These authors contributed equally to this work}
\affiliation{School of Physics and Engineering, ITMO University, Saint  Petersburg 197198, Russia}

\author{Timur Z. Seidov}
\altaffiliation{These authors contributed equally to this work}
\affiliation{School of Physics and Engineering, ITMO University, Saint  Petersburg 197198, Russia}

\author{Daniel A. Bobylev}
\affiliation{School of Physics and Engineering, ITMO University, Saint  Petersburg 197198, Russia}

\author{Maxim A. Gorlach}
\email{m.gorlach@metalab.ifmo.ru}
\affiliation{School of Physics and Engineering, ITMO University, Saint  Petersburg 197198, Russia}

\begin{abstract}

Electromagnetic properties of materials are commonly inferred from how they scatter light. Here we show that this diagnostic is fundamentally incomplete. We predict the existence of an {\it intrinsic electromagnetic property} that remains invisible in any scattering experiment independently of geometry, polarization, far-field or near-field excitation. Still, this property affects the radiation emitted by the sources embedded inside the material. We reveal this phenomenon for a specially designed nonreciprocal magnetoelectric material combining axion and dual axion responses. Our results establish intrinsic electromagnetic properties as a distinct class of material responses and show that scattering experiments alone are insufficient to characterize matter.

%within generalized axion electrodynamics, where coexisting axion and dual-axion couplings suppress all magnetoelectric scattering signatures under a specific condition. 
%As a concrete realization, we demonstrate that in a medium hosting coexisting axion and dual axion magnetoelectric couplings, all scattering signatures of the response can be suppressed under a specific condition, rendering the medium indistinguishable from an ordinary dielectric despite broken time-reversal symmetry and while satisfying Lorentz reciprocity.
%The hidden response is revealed only through radiation emitted by sources embedded within the material.
%Our results establish intrinsic electromagnetic properties as a distinct class of material response and demonstrate that electromagnetic scattering alone is insufficient to characterize matter.

\end{abstract}

\maketitle

%\section*{Introduction}

Since ancient times, people based their perception of the surrounding world on {\it how it looks}. In this way, we typically assess color, transparency and texture of the objects around. In physical terms, electromagnetic properties of materials are understood based on how they scatter incident light. Similar scattering experiments with $\alpha$-particles, electrons, nucleons and other particles provide an experimental foundation of modern particle physics and the Standard Model.

%are at the core of modern physics providing important insights into the structure of the atoms and their nuclei. 

At the same time, it is also understood that the objects are actually not what they look like. This was articulated by the philosopher Immanuel Kant, who introduced the notion of the ``thing in itself" possessing some intrinsic properties which remain fundamentally unobservable. For a long time, this philosophical discourse remained abstract and detached from the scientific agenda.

%Introduction V1
%In this Letter, we uncover an unexpected physical dimension of this long-standing problem by predicting the phenomenon of {\it intrinsic electromagnetic nonreciprocity}. This property of the material is not manifested in any kind of the scattering experiment, regardless of the way the structure is illuminated. However, it does show up when the external sources are placed inside the material by modifying the polarization of emitted light.

%Introduction V2
%In this Letter, we show that electromagnetic scattering does not, in general, provide complete access to material response. We predict the existence of {\it intrinsic electromagnetic properties} that remain entirely invisible to any external scattering experiment, regardless of the geometry or polarization. Despite this invisibility, the response is physically real and manifests itself when electromagnetic sources are embedded within the medium, modifying the polarization of the emitted radiation.

%Remarkably, this hidden electromagnetic property arises in systems that explicitly break time-reversal symmetry, yet satisfy Lorentz reciprocity and appear indistinguishable from ordinary dielectrics in all scattering measurements. This establishes a clear separation between the existence of a material response and its detectability by scattering, revealing a class of electromagnetic properties accessible only through internal excitation.

In this Article, we bring this discourse to the physical ground showing that the experiments on scattering of electromagnetic waves performed in various settings and including both far-field and near-field excitation do not provide complete access to the electromagnetic response of the material. Specifically, we predict the existence of {\it intrinsic electromagnetic properties} that remain entirely invisible in any scattering experiment, but manifest themselves by altering the emitted radiation of electromagnetic sources placed directly inside the medium.

%modifying the polarization of the emitted radiation. Remarkably, such hidden responses may arise in systems that explicitly break $\mathcal{T}$-symmetry while satisfying Lorentz reciprocity and appearing indistinguishable from ordinary dielectrics in all scattering measurements.

%\commentMaxim{There should be some discussion about related topics in photonics. Less philosophy and more connection to modern photonic things: (0) impedance matching; (1) non-radiating current configurations and anapoles; (2) dark modes; (3) BICs; (4) invisibility cloaking. 1 and 2 are different, because can be excited via their near field. 3 are also different because perfect BICs cannot exist in finite structures and only quasi-BICs are available.}
%\commentLalo{Done!}

The concept of intrinsic electromagnetic properties should be distinguished from several other phenomena extensively studied in photonics. A straightforward approach to hide the material and its properties from electromagnetic waves is to put it inside a metallic box as it is done in electromagnetic shielding. However, the shielding prevents wave transmission and causes extra scattering which gives it away. A more advanced technique is invisibility cloaking~\cite{Leonhardt2006,Pendry2006,alu2008plasmonic,Leonhardt2009,fleury2015invisibility} which bends the light rays around the cloaked object causing no extra scattering. In both cases, however, there is a region of space inaccessible to the incident waves, indicating a hidden object. In contrast to that, the material with intrinsic electromagnetic properties can be totally transparent to light having no inaccessible areas.

\begin{figure}[t!]
    \centering   \includegraphics[width=0.95\linewidth]{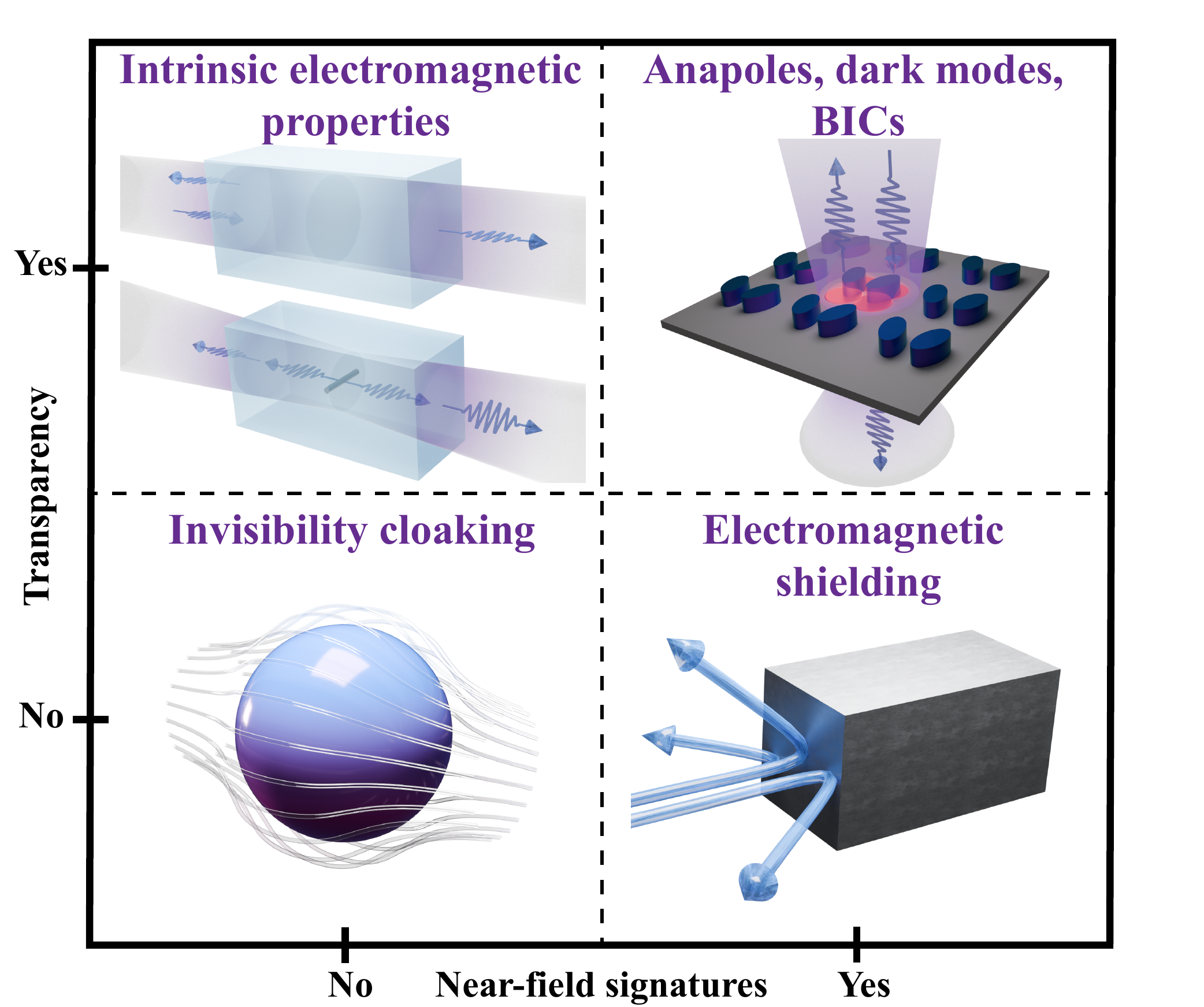}
    \caption{{\bf Conceptual classification of electromagnetic invisibility mechanisms based on wave transmission and near field signatures.} Electromagnetic shielding and invisibility cloaking feature areas inaccessible to electromagnetic waves. Bound states in the continuum (BICs) feature sizable near-field signatures. The material with intrinsic electromagnetic properties has neither inaccessible areas, nor near-field signatures. 
   }
\label{IEPdiagram}
\end{figure}

Another celebrated concept is bound states in the continuum (BIC)~\cite{Hsu2016,azzam2021photonic,Koshelev2023,kang2023applications} which are special modes which do not couple to the plane waves and hence neither radiate to the far field, nor can be excited by the incident plane waves. Despite vanishing far field, BICs feature sizable near fields and allow for the near field excitation which differentiates them from intrinsic electromagnetic properties introduced here. The same differences appear for the related concepts of dark modes~\cite{Gomez2013} and anapoles~\cite{Miroshnichenko2015,yang2019nonradiating,Baryshnikova2019}. The conceptual distinctions between these phenomena are sketched in Fig.~\ref{IEPdiagram}.

% In contrast to the above scenarios, intrinsic electromagnetic properties allow electromagnetic wave transmission through the entire material while remaining undetectable neither in the far-field, nor in the near-field excitation experiments (Fig.~\ref{IEPdiagram}). Though such properties cannot be accessed by any external illumination, they are readily revealed via the sources placed inside the structure.

As a specific material exhibiting intrinsic electromagnetic properties, we consider a nonreciprocal magnetoelectric medium which, in addition to permittivity $\eps$ and permeability $\mu$, possesses two pseudoscalar linear susceptibilities: the axion response $\chi$ and the dual axion response $\tilde{\chi}$.

%As a specific material with intrinsic electromagnetic property we study non-reciprocal magneto-electric medium which in addition to permittivity $\eps$ and permeability $\mu$ features two additional pseudo-scalar susceptibilities: axion response $\chi$ and dual axion response $\cd$.

% To establish the existence of intrinsic electromagnetic properties in a concrete and universal setting, we analyze an extension of Maxwell electrodynamics in which distinct nonreciprocal magnetoelectric couplings, axion and dual axion couplings, coexist and cancel at the level of scattering while remaining operative in the presence of sources. Generalized axion electrodynamics, which we term here biaxionic electrodynamics, provides precisely such a framework. In the particular realization considered here, this cancellation occurs in systems that microscopically break time-reversal symmetry, yet satisfy Lorentz reciprocity and appear indistinguishable from ordinary dielectrics in all scattering measurements.
 
The axion response $\chi$ is a phenomenon in condensed matter~\cite{wilczek1987two,Nenno2020,Sekine2021} and photonics~\cite{Shaposhnikov2023,safaei2024optical,yang2025gigantic} captured by the material equations of the form
\begin{equation*}
    \begin{gathered}
        4\pi\,\mathbf{P} = (\eps-1)\,\mathbf{E} + \chi\,\mathbf{B}, \; 
        4\pi\,\mathbf{M} = (1-\mu^{-1})\,\mathbf{B}+\chi\, \mathbf{E}
    \end{gathered}
\end{equation*}
with real $\chi$, when an applied electric field induces magnetization $\vc{M}$ and an applied magnetic field causes electric polarization $\vc{P}$. This physics has been known for a long time in some antiferromagnetic condensed matter structures~\cite{dzyaloshinskii1960magneto,foner1963magnetoelectric}, where it is termed nonreciprocal magnetoelectric effect, and also in strong three-dimensional topological insulators~\cite{Qi2008,Essin2009,Armitage2016}. The magnitude of $\chi$ in condensed matter is typically in the range $10^{-3}$--$10^{-2}$ and the associated optical effects are very weak.

Recently, it was predicted theoretically~\cite{Shaposhnikov2023}, tested numerically~\cite{safaei2024optical}, and confirmed experimentally~\cite{yang2025gigantic,Asadchy2026} that the same effect can be boosted by $2-3$ orders of magnitude in metamaterials providing a viable platform to probe axion electrodynamics. The axion response can be viewed as a particular case of  bianisotropy~\cite{serdyukov2001electromagnetics,lindell1994electromagnetic} and corresponds to the long-sought Tellegen media~\cite{tretyakov2003artificial,yang2025gigantic}. 

A distinct entity is the dual axion response $\cd$ recently predicted in metamaterials~\cite{seidov2025dual} and in condensed matter systems~\cite{kokurina2026emergencedualaxionresponse}. Heuristically, $\cd$ can be viewed as a magnetic counterpart of the axion response as illustrated in Table~\ref{tab:electric_magnetic_sectors}. This linear response modifies Maxwell's equations in a manner complementary to the axion case and leads to the electrodynamics with magnetic charge. Both $\chi$ and $\cd$ responses are nonreciprocal, violate time reversal $\mathcal{T}$ and spatial inversion $\mathcal{P}$, keeping $\mathcal{PT}$ preserved.

\begin{table}[b!]
\centering
\begin{tabular*}{0.9\linewidth}{@{\extracolsep{\fill}}cc}
\hline
\textbf{Electric sector} & \textbf{Magnetic sector} \\
\hline
Permittivity $\varepsilon$ & Permeability $\mu$ \\
Electric gyrotropy $g_e$ & Magnetic gyrotropy $g_m$ \\
Axion response $\chi$ & Dual axion response $\tilde{\chi}$ \\
\hline
\end{tabular*}
\caption{Duality between electric and magnetic responses of continuous media}
\label{tab:electric_magnetic_sectors}
\end{table}

Here, we combine these entities together and investigate a {\it biaxionic medium} possessing both axion and dual axion responses captured by the following equations~\cite{seidov2025dual}
%\commentMaxim{I suggest using auxiliary $\vc{H}$ and $\vc{D}$ fields to make the dual symmetry of these equations explicit. This will justify why $\cd$ is magnetic counterpart of $\chi$ in Tab.1.}
%
\begin{align}
    \nabla \times \Big ( \mathbf{H} - \chi\, \mathbf{E} \Big) & =  \frac{1}{c}\frac{\partial}{\partial t} \Big( \mathbf{D} + \chi\, \mathbf{B}\Big)+ \frac{4\pi}{c}\,\vc{j}_{\text{e}}^{(\text{ren})} , \label{biaxioniceqs1} \\
     \nabla \cdot \Big ( \mathbf{D} + \chi\,\mathbf{B} \Big) & =  4\pi\, \rho_\text{e}^{(\text{ren})} ,\label{biaxioniceqs2}\\
    \nabla \times \Big( \mathbf{E} + \tilde{\chi}\,\mathbf{H}\Big) & =- \frac{1}{c}\frac{\partial}{\partial t} \Big( \mathbf{B} -\tilde{\chi}\, \mathbf{D} \Big) -\frac{4\pi}{c}\, \mathbf{j}_\text{m}^{(\text{ren})},\label{biaxioniceqs3} \\
      \nabla\cdot \Big( \mathbf{B}-\tilde{\chi}\, \mathbf{D} \Big) & = 4\pi\,\rho_{\text{m}}^{(\text{ren})}. \label{biaxioniceqs4}  
\end{align}
%old form of equations below
% \begin{align}
%      \nabla \cdot \Big ( \varepsilon \mathbf{E} + \chi \mathbf{B} \Big) & =  4\pi \rho_\text{e} ,\label{biaxioniceqs1}\\
%     \nabla \times \Big ( \mu^{-1} \mathbf{B} - \chi \mathbf{E} \Big) -  \frac{1}{c}\frac{\partial}{\partial t} \Big( \varepsilon \mathbf{E} + \chi \mathbf{B}\Big) & = \frac{4\pi}{c} \mathbf{J}_\text{e}, \label{biaxioniceqs2} \\
%       \nabla\cdot \Big( \mathbf{B}-\tilde{\chi} \varepsilon \mathbf{E} \Big) & = 0, \label{biaxioniceqs3} \\
%       \nabla \times \Big( \mathbf{E} + \tilde{\chi} \mu^{-1} \mathbf{B}\Big)+ \frac{1}{c}\frac{\partial}{\partial t} \Big( \mathbf{B} -\tilde{\chi} \varepsilon \mathbf{E} \Big) & = 0.\label{biaxioniceqs4} 
% \end{align}
%
Here, $\vc{j}_{\text{e}}^{\text{(ren)}}=\mathbf{j}_\text{e}+\chi\,\vc{j}_{\text{m}}$, $\rho_\text{e}^{(\text{ren})}=\rho_\text{e}+\chi\,\rho_{\text{m}}$, $\vc{j}_{\text{m}}^{\text{(ren)}}=\mathbf{j}_\text{m}-\cd\,\vc{j}_{\text{e}}$, $\rho_\text{m}^{(\text{ren})}=\rho_{\text{m}}-\cd\,\rho_{\text{e}}$ are renormalized source terms, while $\rho_{\text{e}}$, $\mathbf{j}_\text{e}$ ($\rho_{\text{m}}$, $\mathbf{j}_\text{m}$) are charge and current densities of the known external electric (magnetic) macroscopic sources exciting the structure. The fields $\vc{E}$, $\vc{D}$, $\vc{H}$ and $\vc{B}$ are the averaged (macroscopic) electric and magnetic fields obtained by filtering out rapidly varying contributions from the respective microscopic fields~\cite{Shaposhnikov2023,seidov2025dual} in line with the established theory~\cite{Agranovich1984,Silveirinha2007,Alu2011}. These fields are coupled via $\vc{D}=\eps\,\vc{E}$ and $\vc{B}=\mu\,\vc{H}$.

It should be stressed that the above effective description is rigorously derived from the microscopic Maxwell's equations for a metamaterial composed of gyrotropic layers as discussed in the Supplementary Section~1 and validated by the full-wave numerical simulations. In turn, renormalized source terms entering Eqs.~\eqref{biaxioniceqs1}-\eqref{biaxioniceqs4} correctly capture the interaction of the macroscopic sources with the metamaterial (Supplementary Section~2).

While equations~\eqref{biaxioniceqs1}-\eqref{biaxioniceqs4} look exotic, they lead to the  profound physical consequence. If the sources are absent ($\vc{J}_e=\vc{J}_m=0$, $\rho_e=\rho_m=0$), a redefinition of the fields $\vc{E}$, $\vc{H}$, $\vc{D}$ and $\vc{B}$ {\it inside the material} accompanied by the proper transformation of the material parameters $\eps$, $\mu$, $\chi$, $\cd$ leaves Eqs.~\eqref{biaxioniceqs1}-\eqref{biaxioniceqs4} covariant. This new symmetry relates the media with the same refractive index $n=\sqrt{\eps\mu}$ and holds even if the material parameters depend on time and coordinates.

Since the fields are redefined only inside the medium, this suggests the existence of an entire family of different materials which scatter electromagnetic waves {\it exactly} in the same way. %\red{The detailed derivation of this continuous mapping symmetry, its invariants, and the corresponding transformation group is provided in the Supplementary Section~1.} 
The parameters of such materials are related to each other via the continuous mapping transformations (Supplementary Section~3)
\begin{align}
& \chi(\alpha)=\frac{\chi+Y \tan\alpha}{1-Y\cd\tan\alpha}\:,\label{eq:Chitrans}\\
& \cd(\alpha)=\frac{\cd+Y^{-1} \tan\alpha}{1-Y^{-1}\,\chi\tan\alpha}\:,\label{eq:Cdtrans}\\
& Y(\alpha)=\frac{Y-\chi \tan\alpha}{1-Y\cd\tan\alpha}\:,\label{eq:Ytrans}
\end{align}
where $Y=\sqrt{\eps/\mu}$ is the admittance of the medium, $\chi$, $\cd$ and $Y$ is an initial set of the material parameters and $\alpha\in [0,\pi]$ is a continuous transformation parameter. 

The physical consequences of this symmetry are remarkable.
Equations~\eqref{eq:Chitrans}-\eqref{eq:Ytrans} have a set of invariants which do not depend on $\alpha$. Besides $n=\sqrt{\eps\mu}$ the invariants are
\begin{align}
& \eps_{T}=\frac{\eps+\mu\chi^2}{1+\chi\cd}\:,\mspace{4mu} \mu_T=\frac{\mu+\eps\cd^2}{1+\chi\cd}\:,\label{eq:epsmuinv}\\
& \theta=\frac{\chi\mu-\cd\eps}{1+\chi\cd}\:.\label{eq:extrinsic}
\end{align}
Because of the connection $\eps_T\,\mu_T=n^2+\theta^2$ only three of the invariants are independent. These invariants capture the material properties insensitive to the transformations Eqs.~\eqref{eq:Chitrans}-\eqref{eq:Ytrans}. Therefore, they can in principle be retrieved from the scattering experiments. The first invariant is the refractive index of the medium. $\eps_T$ and $\mu_T$ are some combinations of the permittivity and permeability imprinted in the scattering data. Finally, the parameter $\theta$ is a part of the nonreciprocal magnetoelectric coupling manifested in the scattering. We call it an {\it extrinsic} property.

Physically, this means that any scattering experiment with the above magnetoelectric medium can only measure a combination of axion and dual axion responses, but cannot extract their values separately. An especially interesting situation arises when 
\begin{equation}
    \frac{\cd}{\mu} = \frac{\chi}{\eps}\:.
\label{nonreciprocaldecoycondition}
\end{equation}
In such a case, extrinsic property vanishes: $\theta=0$, $\eps_T=\eps$ and $\mu_T=\mu$ and hence nonreciprocal magneto-electric effect becomes entirely invisible. We call Eq.~\eqref{nonreciprocaldecoycondition} an {\it invisibility condition}.

The analysis above shows that the scattering experiments define at most three independent material parameters out of the four $\eps$, $\mu$, $\chi$ and $\cd$. Therefore, there exists a material characteristic which never shows up in any kind of scattering experiment which we call an {\it intrinsic property}.

A natural choice of the intrinsic property is some function of the continuous transformation parameter $\alpha$. As the point $\alpha=0$ can be chosen arbitrarily, we require that $\alpha=0$ corresponds to $\cd/\mu=-\chi/\eps$. This ensures that the contributions of the axion and dual axion responses to the extrinsic property $\theta$ are equal to each other. With such a convention, a reasonable choice of the intrinsic property is 
\begin{equation}\label{eq:intrinsic}
\psi=\frac{\chi\mu+\cd\eps}{1-\chi\cd}=n\,\tan2\alpha\:,
\end{equation}
as further discussed in the Supplementary Section 3. Though unobservable in the scattering, the intrinsic property is real, as it affects the radiation of the localized sources when they are placed inside the medium.

Notably, the effective theory with $\eps$, $\mu$, $\chi$ and $\cd$ captures the scenario of the source-driven excitation, provided the sources have a macroscopic size exceeding the periodicity of the material and are placed in the region with no gradients of $\chi$ or $\cd$. Special care has to be taken while connecting the renormalized source terms $\vc{j}_{\text{e}}^{\text{(ren)}}$, $\vc{j}_{\text{m}}^{\text{(ren)}}$ to the actual external sources $\vc{j}_{\text{e}}$ and $\vc{j}_{\text{m}}$ exciting the structure. We derive such a connection microscopically in the Supplementary Section~2.

% formulating the boundary conditions for the fields at the surface of the external source immersed in the continuous medium. We derive such conditions directly from the microscopic theory in the Supplementary Materials (Secs.~6,7,9,10).

%\commentMaxim{Need to elaborate a bit more on the role of the sources} \red{In the microscopic multilayer realization, this coupling is sensitive to the position of the source within the unit cell. The source-induced bulk Witten contribution and the response encoded in the biaxionic boundary conditions compensate each other only for special source positions. For a generic position, this balance is violated, leaving a residual magnetoelectric response in the emitted field. Consequently, embedded charges can distinguish scattering-equivalent media and reveal their different axion and dual-axion content.}

To exemplify these general principles and illustrate the physics behind the intrinsic property, we consider the scattering of monochromatic plane wave from a homogeneous isotropic slab with the material parameters $\eps$, $\mu$, $\chi$ and $\cd$ at normal incidence choosing the $z$ axis normal to the slab. Equations~\eqref{biaxioniceqs1}-\eqref{biaxioniceqs4} yield a set of boundary conditions 
\begin{align}
& \vc{H}_{1t}-\chi_1\,\vc{E}_{1t}=\vc{H}_{2t}-\chi_2\,\vc{E}_{2t}\:,\label{eq:BC1}\\
& D_{1n}+\chi_1\,B_{1n}=D_{2n}+\chi_2\,B_{2n}\:,\label{eq:BC2}\\
& \vc{E}_{1t}+\cd_1\,\vc{H}_{1t}=\vc{E}_{2t}+\cd_2\,\vc{H}_{2t}\:,\label{eq:BC3}\\
& B_{1n}-\cd_1\,D_{1n}=B_{2n}-\cd_2\,D_{2n}\:,\label{eq:BC4}
\end{align}
where the subscripts $t$ and $n$ denote the field components tangential and normal to the boundary. Defining the reflection and transmission matrices as $\vc{E}^r=\hat{r}\,\vc{E}^{in}$, $\vc{E}^t=\hat{t}\,\vc{E}^{in}$, we find the cross-polarized scattering coefficients
\begin{align}
& t_{xy}=t_{yx}=0\:,\label{eq:txy}\\
& r_{xy} = - r_{yx}  = \frac{2}{\Delta}Z_0\, \bigg(Z^2\chi-\tilde{\chi}\bigg)\sin(k L)\label{Rxy:mixedaxionslab}
\end{align}
%
%& r_{xy} \hphantom{-} = - r_{yx}  = \frac{2}{\Delta(\chi,\cd)}\sqrt{\dfrac{\mu_0 }{\varepsilon_0}} \bigg(\frac{\mu}{\eps}\chi-\tilde{\chi}\bigg)\sin(k L)\label{Rxy:mixedaxionslab}
with
\begin{equation}
\begin{split}
\Delta = &  2i\,Z_0 Z \bigg(1 + \chi \cd \bigg)\cos(k L) \\
    & +\left( Z_0^2 + Z^2 + Z_0^2 Z^2 \chi^2+ \cd^2\right) \sin(k L). 
\end{split}
\end{equation}
%
%    \Delta(\chi,\cd) = &  2i\sqrt{\frac{\mu_0 \mu}{\varepsilon_0\varepsilon}} \bigg(1 + \chi \cd \bigg)\cos(k L) \\
    % & +\left( \frac{\mu_0}{\varepsilon_0} + \frac{\mu}{\varepsilon} + \frac{\mu_0 \mu}{\varepsilon_0\varepsilon  }\chi^2+ \cd^2\right) \sin(k L). 
Here, $k = q\, \sqrt{\mu \eps}$ is the wave number inside the slab, $q=\omega/c$, $L$ is the thickness of the slab, $Z=\sqrt{\mu/\eps}=1/Y$ is the impedance of the slab material, while $Z_0=\sqrt{\mu_0/\eps_0}$ is the impedance of the host medium. The expressions for the co-polarized scattering coefficients are derived in the same way and presented in the Supplementary Section~4. 

For clarity, we consider a linearly polarized incident wave. Equation~\eqref{eq:txy} then suggests that the transmitted light has the same polarization regardless of the ratio between $\chi$ and $\cd$. As demonstrated in the Supplementary Section~5, vanishing $t_{xy}$ is a consequence of the system $\mathcal{PT}$ symmetry.

At the same time, the cross-polarized reflection $r_{xy}$ is generally nonzero, describing Kerr rotation and providing a tool to experimentally diagnose nonreciprocal magnetoelectric effects~\cite{Armitage2016}. However, if the axion and dual axion responses satisfy the invisibility condition Eq.~\eqref{nonreciprocaldecoycondition}, Kerr rotation vanishes. Moreover, under the same condition, the co-polarized reflection and transmission coefficients become independent of $\chi$ and $\cd$, so that the complete reflection and transmission matrices of the biaxionic slab coincide with those of an ordinary isotropic dielectric with permittivity $\varepsilon$ and permeability $\mu$, embedded in the dielectric host medium. Thus, nonreciprocal  magnetoelectric response becomes invisible.
%(see Supplementary Section~2).

%and nonreciprocal magnetoelectric effect is no longer manifested neither in reflection, nor in  transmission in agreement with the general arguments above.

% More surprisingly, the condition Eq.~\eqref{nonreciprocaldecoycondition} ensures that $\chi$ and $\cd$ are not observable in any other kind of the scattering experiment or geometry including oblique incidence, far-field or near-field excitation. The reason for that is vanishing extrinsic property $\theta$, Eq.~\eqref{eq:extrinsic}. As a result, the medium scatters light similarly to the isotropic dielectric material with permittivity $\eps$ and permeability $\mu$, while its magnetoelectric responses become totally invisible. 

Moreover, the condition Eq.~\eqref{nonreciprocaldecoycondition} guarantees that the magnetoelectric medium satisfies the Lorentz reciprocity theorem in the generalized form, so that {\it all} nonreciprocal optical effects vanish (Supplementary Section 5). This provides an interesting example when the time-reversal symmetry of the medium is explicitly broken by $\chi$ and $\cd$ responses, but the reciprocity holds.

% Yet another striking feature of Eq.~\eqref{nonreciprocaldecoycondition} is generalized Lorentz reciprocity which is fulfilled despite breaking of the time-reversal symmetry by $\chi$ and $\cd$ responses. The reciprocity ensures that the transmission of light through the material in forward and backward direction is exactly the same, masking the signatures of $\chi$ and $\cd$.

% Even though Eq.~\eqref{nonreciprocaldecoycondition} renders nonreciprocal magnetoelectric effect purely intrinsic, still it can be probed experimentally by placing the sources inside the medium. Nontrivial effect in this setting is guaranteed by the nonzero intrinsic property Eq.~\eqref{eq:intrinsic}.

\begin{figure*}[t!]
    \centering
    \includegraphics[width=0.9\linewidth]{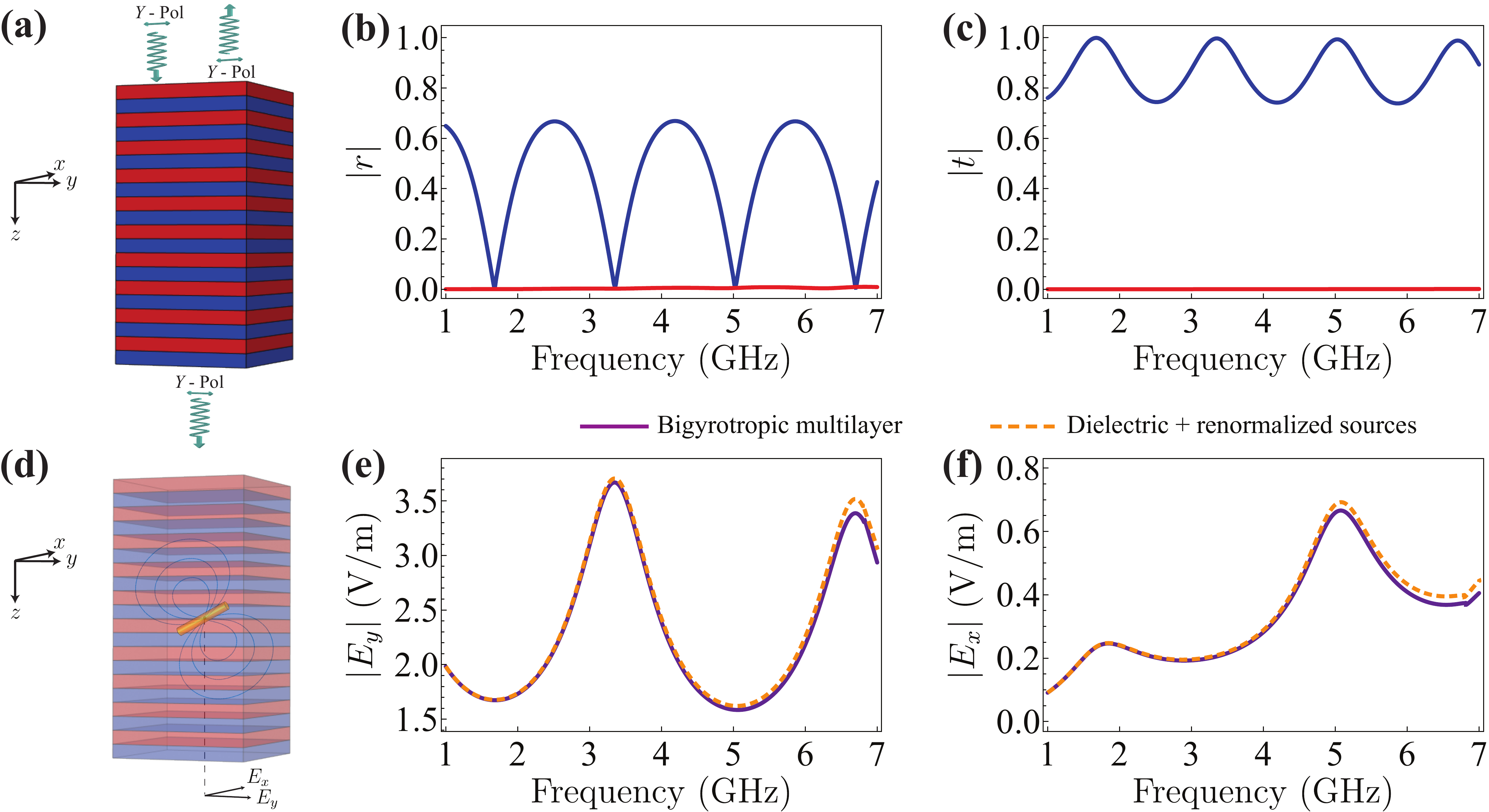}
    \caption{{\bf Probing the intrinsic electromagnetic property in a bigyrotropic multilayered metamaterial.} The structure consists of $20$ layers ($10$ unit cells) with alternating signs of electric and magnetic gyrotropies aligned normal to the layers, periodic boundary conditions of period $20\,\mathrm{mm}$ in the transverse $x$ and $y$ directions and lattice period along the $z$ axis $a=4\,\mathrm{mm}$. Material parameters $\varepsilon=5$, $\mu=1$, $g_e=4$, and $g_m=0.8$ are chosen to satisfy the invisibility condition. Period-to-wavelength ratio $0.013<\xi<0.093$.
    (a) The scheme of plane-wave excitation by $y$-polarized light at normal incidence.
    (b,c) Amplitude reflection and transmission coefficients. Blue and red lines denote co- and cross-polarized components, respectively. Cross-polarized components vanish within numerical accuracy. 
    (d) Excitation by an electric dipole antenna centered at $z=0$ and embedded in the slab. The antenna lies in the $Oyz$ plane, has length $8.94\,\mathrm{mm}$, is tilted by $26.57^\circ$ with respect to the $y$ axis, and carries a current amplitude of $1\,\mathrm{mA}$. The transverse periodic boundary conditions correspond to an infinite
    array of in-phase dipole antennas.
    (e,f) Far-field amplitudes of the co-polarized component $|E_y|$ and the cross-polarized component $|E_x|$, respectively. Purple solid curves show numerical calculation for the actual metamaterial, orange dashed curves are the result of the effective theory. Cross-polarized signal reveals the intrinsic property. 
    }
\label{fig:experimentalscheme3d}
\end{figure*}

Next, we present a specific structure whose effective description is captured by Eqs.~\eqref{biaxioniceqs1}-\eqref{biaxioniceqs4}. We consider a multilayered metamaterial with the layers that combine electric and magnetic gyrotropy, having the constitutive relations of the form
\begin{align}
    \mathbf{D} & = \varepsilon\,\mathbf{E} - i g_e\, \bas \times \mathbf{E}, & \mathbf{B} & = \mu\, \mathbf{H} -  i g_m\, \bas \times \mathbf{H},
\end{align}
where $\bas$ is a unit vector normal to the layers, while the signs of $g_e$ and $g_m$ change from layer to layer ensuring that the overall magnetization is compensated, Fig.~\ref{fig:experimentalscheme3d}(a). Such bigyrotropic response occurs naturally in ferrite garnets, including yttrium
iron garnet (YIG) and Bi-substituted iron garnets, which exhibit simultaneous electric and magnetic gyrotropy in the near-infrared spectral range~\cite{zamani2016transfer,gribova2024inverse}.

%\commentMaxim{To Eduardo: find examples of natural bi-gyrotropic materials and add the relevant references}

%microscopic bigyrotropic multilayer excited by the electric antenna, whereas orange dashed curves show the equivalent dielectric excited by the mapped electric--magnetic source pair. Their agreement validates the source mapping, while the finite cross-polarized emission reveals the intrinsic magnetoelectric response.

When the period $a$ of the structure is much smaller than the wavelength $\lambda$, i.e. $\xi=a/\lambda\ll1$, the electrodynamics of such a metamaterial is captured by the effective permittivity and permeability
\begin{align}
    \varepsilon_{\mathrm{eff}}
    &=
    \varepsilon
    +
    \frac{\pi^2}{12}
    g_e\left(\mu g_e-\varepsilon g_m\right)\xi^2+O(\xi^4), \\ 
    \mu_{\mathrm{eff}}
    &=
    \mu
    -
    \frac{\pi^2}{12}
    g_m\left(\mu g_e-\varepsilon g_m\right)\xi^2+O(\xi^4),
\end{align}
and two emergent responses~\cite{Shaposhnikov2023,seidov2025dual}:
\begin{align}
    \chi & = -\frac{\pi}{2} g_e \,\xi+O(\xi^3), & \cd & =- \frac{\pi}{2} g_m \, \xi+O(\xi^3), 
\end{align}
which realize the physics discussed above. If electric and magnetic gyrotropies are chosen as
\begin{equation}
    \frac{g_m}{\mu} = \frac{g_e}{\eps}\:,
\label{intrinsic_gyrotropy_condition}
\end{equation}
this fulfills the invisibility condition Eq.~\eqref{nonreciprocaldecoycondition} and renders the metamaterial indistinguishable from a dielectric slab with the same $\varepsilon$ and $\mu$ in scattering experiments. Using full-wave numerical simulations (COMSOL Multiphysics), we compute the reflection and transmission coefficients for a finite slab consisting of $10$ unit cells tuned to satisfy the invisibility condition. The results are shown in Figs.~\ref{fig:experimentalscheme3d}(b,c).

In agreement with the effective theory above, the cross-polarized reflection and transmission vanish within numerical accuracy, while the co-polarized scattering coincides with that for an ordinary dielectric slab, even though the individual layers explicitly break the time-reversal symmetry. These scattering properties can be accurately computed using the transfer matrix method presented in the Supplementary Section~6.

%The microscopic bigyrotropic multilayer therefore realizes the scattering invisible biaxionic response predicted analytically. The exact transfer-matrix calculation is presented in Supplementary Section~4.

While the analysis above applies to the metamaterial regime with the period-to-wavelength ratio $\xi=a/\lambda<0.1$, the intrinsic property remains invisible in the photonic crystal regime $\xi\sim 1$ as well. Studying a multilayered photonic crystal satisfying the same condition Eq.~\eqref{intrinsic_gyrotropy_condition}, but operating at higher frequencies, we find that both the cross-polarized transmission and the cross-polarized reflection remain strictly zero showing no signatures of the intrinsic property. As further elaborated in the Supplementary Section~7, this exact compensation comes from the microscopic structure of a single unit cell fine-tuned to satisfy Eq.~\eqref{intrinsic_gyrotropy_condition}.

% To verify that, we study a multilayered photonic crystal satisfying the same condition Eq.~\eqref{intrinsic_gyrotropy_condition} but operating at higher frequencies. Counterintuitively, we find out that both cross-polarized transmission and cross-polarized reflection remain strictly zero showing no signatures of intrinsic property.

% \red{This exact compensation originates at the microscopic level of a complete unit cell: the condition Eq.~\eqref{intrinsic_gyrotropy_condition} makes the layer interfaces polarization independent, while the polarization-dependent propagation phases accumulated in the two oppositely gyrotropic layers cancel over one period.
% Consequently, each complete unit cell is scattering-equivalent to an ordinary
% dielectric layer. The detailed derivation is provided in Supplementary
% Section~5.} 

To reveal the intrinsic property of the metamaterial, we excite the same structure by a periodic array of in-phase electric dipole antennas positioned in the $Oyz$ plane, Fig.~\ref{fig:experimentalscheme3d}(d). The size of the antennas is chosen such that they remain subwavelength, but nevertheless cross several metamaterial layers, ensuring that the effective source is macroscopic. 
%Such relatively large size of the source ensures that the effective medium description remains adequate, while the source can be viewed as  embedded inside the continuous medium. 
Inspecting the fields produced by this array outside the metamaterial, we observe that besides the co-polarized component of the electric field [Fig.~\ref{fig:experimentalscheme3d}(e)] there is a sizable cross-polarized contribution shown in Fig.~\ref{fig:experimentalscheme3d}(f) by the purple solid line. 
Since an electric source embedded
in an isotropic dielectric cannot generate such fields due to symmetry reasons, the appearance of the cross-polarized component in Fig.~\ref{fig:experimentalscheme3d}(f) is the direct evidence of the mirror symmetry breaking in the $Oyz$ plane at the microscopic level. At the same time, our effective description perfectly captures this physics, with the predictions shown by the orange dashed line in Fig.~\ref{fig:experimentalscheme3d}(e,f). Taken together, this evidences the intrinsic property $\psi$, which breaks the mirror symmetry of the problem and reveals itself via source-driven excitation.

The limits of the effective medium description can be probed  further by studying a simpler geometry of the multilayered metamaterial excited by a planar electric source parallel to the $Oxy$ plane [Fig.~\ref{fig:experimentalscheme3d}(d)]. If the source has zero thickness, the radiated field strongly depends on the source position within the unit cell and therefore captures microscopic details beyond the effective theory. However, once the thickness of the source approaches the period of the structure $a$, the dependence on its position within the unit cell disappears, while the cross-polarized signal persists and agrees with our effective theory. This analysis and supporting simulations are presented in the Supplementary Sections~8-10.

The examples above examine the intrinsic electromagnetic property in infinite periodic structures with a planar geometry. However, the same physics arises in finite systems as well. To test that, we examine a spherical scatterer with a void in its center constructed from radially magnetized gyrotropic layers with alternating magnetization sign and the constitutive relations of the form
\begin{align}
    \mathbf{D} &= \varepsilon\, \mathbf{E} \pm i g_e \,\vc{e}_r \times \mathbf{E}, &
    \mathbf{B} &= \mu\, \mathbf{H} \pm i g_m \,\vc{e}_r \times \mathbf{H},
\end{align}
where $\vc{e}_r$ is a unit vector in the radial direction. The gyrotropies of the layers are tuned to satisfy the condition~\eqref{intrinsic_gyrotropy_condition}.

%This system realizes biaxionic response tuned to satisfy the invisibility  condition Eq.~\eqref{nonreciprocaldecoycondition}.

\begin{figure}[t!]
    \centering
    \includegraphics[width=0.90\linewidth]{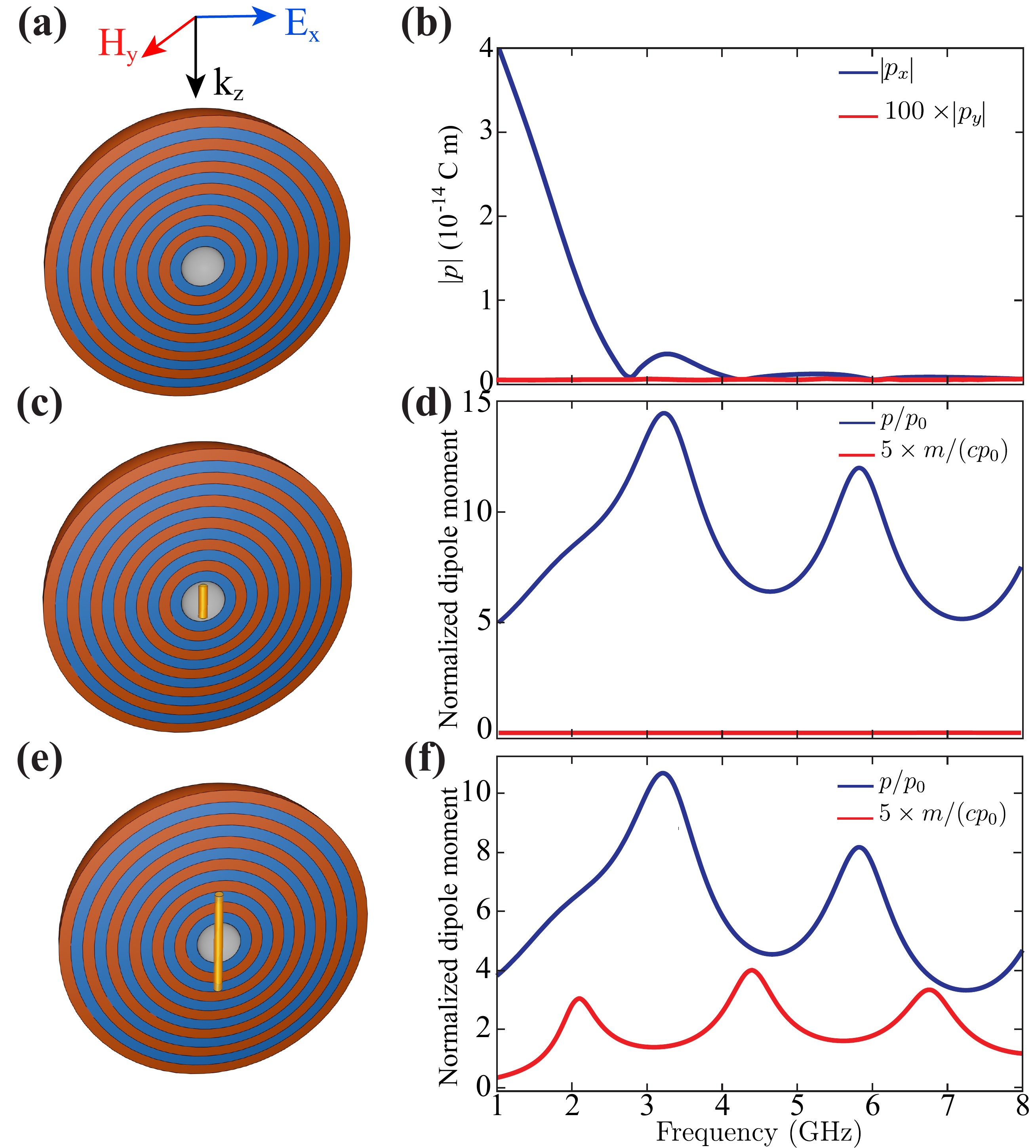}
    \caption{{\bf Probing intrinsic electromagnetic property in finite systems.} The sphere has a void in the center and consists of 12 radially magnetized gyrotropic shells (6 unit cells) with the alternating magnetization and parameters
    $\varepsilon=5$, $\mu=1$, $g_e=4$, and $g_m=0.8$, satisfying the invisibility condition Eq.~(\ref{nonreciprocaldecoycondition}). (a) Excitation by an $x$-polarized plane wave with amplitude $E_0=1~\mathrm{V/m}$ propagating along $Oz$ axis. (b) Multipole content of the scattered field reveals only $x$-oriented electric dipole (blue line) with no electric dipole along $y$ (red line).
    (c,d) Excitation by electric dipole antenna fully confined inside the central void. The far-field radiation shows only collinear electric dipole component (blue line), with no magnetic dipole contribution (red line). 
    (e,f) Excitation by electric dipole antenna partially embedded in the biaxionic sphere: far-field radiation features both electric dipole (blue line) and magnetic dipole (red line) contributions.}
    \label{fig:sphericalexp}
\end{figure}

First, we simulate the scattering of an $x$-polarized plane wave incident at the sphere along the $z$ axis [Fig.~\ref{fig:sphericalexp}(a)] and extract the induced electric dipole moment of the sphere. % We observe that the incident wave excites only $x$-oriented electric dipole moment  [Figs.~\ref{fig:sphericalexp}(b)],  exactly as for an ordinary dielectric sphere. Though the axion or dual axion response normally yield an additional $y$ component of the dipole moment~\cite{Shaposhnikov2023} (\blue{see Supplementary Section 4}), here it vanishes identically.
The incident wave excites only an $x$-oriented dipole moment [Fig.~\ref{fig:sphericalexp}(b)], exactly as for an ordinary dielectric sphere. Although an uncompensated axion or dual-axion response would generally produce a transverse $y$-oriented dipole moment~\cite{Shaposhnikov2023}, this component vanishes under the invisibility condition (see also Supplementary Section 11).

% No cross-polarized component is generated, and no signature of $\mathcal{T}$-symmetry breaking appears in the scattered field. This behavior contrasts sharply with axion spheres or multilayer spheres possessing an effective unbalanced magnetoelectric response, which generically develop a cross-polarized dipole moment under the same illumination (see SM Sec.~4~\cite{Supplement}). Thus, despite its microscopic $\mathcal{T}$-breaking structure, the biaxionic sphere satisfying the scattering invisibility condition remains completely indistinguishable from a reciprocal dielectric in any scattering experiment.

Next we turn to the source-driven excitation. If the dipole antenna is placed entirely inside the central void and does not intersect the
metamaterial boundaries [Figs.~\ref{fig:sphericalexp}(c,d)], the radiated field features only electric dipole component parallel to the antenna and no collinear magnetic dipole contribution. In this configuration, the source does not couple to the intrinsic magneto-electric response remaining effectively  outside of the material.

A very different physics is observed when the antenna  penetrates into the layers and hence can be viewed as being {\it inside} the metamaterial [Figs.~\ref{fig:sphericalexp}(e,f)]. The simulations reveal that the outside field acquires not only an electric dipole component, but also a magnetic dipole contribution in the same frequency range, and both induced dipoles are parallel to the antenna. This provides an instance of the Witten effect studied previously for spherical meta-atoms with the axion response~\cite{Shaposhnikov2023,Seidov2023}.

Thus, the interaction with the sources inside the material reveals the intrinsic electromagnetic property $\psi$, and in this sense the intrinsic property is real. The finite size of the scatterer highlights yet another conceptual distinction of this physics from the celebrated bound states in the continuum (BICs). While the intrinsic property is  geometry-independent effect manifested both in the infinite and finite structures, perfect electromagnetic BICs are supposed to occur in the infinite systems~\cite{Hsu2016}.
%acquiring a finite quality factor in any finite geometry.

The above simulation reveals a strong difference between the situations when an external source is positioned outside of the medium and inside it, which looks counterintuitive. This topological distinction is explained by the fact that both axion and dual axion responses manifest themselves only at the boundaries of the material, and hence the position of the source relative to the boundary becomes crucial.

% a finite magnetic dipole moment emerges in the far field.
% This magnetic response has no counterpart in conventional dielectrics and directly reflects the
% intrinsic magnetoelectric coupling of the biaxionic medium.
% Crucially, it is activated only by sources embedded within the material and remains completely
% inaccessible to any external probe.

% These results demonstrate that even in compact spherical geometry, a biaxionic medium satisfying
% the invisibility condition exhibits no manifestation of $\mathcal{T}$-symmetry breaking in scattering,
% while revealing its intrinsic electromagnetic property exclusively through internal excitation.
% This establishes the effect as geometry-independent and fundamentally distinct from other
% mechanisms of electromagnetic invisibility in photonics.

\begin{figure*}[t!]
  \centering
	\includegraphics[width=0.85\textwidth]{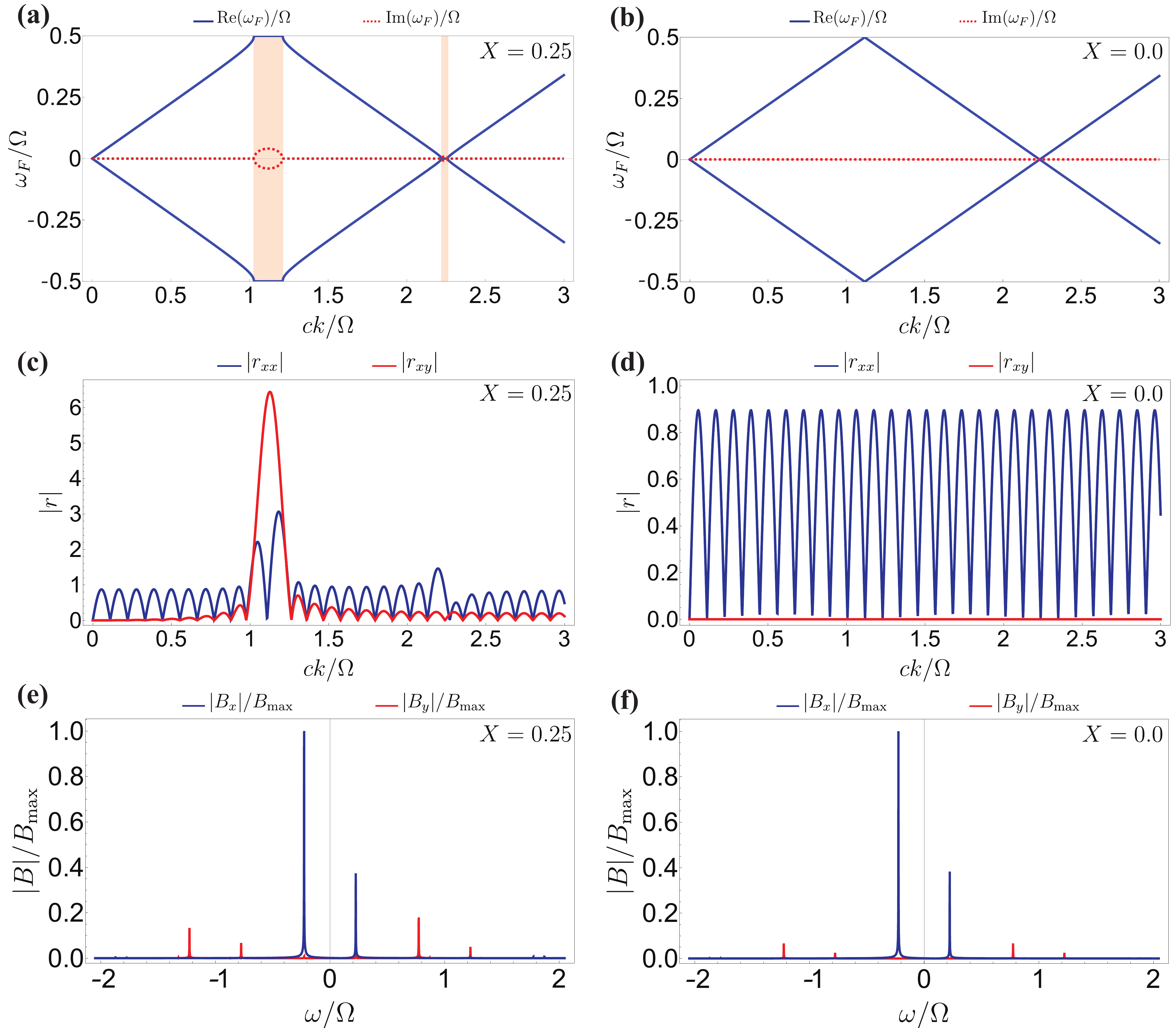}
  \caption{{\bf Photonic band structure, scattering coefficients and the spectrum of the field inside time-modulated biaxionic medium.} (a,c) Photonic band structure (a) and the spectrum of temporal reflection (c) when invisibility condition is not met $(X=0.25)$. The spectrum develops momentum gaps  accompanied by strong cross-polarized scattering. (b,d) When invisibility condition is fulfilled $(X=0.0)$, momentum gaps close and the scattering becomes indistinguishable from the conventional dielectric with vanishing cross-polarized signal. (e,f) Fourier spectrum of the magnetic field inside the medium for an incident $x$-polarized excitation. The calculations are performed for $\varepsilon=5$, $\mu=1$, and $\chi_0=0.3$. The modulated structure in panels (c,d) consists of 10 modulation periods, while the Fourier spectra in panels (e,f) are computed from the bulk evolution over 1000 modulation periods to improve the spectral resolution.}  
\label{Fig:BiaxionicTimeVarying}
\end{figure*}

So far, the intrinsic property which we studied was just a single number protected from the outside measurement, which may look of little use. Much richer physics arises when the intrinsic property becomes time-dependent with a prospect to communicate some information hidden from the outside observer. Such setting connects our problem to time-varying media and photonic time
crystals~\cite{Shapiro2023,Moussa2023,Lustig2023,Wang2024}, which normally deal with time-dependent refractive index. Recent studies~\cite{Xu2025,Shaposhnikov2025} suggest the feasibility of time-dependent axion response as well.

%Conceptually the simplest example of time-dependent properties is a stepwise switching of $\chi$ and $\cd$ on and off. Such change of the material parameters affects the propagation of waves in the medium generating time-transmitted and time-reflected waves. As shown in the Supplementary Section 12, cross-polarized time-scattering vanishes identically under the same invisibility condition Eq.~\eqref{nonreciprocaldecoycondition}.

Conceptually, the simplest example of time-dependent properties is a stepwise switching of $\chi$ and $\tilde{\chi}$ on and off. Such changes generate time-transmitted and time-reflected waves, as studied previously for temporal interfaces in bianisotropic media~\cite{Mirmoosa2024}. As shown in Supplementary Section~12, such stepwise switching of axion and dual axion responses produces no cross-polarized scattering under the same invisibility condition Eq.~\eqref{nonreciprocaldecoycondition}.

% \red{Importantly, rapid oscillations of a spatially uniform magnetization have recently been shown to generate synthetic dynamic axion fields, providing a concrete physical route toward
% time-dependent axionic responses and motivating the biaxionic generalization considered below.}

More interesting situation occurs if the axion and dual axion responses are modulated periodically, for instance as
\begin{equation}
    \chi(t)=\chi_0\cos \Omega t,\qquad
    \tilde{\chi}(t)=\tilde{\chi}_0\cos\Omega t,
\end{equation}
while the permittivity $\eps$ and permeability $\mu$ stay constant.

%In line with that, we consider a homogeneous biaxionic medium with constant $\varepsilon$ and $\mu$, but periodically modulated axion and dual-axion responses,

Depending on the ratio between the amplitudes $\chi_0$ and $\cd_0$, the extrinsic magneto-electric property can be suppressed or eliminated entirely. To quantify the deviation of the parameters from the invisibility condition, we present $\cd_0$ as
\begin{equation}
    \tilde{\chi}_0=\left(\frac{\mu}{\varepsilon}+X\right)\chi_0 ,
\end{equation}
so that $X=0$ corresponds to the invisibility regime Eq.~\eqref{nonreciprocaldecoycondition}, while $X\neq0$ enables an extrinsic magnetoelectric response detectable in the scattering.  

% \red{The same
% invisibility condition remains valid for scattering through finite temporal biaxionic slabs, where the cross-polarized
% scattering amplitudes vanish exactly under the invisibility condition
% (see Supplementary Section~11).}

Even at $X=0$, the equations describing electrodynamics inside the material are nontrivial: the time-dependent intrinsic property produces polarization mixing and temporal sideband generation inside the medium, as analyzed further in the Supplementary Section~13.

%The bulk equations for this time-varying biaxionic medium are nontrivial even for $X=0$: the fields obey the differential equations with the time-dependent coefficients, leading in general to sideband generation and polarization mixing inside the material. \red{The corresponding temporal boundary and finite temporal-slab scattering problems are analyzed in Supplementary Section~8, while the general periodically modulated biaxionic dynamics and Floquet formulation are presented in Supplementary Section~9.} %(\blue{see Supplementary Sec.~9}).

Assuming periodic modulation, we compute the  Floquet spectrum $\omega_F(k)$ for various wave numbers $k$. If $X\not=0$, the modulation opens momentum gaps accompanied by the nonzero imaginary part of $\omega_F$ [Fig.~\ref{Fig:BiaxionicTimeVarying}(a)] and strong cross-polarized temporal reflection [Fig.~\ref{Fig:BiaxionicTimeVarying}(c)]. The enhanced temporal reflection is related to the wave amplification within the momentum gap, drawing energy from the modulation~\cite{Lyubarov2022,wang2023metasurface,Allard2026}. Recently, such modulation-driven optical amplification has been explored in a variety of time-varying photonic systems~\cite{Silveirinha2025Plasmon,verde2026optical,galiffi2026optical}.

% \red{The resulting biaxionic medium behaves as a photonic time crystal whose bulk dynamics can be decomposed into independent polarization channels. For the symmetric modulation considered here, both circular polarizations produce identical Floquet spectra, and therefore only one branch is shown in Fig.~\ref{Fig:BiaxionicTimeVarying}(a,b). For finite detuning $X\neq0$, the modulation opens momentum bandgaps, identified by the nonzero imaginary part of $\omega_F$ [Fig.~\ref{Fig:BiaxionicTimeVarying}(a)]. These instability gaps are accompanied by strong cross-polarized temporal reflection [Fig.~\ref{Fig:BiaxionicTimeVarying}(c)].}

In contrast, at the invisibility condition $X=0$, the Floquet gaps close and the quasifrequencies remain purely real [Fig.~\ref{Fig:BiaxionicTimeVarying}(b)], while the cross-polarized time-reflection vanishes identically [Fig.~\ref{Fig:BiaxionicTimeVarying}(d)]. Consequently, the scattered field outside the medium becomes indistinguishable from that for the conventional unmodulated dielectric, remaining strictly monochromatic despite the temporal modulation of $\chi$ and $\cd$.

%Assuming the periodic modulation, we compute the one-period propagator of the bulk equations and extract the Floquet quasifrequencies $\omega_F(k)$.  When $X\neq0$, the modulation opens Floquet momentum gaps, identified by the nonzero imaginary part of $\omega_F$ [Fig.~\ref{Fig:BiaxionicTimeVarying}(a)], and these gaps are accompanied by the strong cross-polarized reflection [Fig.~\ref{Fig:BiaxionicTimeVarying}(b)]. In contrast, at invisibility condition $X=0$ the Floquet gaps close [Fig.~\ref{Fig:BiaxionicTimeVarying}(d)] and the cross-polarized reflection coefficient vanishes [Fig.~\ref{Fig:BiaxionicTimeVarying}(e)]. It can be verified further that the scattered signal outside of the medium also remains strictly monochromatic showing no sidebands (\blue{Supplementary Section 9}). \red{Correspondingly, the scattered field outside the medium remains monochromatic and indistinguishable from that of a conventional dielectric slab, without observable sideband generation (Supplementary Sec.~9).}

Nevertheless, the disappearance of the observable scattering signatures does not imply trivial internal dynamics. Figs.~\ref{Fig:BiaxionicTimeVarying}(e,f) show the Fourier spectrum of the magnetic field inside the time-modulated medium obtained from bulk propagation over $1000$ modulation periods. Even at $X=0$, where the extrinsic property $\theta$ vanishes, the field inside the medium develops pronounced temporal sidebands together with a finite cross-polarized component $B_y$, revealing ongoing polarization mixing and nontrivial internal evolution.

Physically these results mean that if an outside observer shines a monochromatic light on the medium with the time-varying intrinsic property, but satisfying the invisibility condition Eq.~\eqref{nonreciprocaldecoycondition}, they will never recover any traces of the wave process happening in the medium: neither in the form of the cross-polarized scattering, nor in the form of the sidebands.
%\red{Nevertheless, the electromagnetic field inside the medium continues to exhibit rich temporal evolution and polarization dynamics.}

Thus, the same invisibility condition that hides the static intrinsic property also suppresses the signatures of the time-varying field $\psi(t)$ or, more generally, space-time-dependent field of intrinsic properties $\psi(\vc{r},t)$. 
%Space and temporal modulation provides control over wave propagation and frequency conversion in space--time photonic crystals and metasurfaces~\cite{guo2019nonreciprocal,sisler2024electrically,caballero2026interface}. 
If the modulation of intrinsic electromagnetic property is used to communicate the information, this communication channel is {\it fundamentally protected} from any kind of remote sensing. The only possibility to extract the intrinsic property in this setting is to access the channel physically by inserting there the external probes. The described mechanism is different from the conventional waveguiding, as the probing light can freely enter the modulated medium, but still does not capture its intrinsic property.

In summary, we have predicted the existence of an intrinsic electromagnetic property which is fundamentally unobservable in any kind of scattering experiment and can only be probed by inserting external sources directly inside the material. This theoretical prediction has several important implications.

From the fundamental perspective, this projects the philosophical discourse about Kantian things in themselves onto fruitful physical ground revealing a novel previously unknown class of material responses. While this work introduces only a single example of an intrinsic electromagnetic property in nonreciprocal  magnetoelectric medium, we are convinced that this phenomenon is very general. Similarly to the special relativity rooted in the covariance of the vacuum Maxwell's equations under Lorentz transformations, the notion of intrinsic properties is rooted in the covariance of Maxwell's equations in the medium under the continuous group of mapping transformations. It is very likely that the effective description of other materials features the same kind of symmetry enabling other types of intrinsic properties.

%are convinced that the similar effect arises in the broad class of structures. 

%stemming from the symmetry of the effective medium description.  we are convinced that a similar effect arises in other structures, driven by the covariance of the effective theory under field redefinition. 

Furthermore, magnetoelectric metamaterials studied here are well within the reach of the present-day experiments~\cite{yang2025gigantic,Asadchy2026,Yang2026} so that we are optimistic about the prospects to detect intrinsic electromagnetic properties experimentally in the near future.

From the more applied perspective, the concept of intrinsic electromagnetic properties provides a novel fundamental mechanism of protection against remote sensing which can be harnessed for secure communication providing an alternative principle to existing quantum communication approaches.

Finally, this study pushes us to rethink the role of the scattering experiments in the exploration of nature showing that the information obtained in this way is incomplete at least in some instances. The implications of this conclusion could span well beyond photonics and condensed matter. 
%touching such areas as, for example, astrophysics, where the observational data is the only source of information.

% of a homogeneous material, which does not show up in any scattering experiment: neither for the plane wave illumination, nor for the excitation by the outside  sources. However, this intrinsic axion response is clearly manifested when the material is excited by the sources introduced inside. The predicted effect is distinct from the other concepts in photonics and provides a fundamental protection from remote sensing~[See End Matter section for details]. We anticipate an entire family of other intrinsic electromagnetic properties.

% Our study demonstrates that the information about physical properties of matter retrieved only from the scattering experiments is incomplete, which has vast implications far beyond condensed matter and photonics. On a more abstract level, the discussed response can be viewed as a physical example of the Kantian ``thing in itself". \commentMaxim{Thoughts. 1. Discuss potential experimental implementation. 2. There could be other intrinsic properties, e.g. in reciprocal structures. 3. The information obtained by the observations from the distance with no direct experiments is incomplete~-- interesting parallels with astrophysics.}

\begin{acknowledgments}
%{\it Acknowledgments.~--~}
Numerical simulations were supported by Priority 2030 Federal Academic Leadership Program. Theoretical models were supported by the Russian Science Foundation, grant No.~26-72-10054. E.B.-A. acknowledges partial support from the Foundation for the Advancement of Theoretical Physics and Mathematics ``Basis'' and from SECIHTI (México) doctoral scholarship.
\end{acknowledgments}

\bibliography{decoys.bib}

\pagebreak
\widetext

\clearpage
% The empty curly braces force a clean structural break before page 1


\section*{Supplementary material}

# Supplemental Materials: Intrinsic electromagnetic properties

Eduardo Barredo-Alamilla,[1,*] Timur Z. Seidov,[1,*] Daniel A. Bobylev,[1] and Maxim A. Gorlach[1,†]
[1] School of Physics and Engineering, ITMO University, Saint Petersburg 197198, Russia

## CONTENTS



* These authors contributed equally to this work
† m.gorlach@metalab.ifmo.ru

## S1. EMERGENT BIAXIONIC RESPONSE IN A PERIODIC BIGYROTROPIC MULTILAYER

Spatially periodic multilayered metamaterial with electric and magnetic gyrotropies of the layers and antiferromagnetic magnetization pattern normal to the layers provides a microscopic realization of the medium with effective biaxionic response. The effective description for such structure was derived in Ref. [1] in the first order of the perturbation theory with respect to the small parameter $\xi = a/\lambda$, where $a$ is the period of the structure and $\lambda$ is the wavelength of radiation. Here we revisit that derivation and extend it to the second order in period-to-wavelength ratio, obtaining the corrections to bulk effective permittivity and permeability. Throughout the derivation we assume that the external charges and currents are absent. We first retain arbitrary periodic gyrotropy profiles and then focus on the specific profile corresponding to the layers of equal thickness and magnetization alternating from layer to layer. We restrict the analysis to the normal incidence $\mathbf{k} = k_z \mathbf{e}_z$, so that all fields are transverse.

We use the time dependence $e^{-i\omega t}$ and Maxwell's equations in the form

$$\nabla \times \mathbf{E} = iq\mathbf{B}, \qquad \nabla \times \mathbf{H} = -iq\mathbf{D}, \qquad q = \frac{\omega}{c}. \tag{S1}$$

The microscopic constitutive tensors are

$$\hat{\varepsilon}(z) = \begin{pmatrix} \varepsilon & ig(z) & 0 \\ -ig(z) & \varepsilon & 0 \\ 0 & 0 & \varepsilon \end{pmatrix}, \qquad \hat{\mu}(z) = \begin{pmatrix} \mu & if(z) & 0 \\ -if(z) & \mu & 0 \\ 0 & 0 & \mu \end{pmatrix}. \tag{S2}$$

Here $\varepsilon$ and $\mu$ are independent of coordinate $z$, while $g(z)$ and $f(z)$ are periodic with period $a$ and zero averages. We define

$$b = \frac{2\pi}{a}, \qquad \lambda = \frac{2\pi}{q}, \qquad \xi = \frac{q}{b} = \frac{a}{\lambda} \ll 1. \tag{S3}$$

We introduce the operator

$$\mathbf{e}_z^{\times}\mathbf{v} = \mathbf{e}_z \times \mathbf{v}, \qquad (\mathbf{e}_z^{\times})^2 = -I_t, \tag{S4}$$

where $I_t$ is the identity in the transverse plane. The constitutive relations then read

$$\mathbf{D} = \varepsilon\mathbf{E} - ig(z)\mathbf{e}_z^{\times}\mathbf{E}, \qquad \mathbf{B} = \mu\mathbf{H} - if(z)\mathbf{e}_z^{\times}\mathbf{H}. \tag{S5}$$

We use the Fourier expansions

$$g(z) = \sum_{n\neq 0} g_n e^{inbz}, \qquad g_n = \frac{1}{a}\int_0^a g(z) e^{-inbz}\, dz, \tag{S6}$$

$$f(z) = \sum_{n\neq 0} f_n e^{inbz}, \qquad f_n = \frac{1}{a}\int_0^a f(z) e^{-inbz}\, dz, \tag{S7}$$

together with the Floquet-Bloch expansion of the fields

$$\mathbf{A}(z) = e^{ik_z z}\sum_{n\in\mathbb{Z}} \mathbf{A}_n e^{inbz}, \qquad \mathbf{A} \in \{\mathbf{E}, \mathbf{H}, \mathbf{D}, \mathbf{B}\}. \tag{S8}$$

The subscript zero denotes the zeroth Floquet harmonic, which corresponds to the averaged (macroscopic) field in the metamaterial.

Writing $K_n = k_z + nb$, Maxwell's equations become

$$K_n \mathbf{e}_z^\times \mathbf{E}_n = q\mathbf{B}_n, \qquad K_n \mathbf{e}_z^\times \mathbf{H}_n = -q\mathbf{D}_n. \tag{S9}$$

For $n \neq 0$, these equations yield

$$\mathbf{E}_n = -\frac{q}{K_n}\mathbf{e}_z^\times \mathbf{B}_n, \qquad \mathbf{H}_n = \frac{q}{K_n}\mathbf{e}_z^\times \mathbf{D}_n. \tag{S10}$$

Separating the zeroth-harmonic driving term from the remaining convolution, we obtain

$$\mathbf{B}_n = \mu\mathbf{H}_n - if_n\mathbf{e}_z^\times \mathbf{H}_0 - i\sum_{\ell\neq 0,n} f_{n-\ell}\,\mathbf{e}_z^\times \mathbf{H}_\ell, \tag{S11}$$

$$\mathbf{D}_n = \varepsilon\mathbf{E}_n - ig_n\mathbf{e}_z^\times \mathbf{E}_0 - i\sum_{\ell\neq 0,n} g_{n-\ell}\,\mathbf{e}_z^\times \mathbf{E}_\ell. \tag{S12}$$

Combining these relations with Eq. (S10) and using $(\mathbf{e}_z^\times)^2 = -I_t$, we find

$$\mathbf{E}_n = -\frac{iqf_n}{K_n}\mathbf{H}_0 - \frac{q\mu}{K_n}\mathbf{e}_z^\times \mathbf{H}_n - \frac{iq}{K_n}\sum_{\ell\neq 0,n} f_{n-\ell}\mathbf{H}_\ell, \tag{S13}$$

$$\mathbf{H}_n = \frac{iqg_n}{K_n}\mathbf{E}_0 + \frac{q\varepsilon}{K_n}\mathbf{e}_z^\times \mathbf{E}_n + \frac{iq}{K_n}\sum_{\ell\neq 0,n} g_{n-\ell}\mathbf{E}_\ell. \tag{S14}$$

Expanding up to the second order, we recover

$$\frac{q}{K_n} = \frac{q}{nb} - \frac{qk_z}{n^2b^2} + O(\xi^3). \tag{S15}$$

Then we can calculate the fields as

$$\mathbf{E}_n = -\frac{iqf_n}{nb}\mathbf{H}_0 + \frac{iqk_zf_n}{n^2b^2}\mathbf{H}_0 - \frac{i\mu q^2 g_n}{n^2b^2}\mathbf{e}_z^\times \mathbf{E}_0 + \frac{q^2}{b^2}\sum_{\ell\neq 0,n}\frac{f_{n-\ell}\,g_\ell}{n\ell}\,\mathbf{E}_0 + O(\xi^3). \tag{S16}$$

and

$$\mathbf{H}_n = \frac{iqg_n}{nb}\mathbf{E}_0 - \frac{iqk_zg_n}{n^2b^2}\mathbf{E}_0 - \frac{i\varepsilon q^2 f_n}{n^2b^2}\mathbf{e}_z^\times \mathbf{H}_0 + \frac{q^2}{b^2}\sum_{\ell\neq 0,n}\frac{g_{n-\ell}\,f_\ell}{n\ell}\,\mathbf{H}_0 + O(\xi^3). \tag{S17}$$

The averaged fields $\mathbf{D}_0$ and $\mathbf{B}_0$ satisfy

$$\mathbf{D}_0 = \varepsilon\mathbf{E}_0 - i\sum_{n\neq 0} g_{-n}\mathbf{e}_z^\times \mathbf{E}_n, \tag{S18}$$

$$\mathbf{B}_0 = \mu\mathbf{H}_0 - i\sum_{n\neq 0} f_{-n}\mathbf{e}_z^\times \mathbf{H}_n. \tag{S19}$$

Substituting the second-order harmonics, Eqs. (S16) and (S17) into Eqs. (S18) and (S19), we obtain

$$\mathbf{D}_0 = \left(\varepsilon + \mu\xi^2 S_{gg}\right)\mathbf{E}_0 + \left(-\xi A_{gf} + \frac{qk_z}{b^2}S_{gf}\right)\mathbf{e}_z^\times \mathbf{H}_0 - i\xi^2 S_{gfg}\mathbf{e}_z^\times \mathbf{E}_0 + O(\xi^3), \tag{S20}$$

$$\mathbf{B}_0 = \left(\mu + \varepsilon\xi^2 S_{ff}\right)\mathbf{H}_0 + \left(-\xi A_{gf} - \frac{qk_z}{b^2}S_{gf}\right)\mathbf{e}_z^\times \mathbf{E}_0 - i\xi^2 S_{fgf}\mathbf{e}_z^\times \mathbf{H}_0 + O(\xi^3). \tag{S21}$$

where we introduce the sums

$$S_{gg} = \sum_{n\neq 0}\frac{g_{-n}g_n}{n^2}, \qquad S_{ff} = \sum_{n\neq 0}\frac{f_{-n}f_n}{n^2}, \tag{S22}$$

$$S_{gf} = \sum_{n\neq 0}\frac{g_{-n}f_n}{n^2}, \qquad A_{gf} = \sum_{n\neq 0}\frac{g_{-n}f_n}{n}, \tag{S23}$$

and

$$S_{gfg} = \sum_{n\neq 0}\sum_{\ell\neq 0,n}\frac{g_{-n}f_{n-\ell}g_\ell}{n\ell}, \tag{S24}$$

$$S_{fgf} = \sum_{n\neq 0}\sum_{\ell\neq 0,n}\frac{f_{-n}g_{n-\ell}f_\ell}{n\ell}. \tag{S25}$$

By considering the same modulation profile for $g(z)$ and $f(z)$ averaged to zero

$$g(z) = g\, s(z), \qquad f(z) = f\, s(z), \qquad \langle s\rangle = 0, \tag{S26}$$

with $g$ and $f$ their corresponding ampltiude, we define

$$\mathcal{S} = \sum_{n\neq 0}\frac{s_{-n}s_n}{n^2}. \tag{S27}$$

Hence, Eqs. (S20) and (S21) reduce to

$$\mathbf{D}_0 = \left(\varepsilon + \mu\xi^2 g^2\mathcal{S}\right)\mathbf{E}_0 + \frac{qk_z}{b^2}gf\mathcal{S}\,\mathbf{e}_z^\times \mathbf{H}_0 + O(\xi^3), \tag{S28}$$

$$\mathbf{B}_0 = \left(\mu + \varepsilon\xi^2 f^2\mathcal{S}\right)\mathbf{H}_0 - \frac{qk_z}{b^2}gf\mathcal{S}\,\mathbf{e}_z^\times \mathbf{E}_0 + O(\xi^3). \tag{S29}$$

The zeroth-harmonic Maxwell's equations, Eq. (S9) yield the identities

$$k_z\mathbf{e}_z^\times \mathbf{H}_0 = -q\mathbf{D}_0, \qquad k_z\mathbf{e}_z^\times \mathbf{E}_0 = q\mathbf{B}_0. \tag{S30}$$

Therefore,

$$\left(1 + \xi^2 gf\mathcal{S}\right)\mathbf{D}_0 = \left(\varepsilon + \mu\xi^2 g^2\mathcal{S}\right)\mathbf{E}_0 + O(\xi^3), \tag{S31}$$

$$\left(1 + \xi^2 gf\mathcal{S}\right)\mathbf{B}_0 = \left(\mu + \varepsilon\xi^2 f^2\mathcal{S}\right)\mathbf{H}_0 + O(\xi^3). \tag{S32}$$

Hence, we extract the effective permittivity and permeability as

$$\begin{aligned}\varepsilon_{\text{eff}} &= \varepsilon + \mathcal{S}g\left(\mu g - \varepsilon f\right)\xi^2 + O(\xi^3),\\ \mu_{\text{eff}} &= \mu + \mathcal{S}f\left(\varepsilon f - \mu g\right)\xi^2 + O(\xi^3).\end{aligned} \tag{S33}$$

The leading biaxionic response follows from the boundary conditions for the macroscopic fields derived in Ref. [1]. At an interface $z = z_s$ without external surface currents, continuity of the microscopic tangential fields yields

$$\mathbf{E}^{\text{out}} = \mathbf{E}_0 + \sum_{n\neq 0}\mathbf{E}_n e^{inbz_s}, \tag{S34}$$

$$\mathbf{H}^{\text{out}} = \mathbf{H}_0 + \sum_{n\neq 0}\mathbf{H}_n e^{inbz_s}. \tag{S35}$$

Using the first-order approximation for higher-order Floquet harmonics of the fields, we recover

$$\mathbf{E}^{\text{out}} = \mathbf{E}_0 + \widetilde{\chi}(z_s)\mathbf{H}_0 + O(\xi^2), \tag{S36}$$

$$\mathbf{H}^{\text{out}} = \mathbf{H}_0 - \chi(z_s)\mathbf{E}_0 + O(\xi^2), \tag{S37}$$

with the axion and dual axion responses

$$\chi(z_s) = -i\xi \sum_{n\neq 0} \frac{g_n e^{inbz_s}}{n}, \qquad \widetilde{\chi}(z_s) = -i\xi \sum_{n\neq 0} \frac{f_n e^{inbz_s}}{n}. \tag{S38}$$

These are the axion and dual-axion responses of the metamaterial. They depend on the physical termination, whereas the effective permittivity and permeability are invariant under a shift of the unit-cell origin, as expected. Considering an alternating equal-thickness layer,

$$s(z) = \begin{cases} +1, & 0 < z < a/2, \\ -1, & a/2 < z < a, \end{cases} \qquad s(z+a) = s(z). \tag{S39}$$

gives

$$\mathcal{S} = \frac{4}{\pi^2} \sum_{\substack{n\neq 0 \\ n \text{ odd}}} \frac{1}{n^4} = \frac{\pi^2}{12}. \tag{S40}$$

and the effective bulk parameters become:

$$\begin{aligned} \varepsilon_{\text{eff}} &= \varepsilon + \frac{\pi^2}{12} g\left(\mu g - \varepsilon f\right)\xi^2 + O(\xi^3), \\ \mu_{\text{eff}} &= \mu + \frac{\pi^2}{12} f\left(\varepsilon f - \mu g\right)\xi^2 + O(\xi^3). \end{aligned} \tag{S41}$$

For $z_s = N\,a$ with an integer $N$ and positive gyrotropy $+g$ first, the leading boundary coefficients are

$$\chi = -\frac{\pi}{2} g\xi \qquad \widetilde{\chi} = -\frac{\pi}{2} f\xi. \tag{S42}$$

Exchanging the two layers reverses the sign both for axion and dual axion responses, while leaving the bulk properties unchanged.

The alternating bigyrotropic multilayer therefore realizes both axion and dual axion responses through the modified boundary conditions. The corresponding bulk permittivity and permeability (more precisely, $xx$ and $yy$ components of the permittivity and permeability tensors) are given by Eq. (S41). The boundary corrections arise at the first order in $a/\lambda$, while the bulk modifications appear only in the second order.

## S2. INTERACTION OF BIAXIONIC MEDIUM WITH THE SOURCES: EFFECTIVE THEORY

Above, we analyzed electrodynamics of biaxionic media in the absence of the sources assuming that the medium is excited by the waves incident from outside. Now we analyze a distinct physical situation when the sources are introduced directly inside the structure. In this section, we derive the relevant effective description directly from the microscopic model of a multilayered metamaterial and deduce the expressions for the renormalized source terms.

### A. Matching conditions for external currents in biaxionic medium

We consider the excitation of unbounded periodic layered metamaterial by the thin planar current parallel to $Oxy$ plane carrying the linear densities of electric and magnetic current equal to $\mathbf{J}_e$ and $\mathbf{J}_m$, respectively. For simplicity, both currents are assumed uniform in the $Oxy$ plane.

The displacement of the planar current within the unit cell is denoted by $s$, so that $s = 0$ corresponds to the current inserted between the unit cells, i.e. the distance from the current to the boundary of the material is equal to $N_{\text{cell}}\,a$, where $N_{\text{cell}}$ is an integer. On the contrary, $s = \pm a/2$ corresponds to the current placed in the middle of the unit cell,

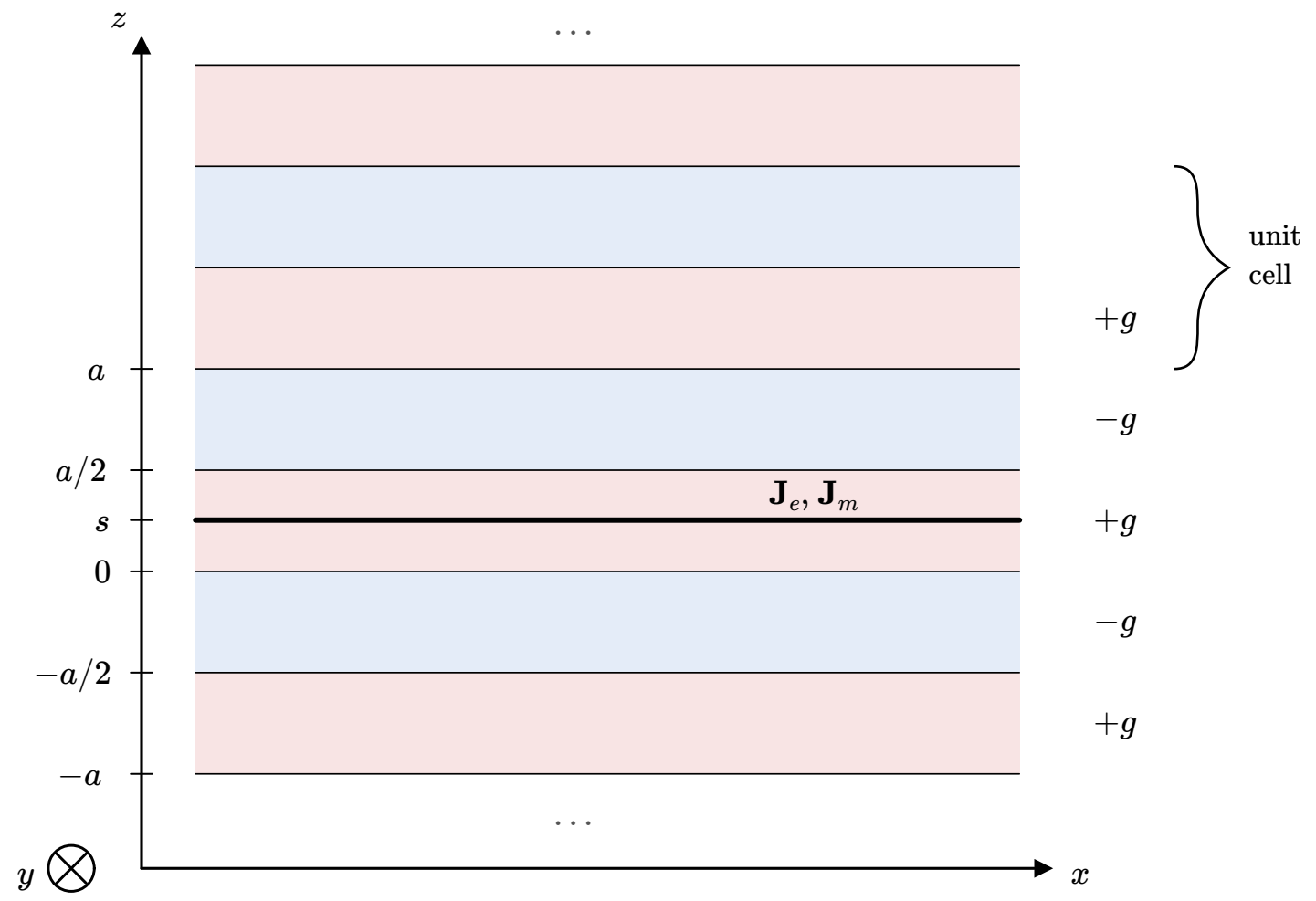


FIG. S1. Periodic multilayered metamaterial consisting of oppositely magnetized layers with electric and magnetic gyrotropy excited by the external planar current. The coordinate of the planar current within the unit cell is denoted as $s$.

so that the distance between the current plane and the boundary of the material is equal to $(N_{\text{cell}} + 1/2)\, a$. Overall, the displacement $s$ is in the interval from $-a/2$ to $a/2$, as illustrated in Fig. S1.

For conceptual clarity, we assume that no reflections happen at the outer boundaries of the material. Therefore, the fields produced by the current, are presented as outgoing Bloch waves:

$$\mathbf{E}^{(+)}(z) = e^{ik(z-s)} \sum_{n=-\infty}^{\infty} \mathbf{E}_n^{(+)}\, e^{inbz},\ z > s\,, \tag{S43}$$

$$\mathbf{E}^{(-)}(z) = e^{-ik(z-s)} \sum_{n=-\infty}^{\infty} \mathbf{E}_n^{(-)}\, e^{inbz},\ z < s\,, \tag{S44}$$

where $b = 2\pi/a$ is reciprocal lattice period. The same expansions are valid for $\mathbf{H}$, $\mathbf{D}$ and $\mathbf{B}$.

The matching conditions for the microscopic fields from both sides of the current are readily obtained from the conventional Maxwell's equations and read

$$\mathbf{H}^{(+)}(s) - \mathbf{H}^{(-)}(s) = -\frac{4\pi}{c}\,\mathbf{e}_z \times \mathbf{J}_e\,, \tag{S45}$$

$$\mathbf{E}^{(+)}(s) - \mathbf{E}^{(-)}(s) = \frac{4\pi}{c}\,\mathbf{e}_z \times \mathbf{J}_m\,. \tag{S46}$$

The microscopic fields entering Eqs. (S45),(S46) contain not only the averaged fields, but also all Floquet harmonics. Our goal is to exclude all higher-order Floquet harmonics deriving the matching conditions for the averaged fields. Similarly to the derivation of effective properties, we keep only the terms of the zero and first order in $\xi = a/\lambda$, while higher-order contributions are neglected.

From both sides of the source, the microscopic fields satisfy the usual Maxwell's equations, namely

$$\operatorname{rot}\mathbf{E} = iq\,\mathbf{B}, \tag{S47}$$

$$\operatorname{rot}\mathbf{H} = -iq\,\mathbf{D}\,, \tag{S48}$$

where $q = \omega/c$. Using these equations and the Floquet expansions Eqs. (S43),(S44), we derive the expressions

$$\mathbf{E}_n^{(\pm)} = -\frac{\xi}{n}\,\mathbf{e}_z \times \mathbf{B}_n^{(\pm)}\,, \tag{S49}$$

$$\mathbf{H}_n^{(\pm)} = \frac{\xi}{n}\,\mathbf{e}_z \times \mathbf{D}_n^{(\pm)}\,, \tag{S50}$$

where $n \neq 0$ and $\xi = a/\lambda = q/b$.

In addition to that, $\mathbf{D}(z) = \hat{\varepsilon}(z)\,\mathbf{E}(z)$ and $\mathbf{B}(z) = \hat{\mu}(z)\,\mathbf{H}(z)$, where $\hat{\varepsilon} = \varepsilon\,\hat{I} - ig(z)\,\mathbf{e}_z^{\times}$ and $\hat{\mu} = \mu\,\hat{I} - if(z)\,\mathbf{e}_z^{\times}$. The functions $g(z)$ and $f(z)$ are periodic and are expanded in the Fourier series as

$$g(z) = \sum_{n\neq 0} g_n\, e^{inbz}\ , \tag{S51}$$

$$f(z) = \sum_{n\neq 0} f_n\, e^{inbz}\ . \tag{S52}$$

Note that the averaged magnetization captured by $g_0$ and $f_0$ is identically zero. Therefore,

$$\mathbf{D}_n^{(\pm)} = \sum_{n'} \hat{\varepsilon}_{n'}\,\mathbf{E}_{n-n'} \approx -ig_n\mathbf{e}_z\times\mathbf{E}_0\ ,$$
$$\mathbf{B}_n^{(\pm)} = \sum_{n'} \hat{\mu}_{n'}\,\mathbf{H}_{n-n'} \approx -if_n\mathbf{e}_z\times\mathbf{H}_0\ ,$$

which yields

$$\mathbf{E}_n^{(\pm)} = -\frac{i\xi}{n}\, f_n\,\mathbf{H}_0^{(\pm)}\ , \tag{S53}$$

$$\mathbf{H}_n^{(\pm)} = \frac{i\xi}{n}\, g_n\,\mathbf{E}_0^{(\pm)}\ . \tag{S54}$$

With these expressions, we return to Eqs. (S45),(S46). Using the notation $[\mathbf{H}] = \mathbf{H}^{(+)} - \mathbf{H}^{(-)}$, we derive the conditions:

$$[\mathbf{H}_0] - \chi(s)\,[\mathbf{E}_0] = -\frac{4\pi}{c}\,\mathbf{e}_z\times\mathbf{J}_e\ , \tag{S55}$$

$$[\mathbf{E}_0] + \tilde{\chi}(s)\,[\mathbf{H}_0] = \frac{4\pi}{c}\,\mathbf{e}_z\times\mathbf{J}_m\ , \tag{S56}$$

where

$$\chi(s) = -i\xi\sum_{n\neq 0}\frac{g_n}{n}\,e^{inbs}\ , \tag{S57}$$

$$\tilde{\chi}(s) = -i\xi\sum_{n\neq 0}\frac{f_n}{n}\,e^{inbs}\ . \tag{S58}$$

Note that $\chi(0)\equiv\chi$ and $\tilde{\chi}(0)\equiv\tilde{\chi}$, where $\chi$ and $\tilde{\chi}$ are the axion and dual axion responses manifested in the boundary conditions. We further call the matching conditions with $\chi(s) = \chi$ and $\tilde{\chi}(s) = \tilde{\chi}$ **"cavity option"**. Note also that the quantities $\chi(s)$ and $\tilde{\chi}(s)$ can be presented as the integrals in the real space as follows

$$\chi(s) = \frac{2\pi}{a^2}\,\xi\int_s^{a+s} g(z)\,z\,dz\ , \tag{S59}$$

$$\tilde{\chi}(s) = \frac{2\pi}{a^2}\,\xi\int_s^{a+s} f(z)\,z\,dz\ . \tag{S60}$$

The explicit dependence on $s$ can be readily evaluated for a model multilayered structure assuming $g(z) = +g$ for $0 < z < a/2$ and $g(z) = -g$ for $-a/2 < z < 0$. The expression for $\chi(s)$ in the interval from $-a/2$ to $a/2$ reads:

$$\chi(s) = \frac{\pi\,g\,\xi}{2}\,\left(\frac{4|s|}{a} - 1\right)\ , \tag{S61}$$

and is illustrated in Fig. S2. Note that $\chi(s)$ is a periodic function of displacement $s$ with the zero average. The same dependence is recovered for $\tilde{\chi}$.

Equations (S55),(S56) together with Eq. (S61) suggest that the matching conditions for the planar current drastically depend on its position within the unit cell of the metamaterial. Such dependence is clearly beyond the scope of the effective medium description. Previously, this motivated many authors to state that the interaction of the structured medium with the sources is not captured by the effective medium model at all.

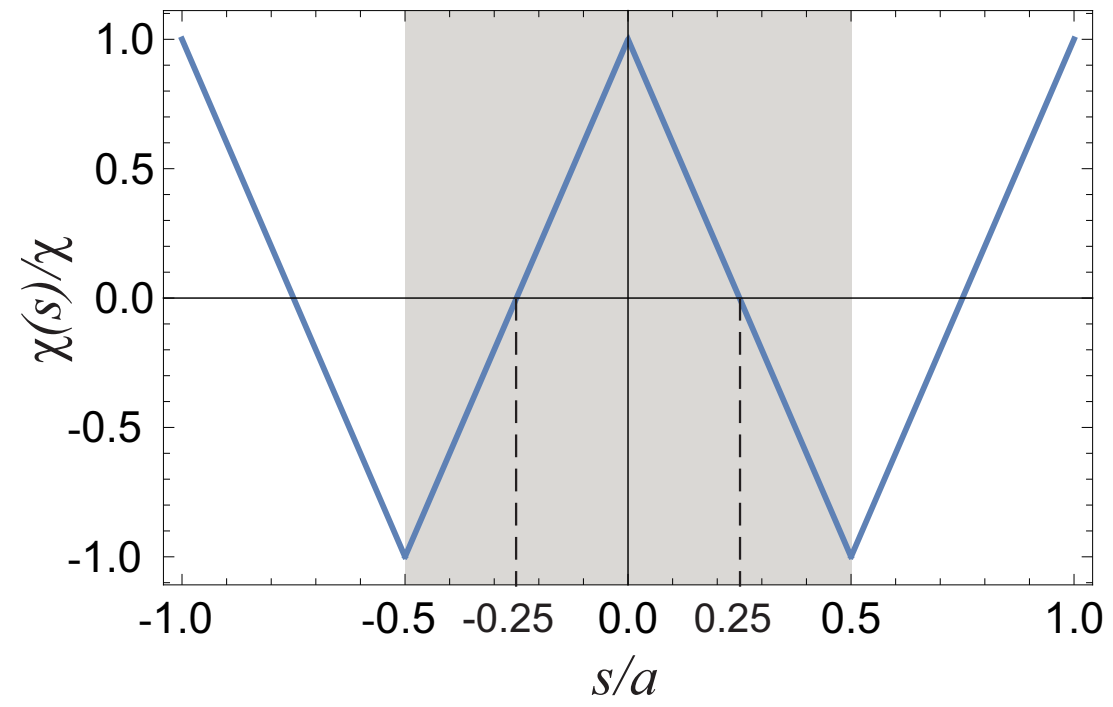


FIG. S2. The dependence of $\chi(s)$ function on the displacement $s$ of the planar current within the unit cell. Shaded area shows one unit cell with $-a/2 < s < a/2$. The function $\chi(s)$ is periodic and normalized to the value of the axion response $\chi = -\pi g \xi/2$.

It should be noted, however, that the above analysis applies to the infinitely thin planar current with the transverse size much smaller than the lattice periodicity. The situation changes qualitatively if the current is macroscopic and has the transverse size comparable or exceeding the lattice periodicity. To derive the matching conditions in the first order in $\xi$ in such case, it is sufficient to integrate Eqs. (S55),(S56) over $s$. More specifically, we assume that the current has a thickness equal to the lattice period $a$. In such case, the result does not depend on the precise position of the current within the unit cell, while the integration of the periodic functions $\chi(s)$ and $\tilde{\chi}(s)$ over their period yields zero result. In the first order in $\xi$ this yields the matching conditions

$$[\mathbf{H}_0] = -\frac{4\pi}{c}\,\mathbf{e}_z \times \mathbf{J}_e\,, \tag{S62}$$

$$[\mathbf{E}_0] = \frac{4\pi}{c}\,\mathbf{e}_z \times \mathbf{J}_m\,, \tag{S63}$$

where $\mathbf{J}_e$ and $\mathbf{J}_m$ are the linear current densities for the macroscopic ("thick") current. Further, we call the conditions Eqs. (S62),(S63) **"no cavity" option**.

Qualitatively, the conditions Eqs. (S62)-(S63) can be understood as follows. The extended current source penetrates into several layers of the metamaterial structure. Therefore, the position of the boundary between the source and the medium is not defined and one has to average over various positions of the boundary. As a result of averaging, the boundary literally disappears, so that the electromagnetic fields in the immediate vicinity of the source coincide with those in the usual dielectric medium.

At the qualitative level, the radiation, emitted by the extended source in the medium, crosses only one outer boundary, and this is exactly the reason why experiments with the sources can distinguish axion and dual axion responses, while scattering experiments can not: in the scattering process the radiation crosses the boundary twice (or, generally, even number of times).

Below, to model the interaction of axion and dual axion media with the macroscopic sources, we apply the conditions Eqs. (S62),(S63).

### B. Equations of biaxionic electrodynamics with the sources

Given the matching conditions Eqs. (S62),(S63) a natural next question is the form of the Maxwell's equations in biaxionic medium in the presence of the sources. Given the form of the Maxwell's equations in biaxionic medium in the absence of the sources [1], we expect the equations of the form

$$\text{rot}\,(\mathbf{H} - \chi\,\mathbf{E}) = \frac{1}{c}\frac{\partial}{\partial t}(\mathbf{D} + \chi\,\mathbf{B}) + \frac{4\pi}{c}\,\mathbf{j}_e^{(\text{ren})}\,, \tag{S64}$$

$$\text{rot}\,(\mathbf{E} + \tilde{\chi}\mathbf{H}) = -\frac{1}{c}\frac{\partial}{\partial t}(\mathbf{B} - \tilde{\chi}\,\mathbf{D}) - \frac{4\pi}{c}\,\mathbf{j}_m^{(\text{ren})}\,, \tag{S65}$$

where the source terms $\mathbf{j}_e^{(\text{ren})}$ and $\mathbf{j}_m^{(\text{ren})}$ are bulk electric and magnetic source terms possibly renormalized due to the magneto-electric coupling in the medium. The matching conditions for the external currents are obtained by

integrating Eqs. (S64),(S65) over a pillbox and by specifying the form of the source terms. We consider the two options below:

**1. Cavity option.** If renormalization is absent, i.e. $\mathbf{j}_e^{(\text{ren})} = \mathbf{j}_e$ and $\mathbf{j}_m^{(\text{ren})} = \mathbf{j}_m$, straightforward calculation yields the matching conditions

$$[\mathbf{H}_0] - \chi\,[\mathbf{E}_0] = -\frac{4\pi}{c}\,\mathbf{e}_z \times \mathbf{J}_e\,, \tag{S66}$$

$$[\mathbf{E}_0] + \tilde{\chi}\,[\mathbf{H}_0] = \frac{4\pi}{c}\,\mathbf{e}_z \times \mathbf{J}_m\,, \tag{S67}$$

which is exactly the "cavity" option discussed above. In such a case, redefinition of the fields inside the metamaterial simultaneously simplifies the conditions at the outer boundary and the matching conditions near the source. As a result, the axion and dual axion responses become completely indistinguishable, as is the case for the sources placed in a vacuum cavity inside the metamaterial. However, Eqs. (S66)-(S67) are different from those expected for the source placed directly inside the metamaterial, see Eqs. (S62)-(S63).

**2. No cavity option.** To recover the conditions Eqs. (S62)-(S63) for the sources inside the medium, the source terms in the equations (S64),(S65) of biaxionic electrodynamics should be properly renormalized. Straightforward analysis yields the renormalization:

$$\mathbf{j}_e^{(\text{ren})} = \mathbf{j}_e + \chi\,\mathbf{j}_m\,, \tag{S68}$$

$$\mathbf{j}_m^{(\text{ren})} = \mathbf{j}_m - \tilde{\chi}\,\mathbf{j}_e\,. \tag{S69}$$

A similar analysis of the divergence equations yields an analogous renormalization for the density of the sources

$$\rho_e^{(\text{ren})} = \rho_e + \chi\,\rho_m\,, \tag{S70}$$

$$\rho_m^{(\text{ren})} = \rho_m - \tilde{\chi}\,\rho_e\,, \tag{S71}$$

where $\rho_e$ and $\rho_m$ are the seed sources, i.e. ones measured in vacuum. Equations (S68)-(S71) is the correct form of the source terms in the equations of biaxionic electrodynamics automatically producing "no cavity" option for the introduced sources.

Note that Eqs. (S64),(S65) with renormalized sources Eqs. (S68),(S69) can be recast as follows:

$$\text{rot}\,\mathbf{H} - \frac{1}{c}\frac{\partial \mathbf{D}}{\partial t} - \frac{4\pi}{c}\,\mathbf{j}_e = \frac{1}{1+\chi\tilde{\chi}}\left\{\nabla\chi \times \mathbf{E} + \frac{1}{c}\frac{\partial \chi}{\partial t}\,\mathbf{B} - \chi\,\nabla\tilde{\chi}\times\mathbf{H} + \frac{\chi}{c}\frac{\partial\tilde{\chi}}{\partial t}\,\mathbf{D}\right\}\,, \tag{S72}$$

$$\text{rot}\,\mathbf{E} + \frac{1}{c}\frac{\partial \mathbf{B}}{\partial t} + \frac{4\pi}{c}\,\mathbf{j}_m = \frac{1}{1+\chi\tilde{\chi}}\left\{-\nabla\tilde{\chi} \times \mathbf{H} + \frac{1}{c}\frac{\partial \tilde{\chi}}{\partial t}\,\mathbf{D} - \tilde{\chi}\,\nabla\chi\times\mathbf{E} - \frac{\tilde{\chi}}{c}\frac{\partial\chi}{\partial t}\,\mathbf{B}\right\}\,. \tag{S73}$$

Thus, only spatial gradients or temporal derivatives of $\chi$ and $\tilde{\chi}$ produce the differences from the canonical Maxwell's equations.

Due to the structure of biaxionic electrodynamics equations, renormalized sources satisfy the standard continuity equations

$$\frac{\partial \rho_e^{(\text{ren})}}{\partial t} + \nabla\cdot\mathbf{j}_e^{(\text{ren})} = 0\,,$$

$$\frac{\partial \rho_m^{(\text{ren})}}{\partial t} + \nabla\cdot\mathbf{j}_m^{(\text{ren})} = 0\,.$$

However, when combined with Eqs. (S68)-(S71), this produces the continuity equations for the actual electric and magnetic charges with the additional corrections of the form $\nabla\chi\cdot\mathbf{j}_m$ and $\nabla\tilde{\chi}\cdot\mathbf{j}_e$. The resolution of this apparent contradiction is as follows: the matching conditions for the currents and the expressions for renormalized sources are obtained under the assumption that *the axion and dual axion responses are homogeneous in the area occupied by the external currents.* Otherwise our analysis above has to be refined, and the expressions Eqs. (S68)-(S71) for renormalized sources will acquire additional corrections.

## S3. SYMMETRY OF BIAXIONIC ELECTRODYNAMICS EQUATIONS AND CONTINUOUS MAPPING GROUP

Biaxionic media are generally characterized by the four material parameters: permittivity $\varepsilon$, permeability $\mu$, axion response $\chi$ and dual axion response $\tilde{\chi}$. It appears that the two biaxionic media with the *different* material parameters

can nevertheless interact with the incident light exactly *in the same way* producing identical scattered near and far field.

Below, we reveal the symmetry responsible for this physics and derive the continuous mapping group connecting all biaxionic media producing the same scattering. In this section, we assume that the external sources are absent, so that $\mathbf{J}_\mathrm{e} = \mathbf{J}_\mathrm{m} = 0$ and $\rho_\mathrm{e} = \rho_\mathrm{m} = 0$ in the Maxwell's equations.

### A. Matrix form of the equations

Equations of biaxionic electrodynamics in the absence of the sources read

$$\nabla\times\Big(\mathbf{H}-\chi\,\mathbf{E}\Big)-\frac{1}{c}\frac{\partial}{\partial t}\Big(\mathbf{D}+\chi\,\mathbf{B}\Big)=0, \tag{S74}$$

$$\nabla\cdot\Big(\mathbf{D}+\chi\,\mathbf{B}\Big)=0, \tag{S75}$$

$$\nabla\times\Big(\mathbf{E}+\tilde{\chi}\,\mathbf{H}\Big)+\frac{1}{c}\frac{\partial}{\partial t}\Big(\mathbf{B}-\tilde{\chi}\,\mathbf{D}\Big)=0, \tag{S76}$$

$$\nabla\cdot\Big(\mathbf{B}-\tilde{\chi}\,\mathbf{D}\Big)=0, \tag{S77}$$

where all four material parameters $(\varepsilon,\mu,\chi,\tilde{\chi})$ can depend on time and coordinates. In the monochromatic case the divergence equations Eqs. (S75),(S77) are a consequence of Eqs. (S74), (S76) and therefore can be omitted. The entire system is reduced to the single matrix equation:

$$\nabla\times\begin{pmatrix}1 & \tilde{\chi}\\ -\chi & 1\end{pmatrix}\begin{pmatrix}\mathbf{E}\\ \mathbf{H}\end{pmatrix}=\frac{1}{c}\frac{\partial}{\partial t}\begin{pmatrix}1 & \tilde{\chi}\\ -\chi & 1\end{pmatrix}\begin{pmatrix}0 & -1\\ 1 & 0\end{pmatrix}\begin{pmatrix}\varepsilon & 0\\ 0 & \mu\end{pmatrix}\begin{pmatrix}\mathbf{E}\\ \mathbf{H}\end{pmatrix}. \tag{S78}$$

Here the fields $\mathbf{D}$ and $\mathbf{H}$ are introduced via $\mathbf{D}=\varepsilon\mathbf{E}$ and $\mathbf{B}=\mu\mathbf{H}$. We further use the following notations:

$$S=\begin{pmatrix}1 & \tilde{\chi}\\ -\chi & 1\end{pmatrix},\quad M=\begin{pmatrix}\varepsilon & 0\\ 0 & \mu\end{pmatrix},\quad \Omega=\begin{pmatrix}0 & -1\\ 1 & 0\end{pmatrix},\quad \Phi=\begin{pmatrix}\mathbf{E}\\ \mathbf{H}\end{pmatrix}. \tag{S79}$$

This way, we write biaxionic Maxwell's equations as a single matrix equation in the form

$$\nabla\times(S\,\Phi)=\frac{1}{c}\frac{\partial}{\partial t}(S\,\Omega\,M\,\Phi)\,. \tag{S80}$$

We immediately see that in the case of constant biaxionic parameters $\chi$ and $\tilde{\chi}$ Eq. (S80) is reduced to the ordinary Maxwell's equations by multiplying Eq. (S80) by $S^{-1}$ from the left, which amounts to taking a linear combination of the two curl equations. Therefore, we state that the bulk properties of a homogeneous time-independent biaxionic medium are equivalent to those of a usual dielectric material. The difference from the dielectric comes due to the boundaries.

### B. Continuous mapping group and its invariants

In many experimental situations, such as scattering experiments, we only have an access to the fields *outside* of the material having no possibility to measure electric and magnetic fields inside. The fields inside the material in our effective description are understood in the averaged sense, when the contributions with high wave vector are filtered out. Hence, such averaged fields can hardly be accessed directly. We consider a redefinition of electric and magnetic fields *inside* the material as

$$\Phi=A\,\Phi'\,, \tag{S81}$$

where $A$ is a transformation matrix. Obviously, such a redefinition induces the change in the Maxwell's equations (S80) which take the form

$$\nabla\times(S\,A\,\Phi')=\frac{1}{c}\frac{\partial}{\partial t}(S\,\Omega\,M\,A\,\Phi')\,. \tag{S82}$$

Next we seek such field transformations $A$ which leave the equations of biaxionic electrodynamics covariant, i.e. produce the equations of the form

$$\nabla \times (S'\,\Phi') = \frac{1}{c}\,\frac{\partial}{\partial t}(S'\,\Omega\,M'\,\Phi')\,, \tag{S83}$$

where $S'$ and $M'$ are the matrices of material parameters. While the values of material parameters may change, the structure of $S'$ and $M'$ matrices should align with Eqs. (S79). We call the transformations (S81) which bring the equations of biaxionic electrodynamics (S80) to the same form (S83) mapping transformations.

Comparing Eqs. (S82) and (S83), we arrive to the following covariance conditions:

$$S' = S\,A\,, \tag{S84}$$

$$S'\,\Omega\,M' = S\,\Omega\,M\,A\,. \tag{S85}$$

The unknown transformation matrix $A$ can be readily excluded yielding

$$S'\,\Omega\,M'\,\left(S'\right)^{-1} = S\,\Omega\,M\,S^{-1}\,. \tag{S86}$$

Hence, the mapping transformations have an invariant matrix which is $Q = S\,\Omega\,M\,S^{-1}$. Taking the determinant of this matrix, we immediately recover that the refractive index $n = \sqrt{\varepsilon\mu}$ is an invariant of the mapping transformations. Examining specific matrix elements, we additionally find the following invariants

$$\theta = \frac{\mu\,\chi - \varepsilon\,\tilde{\chi}}{1 + \chi\,\tilde{\chi}}\,, \tag{S87}$$

$$\varepsilon_T = \frac{\varepsilon + \mu\,\chi^2}{1 + \chi\,\tilde{\chi}}\,, \tag{S88}$$

$$\mu_T = \frac{\mu + \varepsilon\,\tilde{\chi}^2}{1 + \chi\,\tilde{\chi}}\,. \tag{S89}$$

The above invariants are not independent but rather linked to each other via $\varepsilon_T\,\mu_T = \theta^2 + n^2$. Physically, the invariants of the mapping transformation present the quantities which remain insensitive to any redefinition of the field inside the metamaterial which one might apply. They have a direct manifestation in the scattering experiments and can in principle be retrieved from the experimental data. The first of the invariants $\theta$ is an extrinsic electromagnetic property defined in the main text. It defines the polarization rotation for light reflected from magneto-electric medium. The remaining two parameters $\varepsilon_T$ and $\mu_T$ roughly correspond to the permittivity and permeability which can be probed via co-polarized reflection and transmission.

Interestingly, the mapping transformations feature only three independent invariants given four different material parameters $\varepsilon$, $\mu$, $\chi$ and $\tilde{\chi}$. Therefore, there exists a one-parametric continuous group of the mapping transformations which connects different biaxionic media, producing identical scattering signatures.

### C. Derivation of the continuous mapping transformation

To expedite further analysis, we note that the refractive index $n = \sqrt{\varepsilon\mu}$ is invariant and construct the matrix

$$N = \frac{1}{n}\,\Omega\,M = \begin{pmatrix} 0 & -Y^{-1} \\ Y & 0 \end{pmatrix}\,, \tag{S90}$$

where $Y = \sqrt{\varepsilon/\mu}$ is the admittance of the material. Equations (S84)-(S85) are combined to yield

$$N' = A^{-1}\,N\,A\,. \tag{S91}$$

We seek the transformation matrix $A$ in the form of the following ansatz:

$$A = \begin{pmatrix} \cos\alpha & Y^{-1}\,\sin\alpha \\ -Y\,\sin\alpha & \cos\alpha \end{pmatrix}\begin{pmatrix} p & 0 \\ 0 & q \end{pmatrix}\,. \tag{S92}$$

Here, the first (right) matrix describes an independent scaling of electric and magnetic fields, while the second (left) matrix captures a continuous duality rotation. As can be verified, that rotation taken alone leaves the matrix $N$ of the material parameters invariant.

The parameters $p$ and $q$ of the ansatz are found from the requirement that matrix $A$ satisfies the covariance conditions Eqs. (S84), (S91). This yields

$$p = \left(\cos\alpha - \tilde{\chi}\,Y\,\sin\alpha\right)^{-1}\,, \tag{S93}$$

$$q = \left(\cos\alpha - \chi\,Y^{-1}\,\sin\alpha\right)^{-1}\,. \tag{S94}$$

In turn, this allows to calculate how the continuous mapping transformation modifies the material properties of biaxionic medium:

$$\chi(\alpha) = \frac{\chi + Y\,\tan\alpha}{1 - \tilde{\chi}\,Y\,\tan\alpha}\,, \tag{S95}$$

$$\tilde{\chi}(\alpha) = \frac{\tilde{\chi} + Y^{-1}\,\tan\alpha}{1 - \chi\,Y^{-1}\,\tan\alpha}\,, \tag{S96}$$

$$Y(\alpha) = \frac{Y - \chi\tan\alpha}{1 - \tilde{\chi}\,Y\,\tan\alpha}\,. \tag{S97}$$

For $\alpha = 0$ this transformation yields the original material parameters. The transformation is also $\pi$-periodic, i.e. one may always assume that $\alpha$ belongs to the interval $[0, \pi]$.

Note that the transformation parameter $\alpha$ satisfies an addition property. One may start from the set of the material parameters $\chi_0, \tilde{\chi}_0, Y_0$ and apply the mapping transformation Eqs. (S95)-(S97) with the angle $\alpha$ obtaining another set $\chi_1, \tilde{\chi}_1, Y_1$. Then applying to the latter set $\chi_1, \tilde{\chi}_1, Y_1$ yet another mapping transformation with the angle $\beta$ is equivalent to transforming the original parameters $\chi_0, \tilde{\chi}_0, Y_0$ by the angle $\alpha + \beta$. In the other words, $\alpha$ is a good parametrization of the derived continuous transformation group.

While the invariants of the mapping transformation $n$, $\theta$, $\varepsilon_T$ and $\mu_T$ are observable in the scattering, the transformation parameter $\alpha$ cannot be observed this way. To fix the choice of the origin, we request that $\alpha = 0$ corresponds to the following combination of $\chi$ and $\tilde{\chi}$: $\tilde{\chi}/\mu = -\chi/\varepsilon$ or, alternatively, $\chi = -Y^2\,\tilde{\chi}$. Physically, this means that the contributions of axion and dual axion responses to the observable extrinsic property are balanced. With such choice of the origin, it is straightforward to show that

$$\psi = \frac{\chi(\alpha)\,\mu(\alpha) + \tilde{\chi}(\alpha)\,\varepsilon(\alpha)}{1 - \chi(\alpha)\,\tilde{\chi}(\alpha)} \equiv n\,\tan 2\alpha\,. \tag{S98}$$

We call the above combination $\psi$ of the material parameters an *intrinsic property*. As we elaborate in the section below, though unobservable in scattering, the intrinsic property manifests itself in the interaction with the sources placed in the medium.

## D. Particular cases of mapping

Using equations (S95)-(S97) it is straightforward to derive particular cases of the mapping transformation. Choosing $\tan\alpha_1 = -\tilde{\chi}\,Y$, we can map the medium with $\chi, \tilde{\chi}, Y$ onto the axion material with the parameters

$$\chi_1 = \frac{\chi - \tilde{\chi}\,Y^2}{1 + \tilde{\chi}^2\,Y^2}\,, \tag{S99}$$

$$Y_1 = Y\,\frac{1 + \chi\tilde{\chi}}{1 + \tilde{\chi}^2\,Y^2}\,. \tag{S100}$$

This particular case of mapping has previously been derived in Ref. [1] which originally predicted dual axion response.

On the other hand, one may map the same medium onto the dual axion material by choosing $\tan\alpha_2 = -\chi/Y$. This yields the material parameters

$$\tilde{\chi}_2 = \frac{\tilde{\chi} - \chi/Y^2}{1 + \chi^2/Y^2}\,, \tag{S101}$$

$$Y_2 = \frac{Y + \chi^2/Y}{1 + \chi\tilde{\chi}}\,. \tag{S102}$$

### E. Matrix form of biaxionic equations with sources

In the same spirit, the equations of biaxionic electrodynamics in the presence of macroscopic sources (no cavity option) can also be recast in the matrix form

$$\nabla \times (S\,\Phi) = \frac{1}{c}\frac{\partial}{\partial t}\,(S\Omega M\,\Phi) + \frac{4\pi}{c}\,S\Omega\mathbf{J}\,, \tag{S103}$$

where $\Phi = (\mathbf{E}, \mathbf{H})^T$, $\mathbf{J} = (\mathbf{j}_e, \mathbf{j}_m)^T$,

$$\Omega = \begin{pmatrix} 0 & -1 \\ 1 & 0 \end{pmatrix}\,, \qquad S = \begin{pmatrix} 1 & \tilde{\chi} \\ -\chi & 1 \end{pmatrix}\,, \qquad M = \begin{pmatrix} \varepsilon & 0 \\ 0 & \mu \end{pmatrix}\,. \tag{S104}$$

Continuous mapping transformations are generated by the redefinition of the fields inside the metamaterial $\Phi = A\Phi'$. The external source currents $\mathbf{J}$ do not require any specific redefinition, as their magnitude is fixed by the experimental conditions.

Covariance of the sourceless biaxionic equations requires that $S' = SA$ and $S'\Omega M' = S\Omega M A$. However, the last term with the sources explicitly breaks the covariance of biaxionic electrodynamics, because the terms $S'\Omega$ and $S\Omega$ are different for any nontrivial mapping transformation $A$.

This can be summarized as follows. In the absence of the sources, the equations of biaxionic electrodynamics possess a symmetry which allows redefinition of electric and magnetic fields inside the medium accompanied by the transformation of the material parameters which keeps the fields outside the structure unchanged. A consequence of this symmetry is the concept of intrinsic electromagnetic properties which are fundamentally invisible in any scattering experiment.

However, this symmetry breaks down once the source term is added to the equations assuming "no cavity" scenario. As a result, the intrinsic property, previously hidden from the outside observer, becomes manifested in the structure of the outside fields.

## S4. REFLECTION AND TRANSMISSION IN A BIAXIONIC MEDIUM

In this section, we derive analytically the reflection and transmission matrices for planar biaxionic materials in several settings. As shown in Sec. S3, spatially homogeneous and time-independent biaxionic parameters do not modify the source-free bulk propagation. Their scattering signatures therefore originate from the spatial variations of $\chi$ and $\tilde{\chi}$, and in particular from discontinuities at material boundaries.

We begin by reviewing the general boundary conditions at an interface between two biaxionic media and then apply them to the scattering of plane waves at a single planar boundary and in a finite slab.

### A. General biaxionic boundary conditions

We first consider a general interface $\Sigma$ separating two homogeneous biaxionic media, as illustrated in Fig. S3. The two regions are characterized by the material parameters $(\varepsilon_1, \mu_1, \chi_1, \tilde{\chi}_1)$ and $(\varepsilon_2, \mu_2, \chi_2, \tilde{\chi}_2)$, respectively.

Integrating the biaxionic Maxwell equations, Eqs. (S74)–(S77), across the interface $\Sigma$ yields the modified tangential boundary conditions

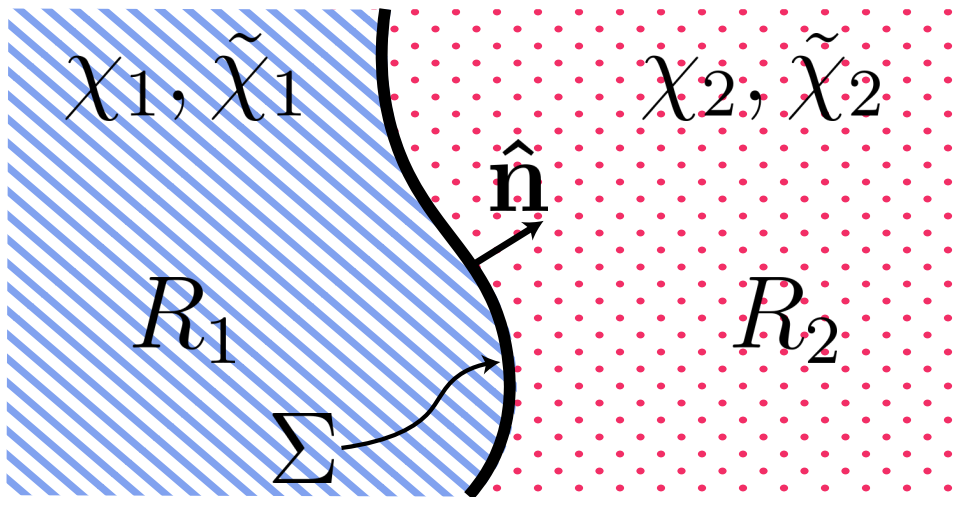


FIG. S3. Interface between two distinct biaxionic media.

$$\mathbf{n}\times\left(\mathbf{E}^{(1)}-\mathbf{E}^{(2)}\right)=-\mathbf{n}\times\left(\tilde{\chi}_1\mathbf{H}^{(1)}-\tilde{\chi}_2\mathbf{H}^{(2)}\right), \tag{S105}$$

$$\mathbf{n}\times\left(\mathbf{H}^{(1)}-\mathbf{H}^{(2)}\right)=\mathbf{n}\times\left(\chi_1\mathbf{E}^{(1)}-\chi_2\mathbf{E}^{(2)}\right), \tag{S106}$$

where the superscripts (1) and (2) denote the fields from the two sides of the interface, and $\mathbf{n}$ is the unit normal directed from $R_1$ toward $R_2$, as shown in Fig. S3.

Equations (S105) and (S106) reveal the complementary roles of the two biaxionic responses. A discontinuity of the axion parameter $\chi$ produces an effective electric surface current, as in the conventional axion electrodynamics [2], whereas a discontinuity of the dual-axion parameter $\tilde{\chi}$ gives rise to the dual effect, namely an effective magnetic surface current. In the absence of the bulk sources, observable biaxionic effects arise solely from these interface currents. These general boundary conditions therefore provide the natural starting point for the reflection and transmission problems considered below.

### B. Reflection and transmission on a single biaxionic boundary

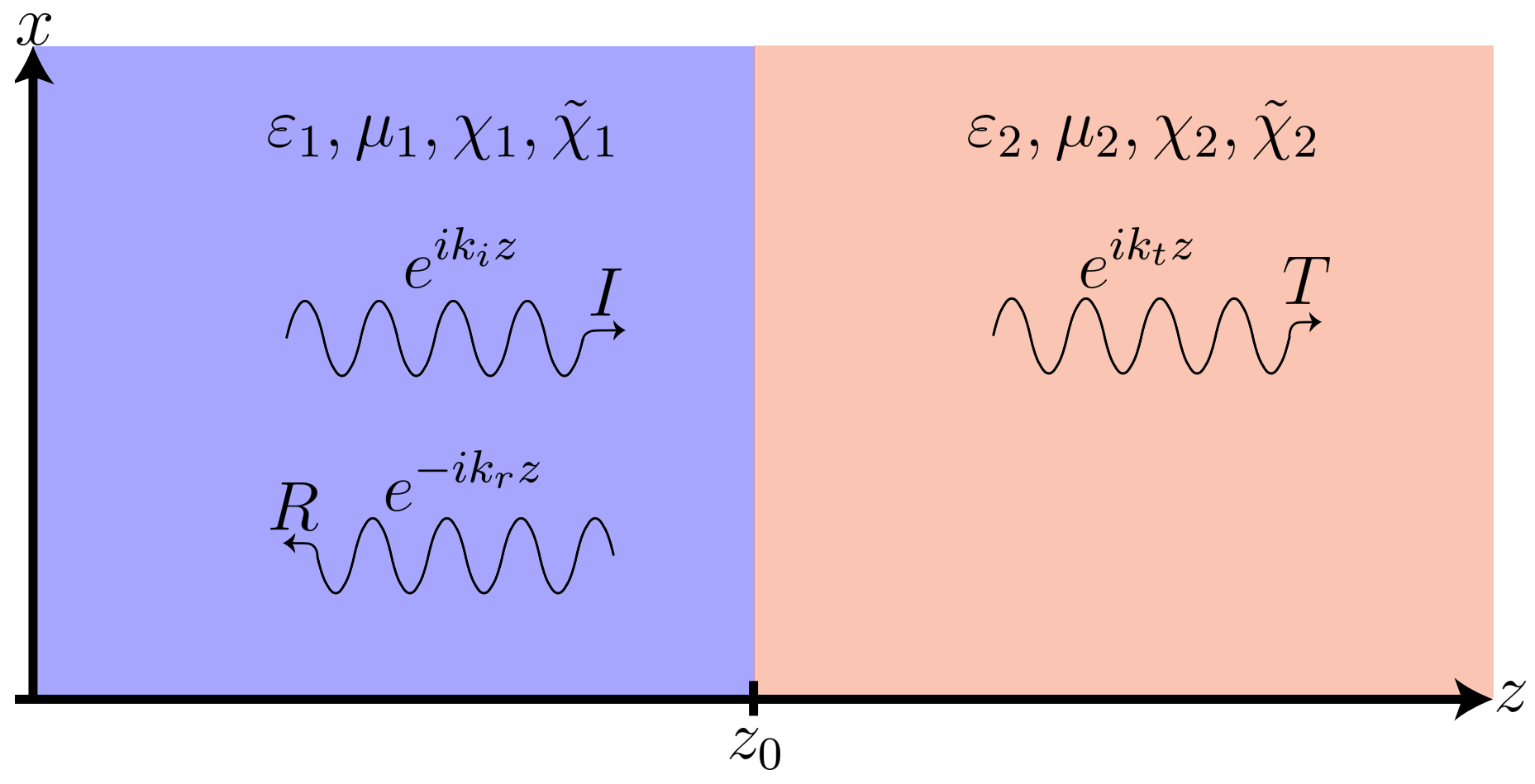


FIG. S4. Reflection and transmission of a plane wave at a planar interface between the two different biaxionic media.

First, we consider two semi-infinite biaxionic media, characterized by the material parameters $\varepsilon_1, \mu_1, \chi_1, \tilde{\chi}_1$ for $z < z_0$ and $\varepsilon_2, \mu_2, \chi_2, \tilde{\chi}_2$ for $z > z_0$. Our goal is to derive the reflection and transmission coefficients for a plane wave incident at the interface $z = z_0$ as shown in Fig. S4.

We start by writing the boundary conditions at a planar interface between the two media with the different axion and dual axion responses:

$$\mathbf{e}_z\times\left(\mathbf{E}^{(1)}-\mathbf{E}^{(2)}\right)=-\,\mathbf{e}_z\times\left(\tilde{\chi}_1\,\mathbf{H}^{(1)}-\tilde{\chi}_2\,\mathbf{H}^{(2)}\right) \tag{S107}$$

$$\mathbf{e}_z\times\left(\mathbf{H}^{(1)}-\mathbf{H}^{(2)}\right)=\mathbf{e}_z\times(\chi_1\,\mathbf{E}^{(1)}-\chi_2\,\mathbf{E}^{(2)}), \tag{S108}$$

where $\mathbf{E}^{(1,2)}$ and $\mathbf{H}^{(1,2)}=\mathbf{B}^{(1,2)}/\mu_{(1,2)}$ are electric and magnetic fields in media 1 and 2, respectively and $\mathbf{e}_z$ is a unit vector normal to the layers.

The above boundary conditions can be recast as a transfer relation between the tangential electric and magnetic fields,

$$\begin{pmatrix}\mathbf{E}^{(2)}\\ \mathbf{H}^{(2)}\end{pmatrix}=\frac{1}{1+\chi_2\tilde{\chi}_2}\begin{pmatrix}(1+\chi_1\tilde{\chi}_2)\,\hat{I} & (\tilde{\chi}_1-\tilde{\chi}_2)\,\hat{I}\\ (\chi_2-\chi_1)\,\hat{I} & (1+\tilde{\chi}_1\chi_2)\,\hat{I}\end{pmatrix}\begin{pmatrix}\mathbf{E}^{(1)}\\ \mathbf{H}^{(1)}\end{pmatrix}, \tag{S109}$$

where $\hat{I}$ is the $2\times 2$ identity matrix acting in the transverse polarization space.

For the subsequent scattering calculation, it is convenient to express the magnetic fields in terms of the electric field amplitudes. At normal incidence,

$$\mathbf{H}_i = \sqrt{\frac{\varepsilon_1}{\mu_1}}\,\mathbf{e}_z \times \mathbf{E}_i, \qquad \mathbf{H}_r = -\sqrt{\frac{\varepsilon_1}{\mu_1}}\,\mathbf{e}_z \times \mathbf{E}_r, \qquad \mathbf{H}_t = \sqrt{\frac{\varepsilon_2}{\mu_2}}\,\mathbf{e}_z \times \mathbf{E}_t. \tag{S110}$$

Substituting these relations into Eq. (S109), the boundary conditions can be written entirely in terms of the electric field amplitudes as

$$\begin{pmatrix} \mathbf{E}_t \\ \sqrt{\frac{\varepsilon_2}{\mu_2}}\mathbf{E}_t \end{pmatrix} = \begin{pmatrix} M_{11} & M_{12} \\ M_{21} & M_{22} \end{pmatrix} \begin{pmatrix} \mathbf{E}_i + \mathbf{E}_r \\ \sqrt{\frac{\varepsilon_1}{\mu_1}}\Big(\mathbf{E}_i - \mathbf{E}_r\Big) \end{pmatrix}, \tag{S111}$$

where

$$M_{11} = \frac{1+\chi_1\tilde{\chi}_2}{1+\chi_2\tilde{\chi}_2}\begin{pmatrix} 1 & 0 \\ 0 & 1 \end{pmatrix}, \qquad M_{22} = \frac{1+\tilde{\chi}_1\chi_2}{1+\chi_2\tilde{\chi}_2}\begin{pmatrix} 1 & 0 \\ 0 & 1 \end{pmatrix}. \tag{S112}$$

$$M_{12} = \frac{\tilde{\chi}_1-\tilde{\chi}_2}{1+\chi_2\tilde{\chi}_2}\begin{pmatrix} 0 & -1 \\ 1 & 0 \end{pmatrix}, \qquad M_{21} = \frac{\chi_1-\chi_2}{1+\chi_2\tilde{\chi}_2}\begin{pmatrix} 0 & -1 \\ 1 & 0 \end{pmatrix}. \tag{S113}$$

This choice of basis simplifies the boundary conditions, enabling a straightforward determination of the reflection and transmission matrices:

$$\hat{r} = -\left[\sqrt{\frac{\varepsilon_2}{\mu_2}}M_{11} - \sqrt{\frac{\varepsilon_1\varepsilon_2}{\mu_1\mu_2}}M_{12} - M_{21} + \sqrt{\frac{\varepsilon_1}{\mu_1}}M_{22}\right]^{-1}\left[\sqrt{\frac{\varepsilon_2}{\mu_2}}M_{11} + \sqrt{\frac{\varepsilon_1\varepsilon_2}{\mu_1\mu_2}}M_{12} - M_{21} - \sqrt{\frac{\varepsilon_1}{\mu_1}}M_{22}\right], \tag{S114}$$

$$\hat{t} = \left[M_{11} + \sqrt{\frac{\varepsilon_1}{\mu_1}}M_{12} + \left(M_{11} - \sqrt{\frac{\varepsilon_1}{\mu_1}}M_{12}\right)\hat{r}\right], \tag{S115}$$

that define the transmitted and reflected fields via $\mathbf{E}_t = \hat{t}\,\mathbf{E}_i$ and $\mathbf{E}_r = \hat{r}\,\mathbf{E}_i$. To simplify the resulting expressions, we introduce the wave admittance of each medium,

$$Y_j = \sqrt{\frac{\varepsilon_j}{\mu_j}}, \qquad j = 1, 2. \tag{S116}$$

The components of the transmission and reflection matrices take the form

$$t_{xx} = t_{yy} = 2Y_1\frac{1+\chi_1\tilde{\chi}_1}{\Delta(\chi,\tilde{\chi})}\left[Y_2\left(1+\chi_1\tilde{\chi}_2\right) + Y_1\left(1+\tilde{\chi}_1\chi_2\right)\right], \tag{S117}$$

$$t_{xy} = -t_{yx} = -2Y_1\frac{1+\chi_1\tilde{\chi}_1}{\Delta(\chi,\tilde{\chi})}\left[\Delta\chi + Y_1Y_2\Delta\tilde{\chi}\right], \tag{S118}$$

$$r_{xx} = r_{yy} = \frac{1}{\Delta(\chi,\tilde{\chi})}\left[Y_1^2Y_2^2\left(\Delta\tilde{\chi}\right)^2 - \left(\Delta\chi\right)^2 - Y_2^2\left(1+\chi_1\tilde{\chi}_2\right)^2 + Y_1^2\left(1+\tilde{\chi}_1\chi_2\right)^2\right], \tag{S119}$$

$$r_{xy} = -r_{yx} = \frac{2Y_1}{\Delta(\chi,\tilde{\chi})}\left[Y_2^2\left(1+\chi_1\tilde{\chi}_2\right)\Delta\tilde{\chi} - \left(1+\tilde{\chi}_1\chi_2\right)\Delta\chi\right]. \tag{S120}$$

with denominator

$$\Delta(\chi,\tilde{\chi}) = \left[Y_2\left(1+\chi_1\tilde{\chi}_2\right) + Y_1\left(1+\tilde{\chi}_1\chi_2\right)\right]^2 + \left[Y_1Y_2\Delta\tilde{\chi} + \Delta\chi\right]^2, \tag{S121}$$

where

$$\Delta\chi = \chi_1 - \chi_2, \qquad \Delta\tilde{\chi} = \tilde{\chi}_1 - \tilde{\chi}_2. \tag{S122}$$

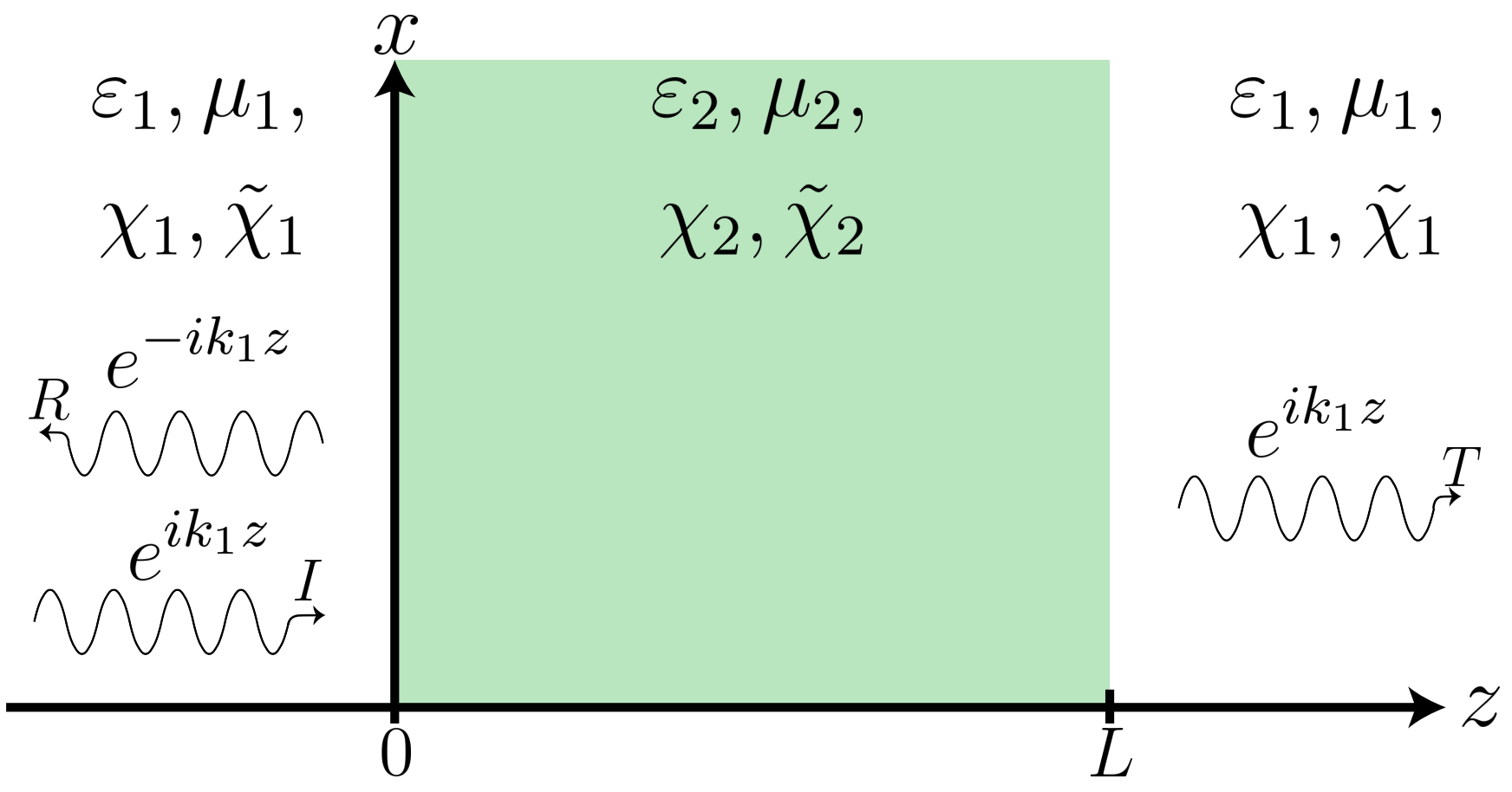


FIG. S5. Reflection and transmission of a plane electromagnetic wave from a slab with axion and dual axion response.

### C. Reflection and transmission in a biaxionic slab

Next as an experimentally relevant situation we consider reflection and transmission of waves through the finite slab of the thickness $L$ as illustrated in Fig. S5. To solve the problem, we use the transfer matrix method defining the transfer matrix as

$$\begin{pmatrix} \mathbf{E}_t(z) \\ \mathbf{H}_t(z) \end{pmatrix} = \hat{M}(z) \begin{pmatrix} \mathbf{E}(0) \\ \mathbf{H}(0) \end{pmatrix} . \tag{S123}$$

Next we present the fields at the input and output facets of the slab in terms of the incident, reflected and transmitted waves and use the connection between $\mathbf{H}$ and $\mathbf{E}$ for the propagating wave:

$$\begin{pmatrix} \mathbf{E}_t(z) \\ \sqrt{\dfrac{\varepsilon}{\mu}}\mathbf{E}_t(z) \end{pmatrix} = \hat{M}(z) \begin{pmatrix} \mathbf{E}_i(0) + \mathbf{E}_r(0) \\ \sqrt{\dfrac{\varepsilon'}{\mu'}}\Big(\mathbf{E}_i(0) - \mathbf{E}_r(0)\Big) \end{pmatrix} . \tag{S124}$$

where $\varepsilon'$ and $\mu'$ are the permittivity and permeability of the medium and $k' = k_0\sqrt{\mu'\varepsilon'}$ corresponds to the wavevector within the medium. The transfer matrix for the slab is presented as:

$$\hat{M} = \hat{M}_{2\to1}\hat{M}_b(L)\hat{M}_{1\to2} , \tag{S125}$$

where $\hat{M}_{1\to2}$ and $\hat{M}_{2\to1}$ capture the effect of boundaries, while $\hat{M}_b(L)$ describes the bulk propagation, which is identical to the usual dielectric medium:

$$\hat{M}_b(z) = \begin{pmatrix} \cos(k'z) & i\sin(k'z)\sqrt{\dfrac{\mu'}{\varepsilon'}} \\ i\sin(k'z)\sqrt{\dfrac{\varepsilon'}{\mu'}} & \cos(k'z) \end{pmatrix} . \tag{S126}$$

The reflection and transmission matrices are computed directly from the transfer matrix Eq. (S125) which yields

$$\hat{r} = -\left[\sqrt{\frac{\varepsilon_1}{\mu_1}}M_{11} - \frac{\varepsilon_1}{\mu_1}M_{12} - M_{21} + \sqrt{\frac{\varepsilon_1}{\mu_1}}M_{22}\right]^{-1} \cdot \left[\sqrt{\frac{\varepsilon_1}{\mu_1}}M_{11} + \frac{\varepsilon_1}{\mu_1}M_{12} - M_{21} - \sqrt{\frac{\varepsilon_1}{\mu_1}}M_{22}\right] , \tag{S127}$$

$$\hat{t} = \left[M_{11} + \frac{\varepsilon_1}{\mu_1}M_{12} + \left(M_{11} - \sqrt{\frac{\varepsilon_1}{\mu_1}}M_{12}\right) \cdot \hat{r}\right] . \tag{S128}$$

By setting $\chi_1 = \tilde{\chi}_1 = 0$, $\chi_2 = \chi$ and $\tilde{\chi}_2 = \tilde{\chi}$, we derive the reflection and transmission matrices for a finite biaxionic slab. We introduce the wave impedance of each medium,

$$Z_j = \sqrt{\frac{\mu_j}{\varepsilon_j}}, \qquad j = 1, 2, \tag{S129}$$

to simplify the expressions for the transmission and reflection coefficients:

$$t_{xx} = t_{yy} = \frac{2i}{\Delta(\chi,\tilde{\chi})}\Big(1+\chi\tilde{\chi}\Big)Z_1Z_2, \tag{S130}$$

$$t_{xy} = t_{yx} = 0, \tag{S131}$$

$$r_{xx} = r_{yy} = \frac{1}{\Delta(\chi,\tilde{\chi})}\left(Z_2^2 - Z_1^2 - Z_1^2Z_2^2\chi^2 + \tilde{\chi}^2\right)\sin(k_2L), \tag{S132}$$

$$r_{xy} = -r_{yx} = \frac{2}{\Delta(\chi,\tilde{\chi})}Z_1\Big(Z_2^2\chi - \tilde{\chi}\Big)\sin(k_2L), \tag{S133}$$

with the denominator

$$\Delta(\chi,\tilde{\chi}) = 2iZ_1Z_2\Big(1+\chi\tilde{\chi}\Big)\cos(k_2L) + \left(Z_1^2 + Z_2^2 + Z_1^2Z_2^2\chi^2 + \tilde{\chi}^2\right)\sin(k_2L), \tag{S134}$$

Generally, in the presence of both axion and dual axion responses, the reflected light still exhibits Kerr rotation while the polarization of transmitted light is unchanged. Remarkably, this nonreciprocal Kerr effect can be completely suppressed (i.e. $r_{xy}=0$) when the axion and dual axion responses are fine-tuned to satisfy the invisibility condition:

$$\tilde{\chi} = Z_2^2\chi = \frac{\mu_2}{\varepsilon_2}\chi\,. \tag{S135}$$

Note also that the results for the semi-infinite geometry in the previous subsection can be readily recovered from the slab solution Eqs. (S130)-(S133) by assuming that $k_2 = k_2' + ik_2''$ with positive $k_2''$ and replacing $\cos k_2L \approx e^{-ik_2L}/2$ and $\sin k_2L \approx ie^{-ik_2L}/2$.

### D. Probing invisibility limit

Substituting the invisibility condition, Eq. (S135), into Eqs. (S130)–(S133), we first find that the cross-polarized reflection coefficient vanishes identically,

$$r_{xy} = -r_{yx} = 0. \tag{S136}$$

Together with ($t_{xy} = t_{yx} = 0$), this shows that the slab does not modify the polarization of the scattered light. Moreover, the terms entering the co-polarized reflection coefficient factorize as

$$\left(Z_2^2 - Z_1^2\right)\left(1 + Z_2^2\chi^2\right), \tag{S137}$$

whereas the denominator becomes

$$\Delta(\chi,\tilde{\chi}) = \left(1+Z_2^2\chi^2\right)\left[2iZ_1Z_2\cos(k_2L) + \left(Z_1^2+Z_2^2\right)\sin(k_2L)\right]. \tag{S138}$$

The same multiplicative factor appears in the numerator of the transmission coefficient through

$$1+\chi\tilde{\chi} = 1+Z_2^2\chi^2. \tag{S139}$$

Consequently, all dependence on the axion and dual-axion parameters drops out, yielding

$$t_{xx} = t_{yy} = \frac{2iZ_1Z_2}{2iZ_1Z_2\cos(k_2L) + (Z_1^2+Z_2^2)\sin(k_2L)}, \qquad t_{xy} = t_{yx} = 0 \tag{S140}$$

$$r_{xx} = r_{yy} = \frac{\left(Z_2^2 - Z_1^2\right)\sin(k_2L)}{2iZ_1Z_2\cos(k_2L) + (Z_1^2+Z_2^2)\sin(k_2L)}, \qquad r_{xy} = r_{yx} = 0. \tag{S141}$$

These expressions coincide exactly with the amplitude reflection and transmission coefficients of an ordinary dielectric slab characterized by $\varepsilon_2$ and $\mu_2$, embedded in a medium with parameters $\varepsilon_1$ and $\mu_1$. Thus, nonreciprocal magneto-electric response is detectable neither through polarization conversion, nor through the corrections to the co-polarized scattering coefficients. The scattering equivalence holds exactly and is not restricted to the perturbative regime of small $\chi$ or $\tilde{\chi}$.

As demonstrated in the main text, this conclusion is much broader and applies not only to the normal incidence and slab geometry, but rather to *any* scattering experiment. The invisibility condition Eq. (S135) ensures vanishing extrinsic property $\theta$, which is the only combination of magneto-electric coefficients $\chi$ and $\tilde{\chi}$ observable in the scattering experiments.

## S5. ELECTROMAGNETIC RECIPROCITY THEOREM FOR BIAXIONIC MEDIA

Electromagnetic reciprocity theorems are differential or integral relations, derived from the time-harmonic Maxwell's equations, that link two current densities, $\mathbf{j}_{1,2}$, to the associated electromagnetic fields, $(\mathbf{E}_{1,2}, \mathbf{B}_{1,2})$, and remain unchanged when the indices 1 and 2 are interchanged. Such a permutation symmetry is referred to as reciprocity, and a medium in which these relations hold is called reciprocal. The concept of reciprocity is typically linked to the time-reversal symmetry [3, 4]. Although the reciprocity theorems themselves do not convey an immediate physical meaning, they provide a powerful framework for deriving such important properties of reciprocal systems as the symmetries of dyadic Green's function and transmission coefficient matrix. In this section, we demonstrate that a biaxionic medium satisfying invisibility condition, Eq. S135, remains reciprocal despite broken time-reversal symmetry. We illustrate the resulting implications on a simple example.

### A. Generalized reciprocity under the invisibility condition

Electromagnetic fields produced by the two independent sets of time-harmonic ($e^{-i\,\omega t}$) external electric and magnetic current densities $\mathbf{j}_{1,2}^{(\mathrm{e,m})}(\mathbf{r})$ in an inhomogeneous isotropic biaxionic medium characterized by material parameters $\varepsilon(\mathbf{r})$, $\mu(\mathbf{r})$, $\chi(\mathbf{r})$, and $\tilde{\chi}(\mathbf{r})$ are governed by the following system of equations (CGS system of units is used):

$$\nabla \times \left(\mu^{-1}\mathbf{B}_{1,2} - \chi\mathbf{E}_{1,2}\right) = -i\,q\left(\varepsilon\mathbf{E}_{1,2} + \chi\mathbf{B}_{1,2}\right) + \frac{4\pi}{c}\left(\mathbf{j}_{1,2}^{(\mathrm{e})} + \chi\,\mathbf{j}_{1,2}^{(\mathrm{m})}\right)\,, \tag{S142a}$$

$$\nabla \times \left(\mathbf{E}_{1,2} + \frac{\tilde{\chi}}{\mu}\mathbf{B}_{1,2}\right) = i\,q\left(\mathbf{B}_{1,2} - \tilde{\chi}\varepsilon\mathbf{E}_{1,2}\right) - \frac{4\pi}{c}\left(\mathbf{j}_{1,2}^{(\mathrm{m})} - \tilde{\chi}\,\mathbf{j}_{1,2}^{(\mathrm{e})}\right)\,, \tag{S142b}$$

where $q = \omega/c$. Equations (S142a)-(S142b) include renormalized source terms derived in the Section S2.

The equations (S142a) dot-multiplied by $\mathbf{E}_{2,1} + \frac{\tilde{\chi}}{\mu}\mathbf{B}_{2,1}$ and the equations (S142b) dot-multiplied by $\mu^{-1}\mathbf{B}_{2,1} - \chi\mathbf{E}_{2,1}$ can be combined using the vector calculus identity $\nabla\cdot(\mathbf{A}\times\mathbf{B}) = \mathbf{B}\cdot(\nabla\times\mathbf{A}) - \mathbf{A}\cdot(\nabla\times\mathbf{B})$ to get the following equation:

$$\begin{aligned}
&\nabla\cdot\left(\left(\mu^{-1}\mathbf{B}_1 - \chi\mathbf{E}_1\right)\times\left(\mathbf{E}_2 + \frac{\tilde{\chi}}{\mu}\mathbf{B}_2\right) - \left(\mu^{-1}\mathbf{B}_2 - \chi\mathbf{E}_2\right)\times\left(\mathbf{E}_1 + \frac{\tilde{\chi}}{\mu}\mathbf{B}_1\right)\right) = \\
&= 2\,i\,q\left(\chi\mathbf{E}_1\cdot\mathbf{B}_2 - \frac{\tilde{\chi}\varepsilon}{\mu}\mathbf{E}_1\cdot\mathbf{B}_2 + \frac{\tilde{\chi}\varepsilon}{\mu}\mathbf{E}_2\cdot\mathbf{B}_1 - \chi\mathbf{E}_2\cdot\mathbf{B}_1\right) + \\
&+\frac{4\pi}{c}\left((1-\chi\tilde{\chi})\left(\mathbf{j}_1^{(\mathrm{e})}\cdot\mathbf{E}_2 - \mu^{-1}\mathbf{j}_1^{(\mathrm{m})}\cdot\mathbf{B}_2 - \mathbf{j}_2^{(\mathrm{e})}\cdot\mathbf{E}_1 + \mu^{-1}\mathbf{j}_2^{(\mathrm{m})}\cdot\mathbf{B}_1\right) + \frac{2\tilde{\chi}}{\mu}\left(\mathbf{j}_1^{(\mathrm{e})}\cdot\mathbf{B}_2 - \mathbf{j}_2^{(\mathrm{e})}\cdot\mathbf{B}_1\right)\right. \\
&\left.+2\chi\left(\mathbf{j}_1^{(\mathrm{m})}\cdot\mathbf{E}_2 - \mathbf{j}_2^{(\mathrm{m})}\cdot\mathbf{E}_1\right)\right)\,.
\end{aligned} \tag{S143}$$

If $\tilde{\chi} = (\mu/\varepsilon)\chi$, the terms proportional to $iq$ on the right-hand side of Eq. (S143) cancel out yielding an analog of the Lorentz reciprocity theorem in its differential form:

$$\begin{aligned}
&\nabla\cdot\left(\left(\mu^{-1}\mathbf{B}_1 - \chi\mathbf{E}_1\right)\times\left(\mathbf{E}_2 + \frac{\chi}{\varepsilon}\mathbf{B}_2\right) - \left(\mu^{-1}\mathbf{B}_2 - \chi\mathbf{E}_2\right)\times\left(\mathbf{E}_1 + \frac{\chi}{\varepsilon}\mathbf{B}_1\right)\right) = \\
&= \frac{4\pi}{c}\left(\left(1 - \frac{\mu\chi^2}{\varepsilon}\right)\left(\mathbf{j}_1^{(\mathrm{e})}\cdot\mathbf{E}_2 - \mu^{-1}\mathbf{j}_1^{(\mathrm{m})}\cdot\mathbf{B}_2 - \mathbf{j}_2^{(\mathrm{e})}\cdot\mathbf{E}_1 + \mu^{-1}\mathbf{j}_2^{(\mathrm{m})}\cdot\mathbf{B}_1\right) + \frac{2\chi}{\varepsilon}\left(\mathbf{j}_1^{(\mathrm{e})}\cdot\mathbf{B}_2 - \mathbf{j}_2^{(\mathrm{e})}\cdot\mathbf{B}_1\right)\right. \\
&\left.+2\chi\left(\mathbf{j}_1^{(\mathrm{m})}\cdot\mathbf{E}_2 - \mathbf{j}_2^{(\mathrm{m})}\cdot\mathbf{E}_1\right)\right)\,.
\end{aligned} \tag{S144}$$

The integration of (S144) over a finite volume $V$ containing no sources leads to an analog of the Lorentz reciprocity theorem in its integral form,

$$\oiint_{\partial V}\left(\left(\mu^{-1}\mathbf{B}_1 - \chi\mathbf{E}_1\right)\times\left(\mathbf{E}_2 + \frac{\chi}{\varepsilon}\mathbf{B}_2\right) - \left(\mu^{-1}\mathbf{B}_2 - \chi\mathbf{E}_2\right)\times\left(\mathbf{E}_1 + \frac{\chi}{\varepsilon}\mathbf{B}_1\right)\right)d\mathbf{S} = 0\,, \tag{S145}$$

whereas the integration over the infinite volume containing both sets of sources leads to an analog of the Rayleigh-

Carson reciprocity theorem,

$$\iiint\limits_{V_1} \left( \left(1-\frac{\mu\chi^2}{\varepsilon}\right)\left(\mathbf{j}_1^{(\mathrm{e})}\cdot\mathbf{E}_2-\mu^{-1}\mathbf{j}_1^{(\mathrm{m})}\cdot\mathbf{B}_2\right)+2\chi\left(\varepsilon^{-1}\mathbf{j}_1^{(\mathrm{e})}\cdot\mathbf{B}_2+\mathbf{j}_1^{(\mathrm{m})}\cdot\mathbf{E}_2\right)\right) dV = \\ = \iiint\limits_{V_2} \left( \left(1-\frac{\mu\chi^2}{\varepsilon}\right)\left(\mathbf{j}_2^{(\mathrm{e})}\cdot\mathbf{E}_1-\mu^{-1}\mathbf{j}_2^{(\mathrm{m})}\cdot\mathbf{B}_1\right)+2\chi\left(\varepsilon^{-1}\mathbf{j}_2^{(\mathrm{e})}\cdot\mathbf{B}_1+\mathbf{j}_2^{(\mathrm{m})}\cdot\mathbf{E}_1\right)\right) dV\,, \tag{S146}$$

where $V_{1,2}$ represent the volumes where the corresponding current densities are nonzero.

The obtained relations (S144), (S145), and (S146) do not change when the indices $1,2$ are interchanged, so they represent reciprocity theorems thereby making the considered medium reciprocal. At this point we stress that the reciprocity holds despite broken time-reversal symmetry: both axion $\chi$ and dual axion $\tilde{\chi}$ responses are odd under time-reversal.

To illustrate the implications of the derived reciprocity theorems, we apply the integral relation (S146) to a biaxionic slab of thickness $L$, characterized by the material parameters $\varepsilon(z)$, $\mu(z)$, and $\tilde{\chi}(z)=(\mu(z)/\varepsilon(z))\chi(z)$, placed in vacuum between the two planar current sources with electric current densities $\mathbf{j}_{1,2}^{(\mathrm{e})}(\mathbf{r})=\mathbf{J}_{1,2}\delta(z-z_{1,2})$, where the positions $z_{1,2}$ are located in vacuum close to the front and rear facets and the vectors $\mathbf{J}_{1,2}=J\,\mathbf{e}_{1,2}$ may have different directions $\mathbf{e}_{1,2}$ but are of equal length $J$. For clarity, we assume that the current sources are outside of the slab, i.e. in vacuum. In such planar geometry, the reciprocity theorem Eq. (S146) reduces to

$$\mathbf{j}_1^{(\mathrm{e})}\cdot\mathbf{E}_2(z_1)=\mathbf{j}_2^{(\mathrm{e})}\cdot\mathbf{E}_1(z_2)\,. \tag{S147}$$

Electric field in the vicinity of the current plane is readily evaluated as

$$\mathbf{E}_1(z_1)=-\frac{2\pi}{c}J\,\mathbf{e}_1\,.$$

To find the field at point $z_2$, this expression should be multiplied by the transmission coefficient $\hat{t}_1$:

$$\mathbf{E}_1(z_2)=-\frac{2\pi}{c}J\,\hat{t}_1\mathbf{e}_1\,.$$

In a similar spirit we find that

$$\mathbf{E}_2(z_1)=-\frac{2\pi}{c}J\,\hat{t}_2\mathbf{e}_2\,.$$

Since the slab has full rotational symmetry with respect to the $z$ axis, the transmission matrix decomposes into co- and cross-polarized parts as follows:

$$\hat{t}_{1,2}=t_{1,2}^{(\mathrm{co})}\,\hat{I}-t_{1,2}^{(\mathrm{cr})}\,\mathbf{e}_z^{\times}\,. \tag{S148}$$

Using Eq. (S147), we arrive to the conclusion that $\mathbf{e}_1\cdot\hat{t}_2\,\mathbf{e}_2=\mathbf{e}_2\cdot\hat{t}_1\,\mathbf{e}_1$. From here, we finally recover that

$$t_1^{(\mathrm{co})}=t_2^{(\mathrm{co})}\,, \tag{S149a}$$

$$t_1^{(\mathrm{cr})}=-t_2^{(\mathrm{cr})}\,. \tag{S149b}$$

These conditions relating transmission in forward and backward direction is a known result for reciprocal media. The obtained result has a clear physical meaning. If invisibility condition is met, biaxionic medium cannot be distinguished from the ordinary dielectric by any scattering experiment. This includes the measurement of co- and cross-polarized transmission coefficients in forward and backward directions. Hence, it is understandable that the transmission coefficients are related exactly in the same way as for the reciprocal medium.

Note also that the cross-polarized transmission $t^{\mathrm{cr}}$ in this general setting could be nonzero, because $z$-dependent axion and dual axion responses may break the $\mathcal{PT}$ symmetry.

### B. Generalized reciprocity relations for biaxionic media

In the subsection above, we examined the connection between the two sets of fields $(\mathbf{E}_1,\mathbf{B}_1)$ and $(\mathbf{E}_2,\mathbf{B}_2)$ in the *same* biaxionic medium. Here, we analyze a different situation when the fields $(\mathbf{E}_1,\mathbf{B}_1)$ correspond to the medium

with parameters $\varepsilon, \mu, \chi, \tilde{\chi}$, while the second set of fields $(\mathbf{E}_2, \mathbf{B}_2)$ is calculated for the complementary medium with material parameters $\varepsilon$, $\mu$, $-\chi$, and $-\tilde{\chi}$. Physically, this means that static magnetization of the metamaterial is flipped. No special connection between $\chi$ and $\tilde{\chi}$ is assumed here.

Repeating the steps leading to Eq. (S143) with the second set of the fields entering the combinations $\mathbf{E}_2 - \frac{\tilde{\chi}}{\mu}\mathbf{B}_2$ and $\mu^{-1}\mathbf{B}_2 + \chi\mathbf{E}_2$, we find that the terms proportional to $i\,q$ cancel out for arbitrary $\chi$ and $\tilde{\chi}$, yielding

$$\nabla \cdot \left( \frac{1+\chi\tilde{\chi}}{\mu} \left( \mathbf{E}_1 \times \mathbf{B}_2 - \mathbf{E}_2 \times \mathbf{B}_1 \right) \right) = \frac{4\pi}{c} \left( 1 + \chi\tilde{\chi} \right) \left( \mathbf{j}_1^{(\mathrm{e})} \cdot \mathbf{E}_2 - \mu^{-1} \mathbf{j}_1^{(\mathrm{m})} \cdot \mathbf{B}_2 - \mathbf{j}_2^{(\mathrm{e})} \cdot \mathbf{E}_1 + \mu^{-1} \mathbf{j}_2^{(\mathrm{m})} \cdot \mathbf{B}_1 \right) . \tag{S150}$$

Since $\chi$ and $\tilde{\chi}$ are odd functions of a time-odd static bias, e.g. static magnetization $\mathbf{M}$, the complementary medium is nothing but the same medium with reversed magnetization. Hence, Eq. (S150) yields the reciprocity theorem in its standard form.

Note also that the axion and dual axion responses satisfy Onsager-Kasimir relations, being odd functions of the static magnetic bias.

### C. $\mathcal{PT}$ symmetry and the absence of cross-polarized transmission

We now apply Eq. (S150) to explain the absence of cross-polarized transmission through a homogeneous biaxionic slab. This derivation builds on $\mathcal{PT}$ symmetry of the slab and its rotational symmetry about the propagation axis and does not assume any special connection between $\chi$ and $\tilde{\chi}$ responses.

Consider the symmetric structure 1|2|1, as in Fig. S5, with the biaxionic slab centered at $z = 0$, surrounded from both sides by the same ordinary medium with parameters $\varepsilon_1, \mu_1$. The slab parameters $\varepsilon_2, \mu_2, \chi, \tilde{\chi}$ are spatially uniform. We denote the physical configuration by $\mathcal{C}_+$ and the complementary configuration, obtained by replacing $(\chi, \tilde{\chi})$ with $(-\chi, -\tilde{\chi})$, by $\mathcal{C}_-$. All transmission matrices below are defined in the same fixed transverse Cartesian basis.

We first apply the complementary-medium reciprocity relation, Eq. (S150), to two electric current sheets located outside the slab:

$$\mathbf{j}_1 = J\mathbf{e}_1 \delta(z - z_1), \qquad \mathbf{j}_2 = J\mathbf{e}_2 \delta(z - z_2), \qquad z_1 < -L/2, \quad z_2 > L/2. \tag{S151}$$

Here $\mathbf{e}_1$ and $\mathbf{e}_2$ are arbitrary transverse polarization vectors. Source 1 generates the fields in $\mathcal{C}_+$, whereas source 2 generates independent fields in $\mathcal{C}_-$. No magnetic currents are required. Since $\chi = \tilde{\chi} = 0$ at both sources, the renormalized source terms coincide with the ordinary electric currents.

Integrating Eq. (S150) along $z$, per unit transverse area, gives

$$\mathbf{e}_1 \cdot \mathbf{E}_2^-(z_1) = \mathbf{e}_2 \cdot \mathbf{E}_1^+(z_2). \tag{S152}$$

The boundary terms vanish for outgoing waves in the identical exterior media. To express the fields at the opposite source planes through the transmission matrices, we first determine the amplitude emitted by an electric current sheet in the exterior medium. Let $Y_1 = \sqrt{\varepsilon_1/\mu_1}$ and $k_1 = q\sqrt{\varepsilon_1\mu_1}$. The directly emitted waves have equal tangential electric-field amplitudes $\mathbf{E}_{\mathrm{em}}$ from both sides of the planar current, while their magnetic fields have opposite signs:

$$\mathbf{H}_{\mathrm{em}}^{(\mathrm{R})} = Y_1 \mathbf{e}_z \times \mathbf{E}_{\mathrm{em}}, \qquad \mathbf{H}_{\mathrm{em}}^{(\mathrm{L})} = -Y_1 \mathbf{e}_z \times \mathbf{E}_{\mathrm{em}}. \tag{S153}$$

The conventional electric jump current boundary condition therefore gives

$$\mathbf{e}_z \times \left( \mathbf{H}_{\mathrm{em}}^{(\mathrm{R})} - \mathbf{H}_{\mathrm{em}}^{(\mathrm{L})} \right) = -2Y_1 \mathbf{E}_{\mathrm{em}} = \frac{4\pi}{c} J\mathbf{e}, \tag{S154}$$

and hence

$$\mathbf{E}_{\mathrm{em}} = -\frac{2\pi J}{cY_1}\mathbf{e}. \tag{S155}$$

This is the directly emitted amplitude. The reflected field from the slab is continuous across the current plane and does not modify the matching condition for the fields. Define the distances from the source planes to the nearest slab interfaces as

$$d_1 = -L/2 - z_1, \qquad d_2 = z_2 - L/2. \tag{S156}$$

For source 1, propagation to the left interface at $z = -L/2$, transmission through the slab, and propagation from the right interface at $z = L/2$ to $z_2$ yield

$$\mathbf{E}_1^+(z_2) = e^{ik_1 d_2}\,\hat{t}_{\rightarrow}^{+}\,e^{ik_1 d_1}\left(-\frac{2\pi J}{cY_1}\mathbf{e}_1\right) = -\frac{2\pi J}{cY_1}e^{ik_1(d_1+d_2)}\hat{t}_{\rightarrow}^{+}\mathbf{e}_1. \tag{S157}$$

Similarly, source 2 produces

$$\mathbf{E}_2^-(z_1) = e^{ik_1 d_1}\,\hat{t}_{\leftarrow}^{-}\,e^{ik_1 d_2}\left(-\frac{2\pi J}{cY_1}\mathbf{e}_2\right) = -\frac{2\pi J}{cY_1}e^{ik_1(d_1+d_2)}\hat{t}_{\leftarrow}^{-}\mathbf{e}_2. \tag{S158}$$

Here both transmission matrices are referenced to the slab facets. Thus, the two fields can be writen as

$$\mathbf{E}_1^+(z_2) = CJ\,\hat{t}_{\rightarrow}^{+}\mathbf{e}_1, \qquad \mathbf{E}_2^-(z_1) = CJ\,\hat{t}_{\leftarrow}^{-}\mathbf{e}_2, \tag{S159}$$

where $C = -2\pi e^{ik_1(d_1+d_2)}/(cY_1)$ contains the common amplitude and the propagation phases between the source planes and the slab interfaces. Hence Eq. (S152) yields

$$\hat{t}_{\rightarrow}^{+} = \left(\hat{t}_{\leftarrow}^{-}\right)^{\mathsf{T}}. \tag{S160}$$

Next, reflection with respect to the slab midplane, $\mathcal{P}_z : z \mapsto -z$, reverses the propagation direction. Both biaxionic parameters are pseudoscalars and therefore transform as

$$\chi(z) \mapsto -\chi(-z), \qquad \tilde{\chi}(z) \mapsto -\tilde{\chi}(-z). \tag{S161}$$

Consequently, this reflection maps $\mathcal{C}_-$ onto $\mathcal{C}_+$ for the centered homogeneous slab. Since the transverse electric-field components are unchanged under $\mathcal{P}_z$, we obtain

$$\hat{t}_{\leftarrow}^{-} = \hat{t}_{\rightarrow}^{+}. \tag{S162}$$

For the lossless media considered here, $\chi$ and $\tilde{\chi}$ are also odd under time reversal, so this spatial relation expresses the $\mathcal{P}_z\mathcal{T}$ symmetry of the material configuration. Combining Eqs. (S160) and (S162) gives

$$\hat{t}_{\rightarrow}^{+} = \left(\hat{t}_{\rightarrow}^{+}\right)^{\mathsf{T}}. \tag{S163}$$

Finally, at normal incidence, rotational symmetry about the $z$ axis restricts the transmission matrix to the form

$$\hat{t} = t^{(\mathrm{co})}\hat{I} - t^{(\mathrm{cr})}\mathbf{e}_z^{\times} = \begin{pmatrix} t^{(\mathrm{co})} & t^{(\mathrm{cr})} \\ -t^{(\mathrm{cr})} & t^{(\mathrm{co})} \end{pmatrix}. \tag{S164}$$

Its symmetry under transposition therefore requires

$$t_{xy} = t_{yx} = 0. \tag{S165}$$

The same argument applies to transmission in the opposite direction. Thus, a homogeneous biaxionic slab preserves the polarization of normally incident transmitted light for the arbitrary homogeneous $\chi$ and $\tilde{\chi}$. The same conclusion holds if $\chi(z)$ and $\tilde{\chi}(z)$ are even functions of the $z$ coordinate relative to the slab center.

## S6. REFLECTION AND TRANSMISSION IN A MULTILAYERED BIGYROTROPIC METAMATERIAL

In this section, we apply transfer matrix method to a multilayered metamaterial whose layers combine gyrotropy of both permittivity and permeability. As was shown previously [1], in the effective medium limit such structure features a combination of axion and dual axion responses. The magnitudes of those responses can be tuned to satisfy an invisibility condition Eq. (S135). Below, we investigate the optical properties of the structure including reflection and transmission coefficients, numerically.

We consider a one-dimensional photonic crystal composed of alternating bi-gyrotropic layers A and B with the same thickness $a/2$, forming a periodic unit cell of size $a$. Each layer exhibits gyroelectric and gyromagnetic responses, described by spatially periodic permittivity and permeability tensors:

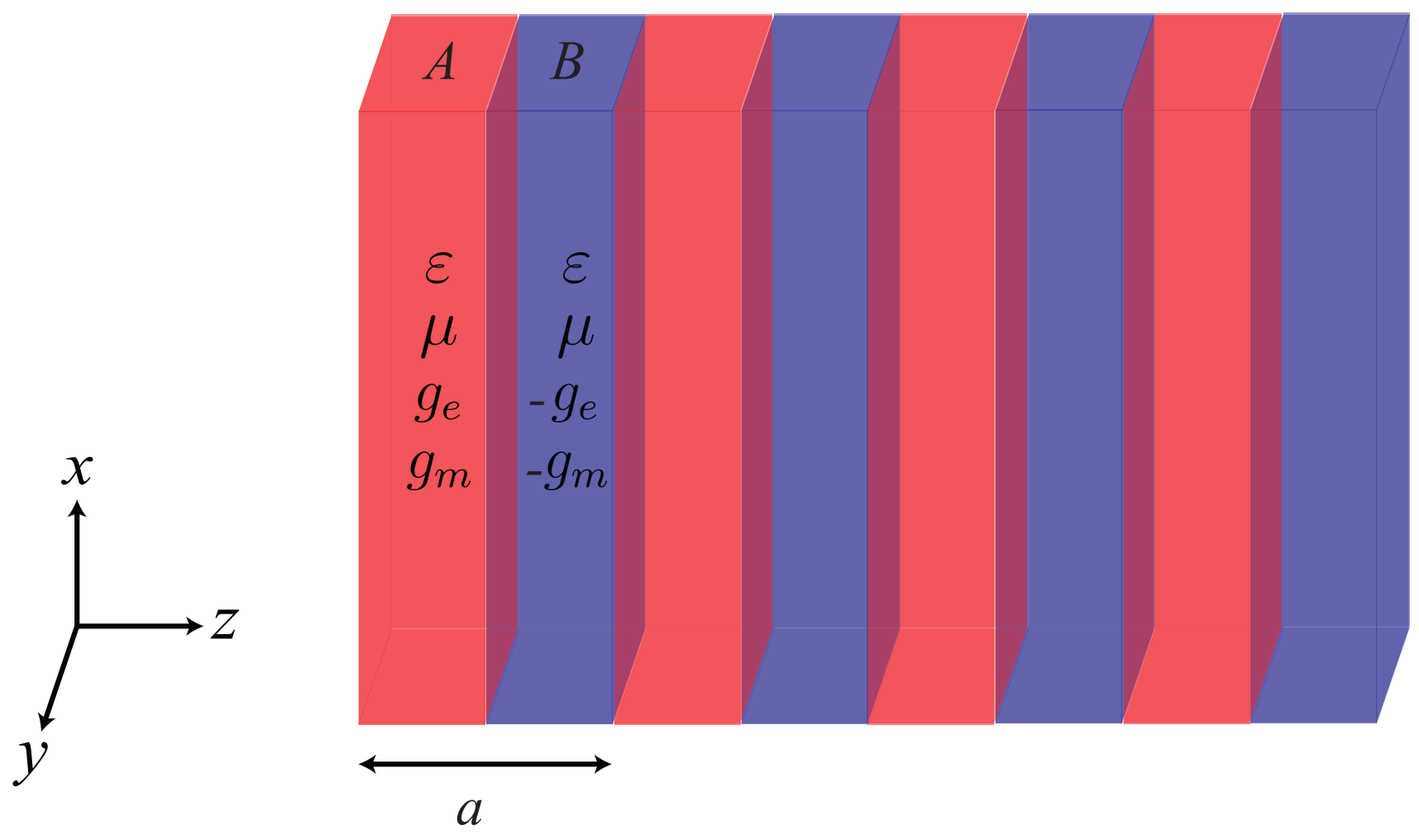


FIG. S6. Schematic of the finite one-dimensional photonic crystal composed of alternating bi-gyrotropic layers, each of thickness $a/2$, forming a unit cell of size $a$. Adjacent layers have opposite electric and magnetic gyrotropy, $g_e$ and $g_m$. The multilayer slab is surrounded by vacuum.

$$\hat{\varepsilon}(z+a)=\hat{\varepsilon}(z)=\begin{pmatrix}\varepsilon & i\,g(z) & 0\\ -i\,g(z) & \varepsilon & 0\\ 0 & 0 & \varepsilon\end{pmatrix}, \qquad \hat{\mu}(z+a)=\hat{\mu}(z)=\begin{pmatrix}\mu & i\,f(z) & 0\\ -i\,f(z) & \mu & 0\\ 0 & 0 & \mu\end{pmatrix}. \tag{S166}$$

The geometry of the multilayered bi-gyrotropic metamaterial placed in vacuum is illustrated schematically in Fig. S6. The constitutive relations for each layer $\alpha = A, B$ are given by:

$$B^i_\alpha = \mu^{ij}_\alpha\, H^j_\alpha, \qquad \hat{\mu}_\alpha = \begin{pmatrix}\mu_\alpha & if_\alpha & 0\\ -if_\alpha & \mu_\alpha & 0\\ 0 & 0 & \mu_\alpha\end{pmatrix}, \tag{S167}$$

$$D^i_\alpha = \varepsilon^{ij}_\alpha\, E^j_\alpha, \qquad \hat{\varepsilon}_\alpha = \begin{pmatrix}\varepsilon_\alpha & ig_\alpha & 0\\ -ig_\alpha & \varepsilon_\alpha & 0\\ 0 & 0 & \varepsilon_\alpha\end{pmatrix}, \tag{S168}$$

We consider opposite gyrotropy in the adjacent layers: $g_A = g_e = -g_B$ and $f_A = g_m = -f_B$. Using the circular polarization basis, $\mathbf{E}_\alpha = \sum_{\eta=\pm1} E_{\alpha,\eta}\,(\mathbf{e}_x + i\eta\,\mathbf{e}_y)$, where $\eta = \pm1$ for right- and left-handed polarization, we can write the $\mathbf{D}$ and $\mathbf{B}$ fields as

$$D_{\alpha,\eta} = \varepsilon_{\alpha,\eta} E_{\alpha,\eta}, \qquad \varepsilon_{\alpha,\eta} = \varepsilon_\alpha - \eta g_e, \tag{S169}$$

$$B_{\alpha,\eta} = \mu_{\alpha,\eta} H_{\alpha,\eta}, \qquad \mu_{\alpha,\eta} = \mu_\alpha - \eta g_m \tag{S170}$$

Therefore, we can construct the electromagnetic fields propagating in a layer $\alpha$ as [5]

$$\mathbf{E}_{\alpha,\eta} = E_{\alpha,\eta}\, e^{ik_{\alpha,\eta}z}\mathbf{v}_\eta + E'_{\alpha,\eta}\, e^{-ik_{\alpha,\eta}z}\mathbf{v}_\eta \tag{S171}$$

and

$$\mathbf{H}_{\alpha,\eta} = -i\eta\sqrt{\frac{\varepsilon_{\alpha,\eta}}{\mu_{\alpha,\eta}}}E_{\alpha,\eta}\, e^{ik_{\alpha,\eta}z}\mathbf{v}_\eta + i\eta\sqrt{\frac{\varepsilon_{\alpha,\eta}}{\mu_{\alpha,\eta}}}E'_{\alpha,\eta}\, e^{-ik_{\alpha,\eta}z}\mathbf{v}_\eta, \tag{S172}$$

where

$$k_{\alpha,\eta} = k_0\sqrt{\varepsilon_{\alpha,\eta}\mu_{\alpha,\eta}}, \tag{S173}$$

is the propagation constant and $k_0 = \omega/c$ is a vacuum wave number, $E_{\alpha,\eta}$ and $E'_{\alpha,\eta}$ are the amplitudes of the wave travelling to the right (left). Here, $\mathbf{v}_\eta = \mathbf{v}_\pm$ vector is given by

$$\mathbf{v}_\eta = \frac{1}{\sqrt{2}} \begin{pmatrix} 1 \\ i\eta \\ 0 \end{pmatrix}. \tag{S174}$$

By applying the conventional boundary conditions, i.e., the continuity of the tangential components of $\mathbf{H}$ and $\mathbf{E}$ fields at the boundaries of the layers, to Eqs. (S171),(S172) we obtain

$$E_A + E'_A = E_B + E'_B, \tag{S175}$$
$$\xi_A E_A - \xi_A E'_A = \xi_B E_B - \xi_B E'_B, \tag{S176}$$

where $\xi_{\alpha,\eta} = -i\eta\sqrt{\varepsilon_{\alpha,\eta}/\mu_{\alpha,\eta}}$. From the latter set of equations, (S175),(S176) we can write the $2\times 2$ interface matrix connecting the fields in layers $A$ to $B$:

$$\begin{pmatrix} E_B \\ E'_B \end{pmatrix} = M_{AB} \begin{pmatrix} E_A \\ E'_A \end{pmatrix}, \qquad M_{AB} = \frac{1}{2} \begin{pmatrix} 1 + \frac{Y_A}{Y_B} & 1 - \frac{Y_A}{Y_B} \\ 1 - \frac{Y_A}{Y_B} & 1 + \frac{Y_A}{Y_B} \end{pmatrix}. \tag{S177}$$

We adopt a basis with the amplitude vector constructed as $(E_{\alpha,+}, E'_{\alpha,+}, E_{\alpha,-}, E'_{\alpha,-})^T$, where $+(-)$ describes the right- (left-) polarization. The $4\times 4$ interface matrix is given by

$$\begin{pmatrix} E_{B,+} \\ E'_{B,+} \\ E_{B,-} \\ E'_{B,-} \end{pmatrix} = \frac{1}{2} \begin{pmatrix} 1 + \frac{Y_{A,+}}{Y_{B,+}} & 1 - \frac{Y_{A,+}}{Y_{B,+}} & 0 & 0 \\ 1 - \frac{Y_{A,+}}{Y_{B,+}} & 1 + \frac{Y_{A,+}}{Y_{B,+}} & 0 & 0 \\ 0 & 0 & 1 + \frac{Y_{A,-}}{Y_{B,-}} & 1 - \frac{Y_{A,-}}{Y_{B,-}} \\ 0 & 0 & 1 - \frac{Y_{A,-}}{Y_{B,-}} & 1 + \frac{Y_{A,-}}{Y_{B,-}} \end{pmatrix} \begin{pmatrix} E_{A,+} \\ E'_{A,+} \\ E_{A,-} \\ E'_{A,-} \end{pmatrix}, \tag{S178}$$

where we define $Y_{\alpha,\eta} = \sqrt{\varepsilon_{\alpha,\eta}/\mu_{\alpha,\eta}}$. Finally, for the case of an electromagnetic wave propagating inside a layer $\alpha = A, B$, with thickness $d = a/2$ and wavevector $k_{\alpha,\eta}$, the $4\times 4$ bulk-propagation matrix is given by

$$M_\alpha = M_\alpha(d) = \begin{pmatrix} e^{ik_{\alpha,+}d} & 0 & 0 & 0 \\ 0 & e^{-ik_{\alpha,+}d} & 0 & 0 \\ 0 & 0 & e^{ik_{\alpha,-}d} & 0 \\ 0 & 0 & 0 & e^{-ik_{\alpha,-}d} \end{pmatrix}. \tag{S179}$$

The resultant transfer matrix describing the gyrotropic multilayered structure with even number of $2N$ layers then reads

$$M_{\text{slab}} = M_{B-\text{air}} M_{AB} (M_{BA} M_B M_{AB} M_A)^{2N} M_{\text{air}-A} = \begin{pmatrix} M^+_{\text{slab}} & 0 \\ 0 & M^-_{\text{slab}} \end{pmatrix}. \tag{S180}$$

This matrix connects incident $E^{in}$, reflected $E^r$ and transmitted $E^t$ fields in the following way:

$$\begin{pmatrix} E^t_\eta \\ 0 \end{pmatrix} = M^\eta_{\text{slab}} \begin{pmatrix} E^i_\eta \\ E^r_\eta \end{pmatrix}, \qquad M^\eta_{\text{slab}} = \begin{pmatrix} M^\eta_{11} & M^\eta_{12} \\ M^\eta_{21} & M^\eta_{22} \end{pmatrix}. \tag{S181}$$

Using this, we deduce the amplitudes of the transmitted and reflected wave for both polarizations

$$E^t_\eta = \frac{\det(M^\eta)}{M^\eta_{22}} E^i_\eta \qquad E^r_\eta = -\frac{M^\eta_{21}}{M^\eta_{22}} E^i_\eta. \tag{S182}$$

It is straightforward to check that $\det(M^\eta) = 1$. For incident $x$-polarized light, the reflected and transmitted fields are:

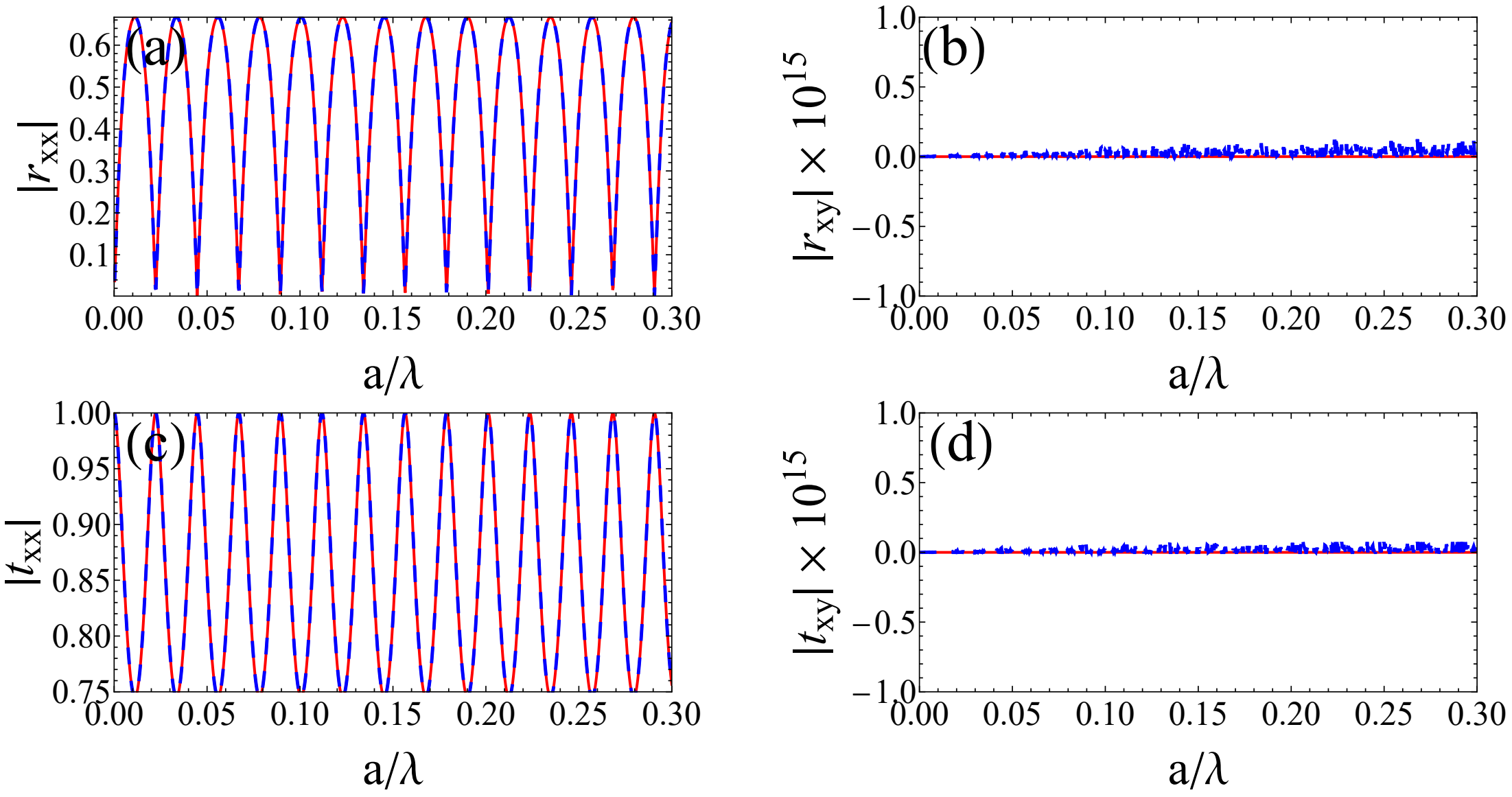


FIG. S7. Validation of the effective biaxionic response. (a-f) The dashed blue lines correspond to the effective medium approximation and red lines to the results of the transfer-matrix method applied to the multilayered structure. Panel (a) corresponds to the co-polarized reflection, panel (b) shows the cross-polarized reflection coefficient which is zero. Panel (c) illustrates the co-polarized transmission, and panel (d) illustrates the cross-polarized transmission which is zero. Simulation parameters: number of periods $N = 10$, $\varepsilon_0 = 1, \mu_0 = 1, \varepsilon = 5, \mu = 1, g_e = 4, g_m = 0.8$.

$$\mathbf{E}^t = \frac{1}{2}\left(\frac{1}{M_{22}^+} + \frac{1}{M_{22}^-}\right) E^i\,\hat{\mathbf{x}} + \frac{i}{2}\left(\frac{1}{M_{22}^+} - \frac{1}{M_{22}^-}\right) E^i\,\hat{\mathbf{y}}, \tag{S183}$$

$$\mathbf{E}^r = -\frac{1}{2}\left(\frac{M_{21}^+}{M_{22}^+} + \frac{M_{21}^-}{M_{22}^-}\right) E^i\,\hat{\mathbf{x}} - \frac{i}{2}\left(\frac{M_{21}^+}{M_{22}^+} - \frac{M_{21}^-}{M_{22}^-}\right) E^i\,\hat{\mathbf{y}}, \tag{S184}$$

leading to the amplitude transmission and reflection coefficients:

$$t_{xx} = \frac{1}{2}\left(\frac{1}{M_{22}^+} + \frac{1}{M_{22}^-}\right), \qquad t_{xy} = \frac{i}{2}\left(\frac{1}{M_{22}^+} - \frac{1}{M_{22}^-}\right), \tag{S185}$$

$$r_{xx} = -\frac{1}{2}\left(\frac{M_{21}^+}{M_{22}^+} + \frac{M_{21}^-}{M_{22}^-}\right), \qquad r_{xy} = -\frac{i}{2}\left(\frac{M_{21}^+}{M_{22}^+} - \frac{M_{21}^-}{M_{22}^-}\right). \tag{S186}$$

Having the microscopic description of the multilayered structure, we now validate the effective description in terms of emergent axion and dual axion fields. We consider a multilayered gyrotropic metamaterial composed of $N = 10$ unit cells, although this number can be easily increased. Each unit cell contains two gyrotropic layers of thickness $a/2$, with $\varepsilon_A = \varepsilon_B = \varepsilon = 5$, $\mu_A = \mu_B = \mu = 1$, $g_e = 4$, and $g_m = 0.8$. The surrounding medium is vacuum ($\varepsilon_0 = \mu_0 = 1$). The effective axion response describes the boundary conditions, and the boundary conditions depend on the termination of the layers [6]. If the first layer has positive gyrotropies, $g_e, g_m$, then the effective axion response and effective dual axion response, $\tilde{\chi}_{\text{eff}}$ are given by [1, 7]

$$\tilde{\chi}_{\text{eff}} = -\frac{\pi}{2}\, g_m\, \xi, \qquad \chi_{\text{eff}} = -\frac{\pi}{2}\, g_e\, \xi \tag{S187}$$

with

$$\varepsilon_{\text{eff}} = \varepsilon + \frac{\pi^2}{12} g_e\left(\mu g_e - \varepsilon g_m\right)\xi^2, \qquad \mu_{\text{eff}} = \mu - \frac{\pi^2}{12} g_m\left(\mu g_e - \varepsilon g_m\right)\xi^2. \tag{S188}$$

up to second-order corrections in period-to-wavelength ratio $\xi = q/b = a/\lambda$ and $\lambda$ is the vacuum wavelength. In this case, the invisibility condition translates into

$$g_e = \frac{\varepsilon}{\mu} g_m, \tag{S189}$$

as it produces the $\tilde{\chi}_{\text{eff}} = \chi_{\text{eff}}(\mu/\varepsilon)$, and $\varepsilon_{\text{eff}} = \varepsilon$, $\mu_{\text{eff}} = \mu$. Therefore, since $\varepsilon = 5$ and $\mu = 1$, we take $g_e = 4$, and $g_m = 0.8$. The transmission and reflection coefficients for the biaxionic slab can be computed analytically as given by Eqs. (S130)-(S133). The optical path inside the slab is defined as:

$$k_2 L = 2\pi\sqrt{\mu_{\text{eff}}\,\varepsilon_{\text{eff}}}\,N\xi, \tag{S190}$$

where $N = 10$ represents the number of periods in the multilayered structure.

Fig. S7 compares the transfer-matrix calculation of the full multilayer structure (solid red) with the analytic effective medium model (dashed blue). The agreement persists up to $\xi$ from 0 to 0.3 confirming the validity of the effective biaxionic description in the long-wavelength regime.

Most importantly, both cross-polarized transmission $t_{xy}$ and cross-polarized reflection $r_{xy}$ vanish, demonstrating that the magneto-electric response is invisible to the plane-wave probes, even though the time reversal symmetry is broken locally inside the structure.

## S7. MICROSCOPIC ORIGIN OF EXACT SCATTERING INVISIBILITY

A natural next question is whether the scattering invisibility persists when the effective medium approximation breaks down, i.e., for the larger values of the period-to-wavelength ratio $\xi = a/\lambda$. Figure S8 compares the effective biaxionic description with the exact transfer-matrix calculation in the interval

$$1.1 \leq \frac{a}{\lambda} \leq 1.3. \tag{S191}$$

Despite the large period-to-wavelength ratio, the co-polarized reflection and transmission coefficients obtained from the two descriptions remain indistinguishable, including frequencies close to Fabry–Pérot resonances. At the same time, the cross-polarized scattering coefficients remain zero.

As shown below, this agreement is not a manifestation of an unexpectedly accurate effective-medium approximation. Instead, it follows from an exact property of the microscopic transfer matrix. Under the microscopic invisibility condition Eq. (S189), the interfaces of the gyrotropic layers are indistinguishable from ordinary dielectric interfaces, while the polarization-dependent propagation phases cancel exactly over one complete unit cell.

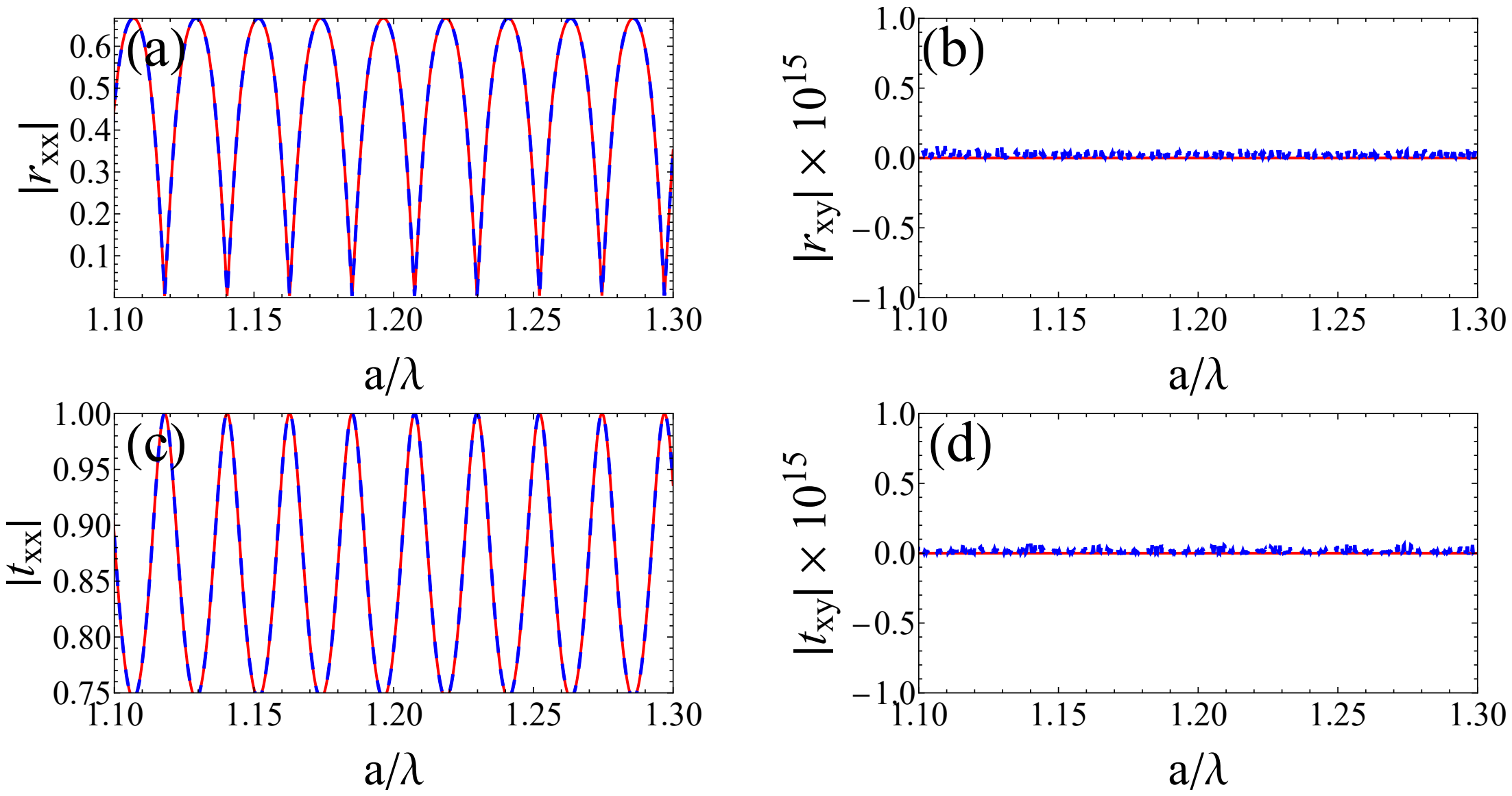


FIG. S8. Exact scattering equivalence beyond the effective-medium regime. The dashed blue curves show effective medium result, while the solid red curves show the exact transfer-matrix calculation for the multilayered gyrotropic structure. Panels (a) and (b) show the co- and cross-polarized reflection coefficients, $|r_{xx}|$ and $|r_{xy}|$, respectively, whereas panels (c) and (d) show the corresponding transmission coefficients, $|t_{xx}|$ and $|t_{xy}|$. The exact agreement of the co-polarized coefficients and the suppression of the cross-polarized coefficients to numerical precision persist for $a/\lambda > 1$. As demonstrated analytically in the text, this behavior does not originate from long-wavelength homogenization, but from an exact scattering equivalence established at the level of a single unit cell.

Let $\eta = \pm 1$ denote the two circularly polarized eigenstates. For the two layers $A$ and $B$, whose electric and magnetic gyrotropies have opposite signs, the corresponding constitutive parameters can be written as

$$\varepsilon_{A,\eta} = \varepsilon + \eta g_e, \qquad \mu_{A,\eta} = \mu + \eta g_m, \tag{S192}$$
$$\varepsilon_{B,\eta} = \varepsilon - \eta g_e, \qquad \mu_{B,\eta} = \mu - \eta g_m. \tag{S193}$$

The microscopic compensation condition is

$$g_e = \frac{\varepsilon}{\mu} g_m, \tag{S194}$$

which is equivalent to

$$\frac{g_e}{\varepsilon} = \frac{g_m}{\mu} \equiv \gamma. \tag{S195}$$

Consequently,

$$\varepsilon_{A,\eta} = \varepsilon(1 + \eta\gamma), \qquad \mu_{A,\eta} = \mu(1 + \eta\gamma), \tag{S196}$$
$$\varepsilon_{B,\eta} = \varepsilon(1 - \eta\gamma), \qquad \mu_{B,\eta} = \mu(1 - \eta\gamma). \tag{S197}$$

Using the admittances introduced in the previous section,

$$Y_{\alpha,\eta} = \sqrt{\frac{\varepsilon_{\alpha,\eta}}{\mu_{\alpha,\eta}}}, \tag{S198}$$

one immediately obtains

$$Y_{A,\eta} = Y_{B,\eta} = \sqrt{\frac{\varepsilon}{\mu}} \equiv Y_{\rm d}, \tag{S199}$$

independently of $\eta$. Hence,

$$Y_{A,+} = Y_{A,-} = Y_{B,+} = Y_{B,-}. \tag{S200}$$

Substitution of Eq. (S199) into the interface matrices, (S178), derived in the previous section shows that the internal interfaces are trivial

$$M_{AB} = M_{BA} = \mathbb{I}_4. \tag{S201}$$

Hence there is no polarization conversion at an $A$–$B$ interface. Moreover, the air–$A$ and $B$–air interface matrices are identical for the two circular polarizations and coincide with the interface matrices of an ordinary isotropic dielectric with admittance $Y_{\rm d}$.

Equation (S201) establishes that every interface behaves like dielectric. Nevertheless, an individual gyrotropic layer is not, by itself, fully equivalent to a dielectric slab because the propagation constants remain polarization dependent. Defining

$$k_{\rm d} = k_0\sqrt{\varepsilon\mu}, \tag{S202}$$

the wavevectors inside the two layers are, for the present parameters $0 < \gamma < 1$,

$$k_{A,\eta} = k_0\sqrt{\varepsilon_{A,\eta}\mu_{A,\eta}} = k_{\rm d}(1 + \eta\gamma), \tag{S203}$$
$$k_{B,\eta} = k_0\sqrt{\varepsilon_{B,\eta}\mu_{B,\eta}} = k_{\rm d}(1 - \eta\gamma). \tag{S204}$$

Thus, the two circular polarizations generally accumulate different phases within one isolated layer.

The phase splitting, however, cancels exactly over one unit cell. For equal layer thicknesses $d = a/2$,

$$k_{A,\eta}d + k_{B,\eta}d = k_{\rm d}a, \tag{S205}$$

which is independent of $\eta$. Using the propagation matrices defined in the previous section, (S179), Eq. (S205) gives

$$\begin{aligned} M_{B,\eta}(d)M_{A,\eta}(d) &= \begin{pmatrix} e^{i(k_{A,\eta}+k_{B,\eta})d} & 0 \\ 0 & e^{-i(k_{A,\eta}+k_{B,\eta})d} \end{pmatrix} \\ &= \begin{pmatrix} e^{ik_{\rm d}a} & 0 \\ 0 & e^{-ik_{\rm d}a} \end{pmatrix} \equiv P_{\rm d}(a). \end{aligned} \tag{S206}$$

Therefore, the exact $4 \times 4$ transfer matrix of one compensated unit cell takes the form

$$M_{\text{cell}} = \begin{pmatrix} P_{\text{d}}(a) & 0 \\ 0 & P_{\text{d}}(a) \end{pmatrix}. \tag{S207}$$

Equation (S207) is precisely the propagation matrix of an isotropic dielectric layer of thickness $a$, permittivity $\varepsilon$, and permeability $\mu$. In particular, this identity is exact and does not require $a/\lambda \ll 1$.

For a slab containing $N$ compensated unit cells,

$$M_{\text{cell}}^{N} = \begin{pmatrix} P_{\text{d}}(Na) & 0 \\ 0 & P_{\text{d}}(Na) \end{pmatrix}. \tag{S208}$$

After including the two external interfaces, which are also polarization independent, the complete slab transfer matrix becomes

$$M_{\text{slab}} = \begin{pmatrix} M_{\text{diel}} & 0 \\ 0 & M_{\text{diel}} \end{pmatrix}, \tag{S209}$$

where $M_{\text{diel}}$ is the transfer matrix of an ordinary dielectric slab of thickness $Na$. Consequently,

$$M_{\text{slab}}^{+} = M_{\text{slab}}^{-}, \tag{S210}$$

and the reflection and transmission amplitudes in the two circular sectors satisfy

$$r_{+} = r_{-} = r_{\text{diel}}, \qquad t_{+} = t_{-} = t_{\text{diel}}. \tag{S211}$$

Transforming back to the linear polarization basis gives

$$r_{xx} = \frac{r_{+} + r_{-}}{2} = r_{\text{diel}}, \qquad r_{xy} = \frac{r_{+} - r_{-}}{2i} = 0, \tag{S212}$$

$$t_{xx} = \frac{t_{+} + t_{-}}{2} = t_{\text{diel}}, \qquad t_{xy} = \frac{t_{+} - t_{-}}{2i} = 0. \tag{S213}$$

In the long-wavelength regime, the same microscopic compensation condition produces effective axion and dual-axion responses satisfying

$$\chi_{\text{eff}} = \tilde{\chi}_{\text{eff}} \frac{\varepsilon_{\text{eff}}}{\mu_{\text{eff}}}, \tag{S214}$$

as derived in Eq. (S187). The corresponding effective biaxionic slab is therefore also scattering-equivalent to the same ordinary dielectric slab. This explains the precise matching between the effective model and transfer-matrix curves in Fig. S8.

The numerical results at large $a/\lambda$ should thus be regarded as a visual confirmation of the exact microscopic identity Eq. (S209). The hidden character of the biaxionic response originates from the two complementary microscopic properties: the compensation condition makes every interface polarization independent, and the opposite gyrotropies cancel the circular phase splitting over one complete unit cell. As a result, the absence of polarization conversion holds for an arbitrary number of unit cells and for the arbitrary values of period-to-wavelength ratio.

## S8. BIAXIONIC SLAB EXCITED BY THE EXTERNAL PLANAR CURRENT: EFFECTIVE MEDIUM ANALYSIS

In this section, we apply the developed effective theory to derive analytically the fields from biaxionic slab with the combination of $\chi$ and $\tilde{\chi}$ responses excited by the macroscopic (i.e. "thick") electric and magnetic planar currents placed in the middle of the slab. These predictions are compared to the rigorous transfer matrix calculation. Note that the particular cases of this problem for the pure axion or dual axion response were solved in Ref. [8].

### A. Transfer matrix in the circular basis

We first derive the transfer matrix for a planar biaxionic interface in the circular basis. For each circular polarization sector $\eta = \pm 1$, we use

$$\mathbf{v}_\eta = \frac{1}{\sqrt{2}}\begin{pmatrix}1\\ i\eta\\ 0\end{pmatrix}, \qquad \mathbf{e}_z\times\mathbf{v}_\eta = -i\eta\mathbf{v}_\eta. \tag{S215}$$

Inside a homogeneous biaxionic region $\alpha$, the bulk Maxwell equations are not modified by constant $\chi_\alpha$ and $\tilde{\chi}_\alpha$. Thus,

$$\mathbf{E}_{\alpha,\eta} = E_{\alpha,\eta}e^{ik_\alpha z}\mathbf{v}_\eta + E'_{\alpha,\eta}e^{-ik_\alpha z}\mathbf{v}_\eta, \tag{S216}$$

and

$$\mathbf{H}_{\alpha,\eta} = \xi_{\alpha,\eta}E_{\alpha,\eta}e^{ik_\alpha z}\mathbf{v}_\eta - \xi_{\alpha,\eta}E'_{\alpha,\eta}e^{-ik_\alpha z}\mathbf{v}_\eta, \tag{S217}$$

where

$$k_\alpha = \frac{\omega}{c}\sqrt{\varepsilon_\alpha\mu_\alpha}, \qquad \xi_{\alpha,\eta} = -i\eta\sqrt{\frac{\varepsilon_\alpha}{\mu_\alpha}}. \tag{S218}$$

The biaxionic boundary conditions at an interface between media 1 and 2 are

$$\mathbf{n}\times\left(\mathbf{E}^{(1)}-\mathbf{E}^{(2)}\right) = -\mathbf{n}\times\left(\tilde{\chi}_1\mathbf{H}^{(1)}-\tilde{\chi}_2\mathbf{H}^{(2)}\right), \tag{S219}$$

$$\mathbf{n}\times\left(\mathbf{H}^{(1)}-\mathbf{H}^{(2)}\right) = \mathbf{n}\times\left(\chi_1\mathbf{E}^{(1)}-\chi_2\mathbf{E}^{(2)}\right), \tag{S220}$$

where $\mathbf{n}$ is a unit vector pointing from medium 1 to medium 2. For each circular sector, these conditions can be written as

$$A_1^{(\eta)}\mathbf{e}_{1,\eta} = A_2^{(\eta)}\mathbf{e}_{2,\eta}, \qquad \mathbf{e}_{\alpha,\eta} = \begin{pmatrix}E_{\alpha,\eta}\\ E'_{\alpha,\eta}\end{pmatrix}, \tag{S221}$$

with

$$A_\alpha^{(\eta)} = \begin{pmatrix}1+\tilde{\chi}_\alpha\xi_{\alpha,\eta} & 1-\tilde{\chi}_\alpha\xi_{\alpha,\eta}\\ \xi_{\alpha,\eta}-\chi_\alpha & -\left(\xi_{\alpha,\eta}+\chi_\alpha\right)\end{pmatrix}. \tag{S222}$$

Thus, the interface transfer matrix connecting the amplitudes across the biaxionic boundary is

$$\begin{pmatrix}E_{2,\eta}\\ E'_{2,\eta}\end{pmatrix} = M_{1\to 2}^{(\eta)}\begin{pmatrix}E_{1,\eta}\\ E'_{1,\eta}\end{pmatrix}, \qquad M_{1\to 2}^{(\eta)} = \left(A_2^{(\eta)}\right)^{-1}A_1^{(\eta)}. \tag{S223}$$

After multiplying the matrices explicitly, one obtains

$$M_{1\to 2}^{(\eta)} = \frac{1}{2\xi_{2,\eta}\left(1+\chi_2\tilde{\chi}_2\right)}\begin{pmatrix}m_{11}^{(\eta)} & m_{12}^{(\eta)}\\ m_{21}^{(\eta)} & m_{22}^{(\eta)}\end{pmatrix}, \tag{S224}$$

with

$$m_{11}^{(\eta)} = \xi_{1,\eta}+\xi_{2,\eta}+\chi_2-\chi_1+\chi_2\tilde{\chi}_1\xi_{1,\eta}+\chi_1\tilde{\chi}_2\xi_{2,\eta}+\left(\tilde{\chi}_1-\tilde{\chi}_2\right)\xi_{1,\eta}\xi_{2,\eta}, \tag{S225}$$

$$m_{12}^{(\eta)} = \xi_{2,\eta}-\xi_{1,\eta}+\chi_2-\chi_1-\chi_2\tilde{\chi}_1\xi_{1,\eta}+\chi_1\tilde{\chi}_2\xi_{2,\eta}+\left(\tilde{\chi}_2-\tilde{\chi}_1\right)\xi_{1,\eta}\xi_{2,\eta}, \tag{S226}$$

$$m_{21}^{(\eta)} = \xi_{2,\eta}-\xi_{1,\eta}+\chi_1-\chi_2-\chi_2\tilde{\chi}_1\xi_{1,\eta}+\chi_1\tilde{\chi}_2\xi_{2,\eta}+\left(\tilde{\chi}_1-\tilde{\chi}_2\right)\xi_{1,\eta}\xi_{2,\eta}, \tag{S227}$$

$$m_{22}^{(\eta)} = \xi_{1,\eta}+\xi_{2,\eta}+\chi_1-\chi_2+\chi_2\tilde{\chi}_1\xi_{1,\eta}+\chi_1\tilde{\chi}_2\xi_{2,\eta}+\left(\tilde{\chi}_2-\tilde{\chi}_1\right)\xi_{1,\eta}\xi_{2,\eta}. \tag{S228}$$

As a consistency check, in the absence of axion and dual axion responses, $\chi_1=\chi_2=\tilde{\chi}_1=\tilde{\chi}_2=0$, the interface matrix reduces to

$$M_{1\to 2}^{(\eta)} = \frac{1}{2}\begin{pmatrix}1+\dfrac{\xi_{1,\eta}}{\xi_{2,\eta}} & 1-\dfrac{\xi_{1,\eta}}{\xi_{2,\eta}}\\ 1-\dfrac{\xi_{1,\eta}}{\xi_{2,\eta}} & 1+\dfrac{\xi_{1,\eta}}{\xi_{2,\eta}}\end{pmatrix}. \tag{S229}$$

Since

$$\xi_{\alpha,\eta} = -i\eta\sqrt{\frac{\varepsilon_\alpha}{\mu_\alpha}}, \tag{S230}$$

one has $\xi_{1,\eta}/\xi_{2,\eta} = Y_1/Y_2$, with $Y_\alpha = \sqrt{\varepsilon_\alpha/\mu_\alpha}$, recovering the standard interface matrix in the circular basis.

### B. Planar current in a finite biaxionic slab

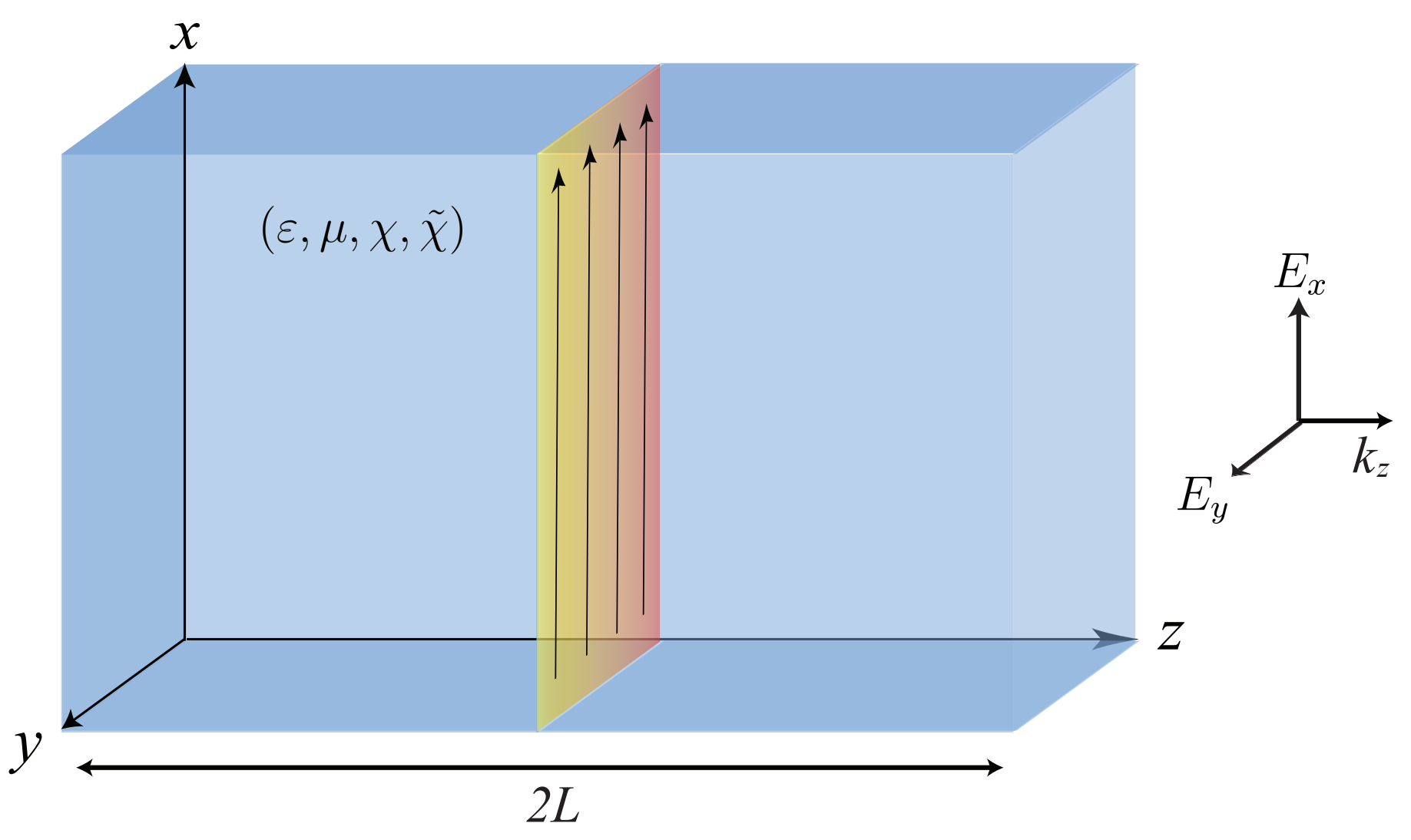


FIG. S9. Macroscopic planar current in a finite biaxionic slab. Schematic of a biaxionic slab of thickness $2L$, characterized by $(\varepsilon, \mu, \chi, \tilde{\chi})$, with a planar current placed at its center. The emitted fields are analyzed on the right side of the slab in terms of co- and cross-polarized components.

We study the excitation of a finite biaxionic slab of thickness $2L$ by the thin planar current placed it its center ($z = 0$), as depicted in Fig. S9. The slab is surrounded by vacuum. For each circular polarization sector, the fields immediately to the right and left of the current plane are denoted by

$$\mathbf{V}_{+,\eta} = \begin{pmatrix} E_{+,\eta} \\ E'_{+,\eta} \end{pmatrix}, \qquad \mathbf{V}_{-,\eta} = \begin{pmatrix} E_{-,\eta} \\ E'_{-,\eta} \end{pmatrix}. \tag{S231}$$

The outgoing amplitudes outside the slab are

$$\mathbf{V}_{R,\eta} = \begin{pmatrix} E_{\eta} \\ 0 \end{pmatrix}, \qquad \mathbf{V}_{L,\eta} = \begin{pmatrix} 0 \\ E'_{\eta} \end{pmatrix}. \tag{S232}$$

The matching condition for the fields $\mathbf{E}_{\pm}$ and $\mathbf{H}_{\pm}$ in the immediate vicinity of the planar source are derived above and correspond to the so-called "no cavity" option:

$$\mathbf{e}_z \times (\mathbf{H}_+ - \mathbf{H}_-) = \frac{4\pi}{c}\mathbf{J}^s_e, \qquad \mathbf{e}_z \times (\mathbf{E}_+ - \mathbf{E}_-) = -\frac{4\pi}{c}\mathbf{J}^s_m. \tag{S233}$$

By decomposing $x$-oriented currents $J^s_e$ and $J^s_m$ into the circular basis, we recover

$$\mathbf{V}_{-,\eta} - \mathbf{V}_{+,\eta} = \mathbf{Q}_{\eta}, \tag{S234}$$

where

$$\mathbf{Q}_{\eta} = \frac{2\pi}{c\sqrt{2}} \begin{pmatrix} ZJ^s_e + i\eta J^s_m \\ i\eta J^s_m - ZJ^s_e \end{pmatrix}, \qquad Z = \sqrt{\frac{\mu}{\varepsilon}}. \tag{S235}$$

The propagation matrix inside the slab coincides with that in the usual dielectric:

$$D_{\eta}(z) = \begin{pmatrix} e^{ikz} & 0 \\ 0 & e^{-ikz} \end{pmatrix}, \qquad k = \frac{\omega}{c}\sqrt{\varepsilon\mu}. \tag{S236}$$

Then

$$\mathbf{V}_{R,\eta} = B_{\eta} D_{\eta}(L)\mathbf{V}_{+,\eta}, \qquad \mathbf{V}_{L,\eta} = B_{\eta} D_{\eta}(-L)\mathbf{V}_{-,\eta}. \tag{S237}$$

where $B_\eta = M^{(\eta)}_{\text{slab}\to\text{vac}}$ is the interface matrix from the slab material to vacuum.

$$M^{(\eta)}_{\text{slab}\to\text{vac}} = \left[A^{(\eta)}_{\text{vac}}\right]^{-1} A^{(\eta)}_{\text{slab}}. \tag{S238}$$

which takes the explicit form

$$B_\eta = \frac{1}{2}\begin{pmatrix} 1+Y-i\eta(\chi+\tilde{\chi}Y) & 1-Y-i\eta(\chi-\tilde{\chi}Y) \\ 1-Y+i\eta(\chi-\tilde{\chi}Y) & 1+Y+i\eta(\chi+\tilde{\chi}Y) \end{pmatrix}. \tag{S239}$$

Equivalently,

$$\mathbf{V}_{+,\eta} = \mathcal{R}_\eta \begin{pmatrix} E_\eta \\ 0 \end{pmatrix}, \qquad \mathbf{V}_{-,\eta} = \mathcal{L}_\eta \begin{pmatrix} 0 \\ E'_\eta \end{pmatrix}, \tag{S240}$$

where

$$\mathcal{R}_\eta = \left[B_\eta D_\eta(L)\right]^{-1}, \qquad \mathcal{L}_\eta = \left[B_\eta D_\eta(-L)\right]^{-1}. \tag{S241}$$

Substitution into the matching condition yields

$$\mathbf{Q}_\eta = \mathcal{L}_\eta \begin{pmatrix} 0 \\ E'_\eta \end{pmatrix} - \mathcal{R}_\eta \begin{pmatrix} E_\eta \\ 0 \end{pmatrix}. \tag{S242}$$

Writing

$$\mathcal{R}_\eta = \begin{pmatrix} r^{(\eta)}_{11} & r^{(\eta)}_{12} \\ r^{(\eta)}_{21} & r^{(\eta)}_{22} \end{pmatrix}, \qquad \mathcal{L}_\eta = \begin{pmatrix} \ell^{(\eta)}_{11} & \ell^{(\eta)}_{12} \\ \ell^{(\eta)}_{21} & \ell^{(\eta)}_{22} \end{pmatrix}, \tag{S243}$$

one obtains

$$E_\eta = \frac{\ell^{(\eta)}_{12} Q_{2,\eta} - \ell^{(\eta)}_{22} Q_{1,\eta}}{r^{(\eta)}_{11}\ell^{(\eta)}_{22} - r^{(\eta)}_{21}\ell^{(\eta)}_{12}}. \tag{S244}$$

In turn, the Cartesian components of the right-emitted field are

$$E_x = \frac{E_+ + E_-}{\sqrt{2}}, \qquad E_y = \frac{i(E_+ - E_-)}{\sqrt{2}}. \tag{S245}$$

We now evaluate the result for a homogeneous biaxionic slab excited by the independent electric and magnetic planar currents with the amplitudes $J^s_e$ and $J^s_m$. Define

$$Y = \sqrt{\frac{\varepsilon}{\mu}}, \qquad Z = \frac{1}{Y}, \qquad \phi = kL, \qquad \Gamma = 1 + \chi\tilde{\chi}, \tag{S246}$$

where $L$ is the half-thickness of the slab. The common denominator is

$$\mathcal{D} = -2Y\Gamma\cos(2\phi) + i\left(1 + Y^2 + \chi^2 + Y^2\tilde{\chi}^2\right)\sin(2\phi). \tag{S247}$$

The right-emitted fields can be written as

$$E_x = \frac{4\pi\Gamma}{c}\,\frac{J^s_e\left(Y\cos\phi - i\sin\phi\right) + J^s_m\left(Y\chi\cos\phi - iY^2\tilde{\chi}\sin\phi\right)}{\mathcal{D}}, \tag{S248}$$

and

$$E_y = \frac{4\pi\Gamma}{c}\,\frac{J^s_e\left(Y\tilde{\chi}\cos\phi - i\chi\sin\phi\right) + J^s_m\left(-Y\cos\phi + iY^2\sin\phi\right)}{\mathcal{D}}. \tag{S249}$$

The same equations can be recast in the vector form valid for the arbitrary orientation of $\mathbf{J}^s_e$ and $\mathbf{J}^s_m$ currents:

$$\begin{aligned} \mathbf{E} = \frac{4\pi\Gamma}{c\mathcal{D}} \Big\{ &\left[Y\cos\phi - i\sin\phi\right]\mathbf{J}^s_e + \left[\tilde{\chi}\,Y\cos\phi - i\chi\sin\phi\right]\mathbf{e}^{\times}_z\mathbf{J}^s_e \\ &+ \left[\chi Y\cos\phi - i\tilde{\chi}Y^2\,\sin\phi\right]\mathbf{J}^s_m + \left[-Y\cos\phi + iY^2\,\sin\phi\right]\mathbf{e}^{\times}_z\mathbf{J}^s_m \Big\}\,. \end{aligned} \tag{S250}$$

### C. Invisibility condition and equivalent dielectric excitation

Next we inspect Eqs. (S248),(S249) under invisibility condition,

$$\tilde{\chi} = \frac{\mu}{\varepsilon}\chi = \frac{\chi}{Y^2}. \tag{S251}$$

The denominator in Eq. (S247) factorizes as

$$\mathcal{D} = \Gamma \mathcal{D}_0, \qquad \Gamma = 1 + \frac{\chi^2}{Y^2}, \qquad \mathcal{D}_0 = -2Y\cos(2\phi) + i(1+Y^2)\sin(2\phi), \tag{S252}$$

while Eqs. (S248) and (S249) simplify to

$$E_x = \frac{4\pi}{c}\left(J_e^s + \chi J_m^s\right)\frac{Y\cos\phi - i\sin\phi}{\mathcal{D}_0}, \tag{S253}$$

$$E_y = \frac{4\pi}{c}\left(J_m^s - \tilde{\chi} J_e^s\right)\frac{-Y\cos\phi + iY^2\sin\phi}{\mathcal{D}_0}. \tag{S254}$$

The source combinations appearing in these expressions are precisely the renormalized currents introduced in Eqs. (S68) and (S69):

$$J_e^{(\mathrm{ren})} = J_e^s + \chi J_m^s, \qquad J_m^{(\mathrm{ren})} = J_m^s - \tilde{\chi} J_e^s. \tag{S255}$$

We therefore conclude that the excitation of biaxionic slab under invisibility condition by the seed planar currents $J_e^s$ and $J_m^s$ produces the same outside fields as an excitation of a dielectric slab with permittivity $\varepsilon$ and permeability $\mu$ by the renormalized currents Eq. (S255). Here we derived this result for the particular geometry. However, as we will show in Sec. S10, this result is general and provides a direct route to simulate numerically continuous biaxionic medium in Comsol Multiphysics for those situations, where analytical solution is not available.

For a purely electric excitation, $J_m^s = 0$, the equivalent dielectric source consists of the correlated pair

$$J_e^{(\mathrm{ren})} = J_e^s, \qquad J_m^{(\mathrm{ren})} = -\tilde{\chi} J_e^s. \tag{S256}$$

The co- and cross-polarized emitted fields are therefore

$$E_x = \frac{4\pi J_e^s}{c}\frac{Y\cos\phi - i\sin\phi}{\mathcal{D}_0}, \tag{S257}$$

$$E_y = \frac{4\pi\tilde{\chi} J_e^s}{c}\frac{Y\cos\phi - iY^2\sin\phi}{\mathcal{D}_0}. \tag{S258}$$

Although the biaxionic slab with invisibility condition has the same scattering response as an ordinary dielectric, its radiation under excitation by the introduced electric source contains a cross-polarized component. In this description, it is linked to the nonzero renormalized magnetic current.

## S9. MULTILAYERED BIGYROTROPIC METAMATERIAL EXCITED BY EXTERNAL CURRENT

In this section, we consider the excitation of the layered metamaterial by external electric and magnetic currents. The problem is solved directly using the transfer matrix method, and the resulting fields are compared with the predictions of the effective medium model introduced in the previous section.

The emitted field is calculated by incorporating the jump conditions generated by the electric and magnetic currents into the transfer matrix of the individual gyrotropic layers. A current plane localized on a scale smaller than the unit-cell period probes its immediate microscopic environment. Consequently, its radiation depends on whether it is placed inside layer $A$, inside layer $B$, or at one of the interfaces. We first derive this position-dependent response. We then construct a finite-width current by coherently superposing planar currents over one complete unit cell. The latter suppresses the dependence on an arbitrary microscopic source position and provides the appropriate quantity for comparison with the effective medium model derived in Sec. S8.

### A. Planar current inside a bi-gyrotropic layer

Let electric and magnetic planar currents be placed at a given position inside the bi-gyrotropic multilayer. The two sides of the current plane are labeled by 1 and 2, with region 1 immediately to the left of the current and region 2 immediately to its right. In general, these regions may have different gyrotropic parameters. If the source lies inside a homogeneous layer, regions 1 and 2 are identical. If it lies at an interface between two adjacent layers, the two regions are different.

For each circular polarization sector $\eta = \pm 1$, we write the fields in region $\alpha = 1, 2$ as

$$\mathbf{E}_{\alpha,\eta} = E_{\alpha,\eta} e^{ik_{\alpha,\eta} z} \mathbf{v}_\eta + E'_{\alpha,\eta} e^{-ik_{\alpha,\eta} z} \mathbf{v}_\eta, \tag{S259}$$

and

$$\mathbf{H}_{\alpha,\eta} = \xi_{\alpha,\eta} E_{\alpha,\eta} e^{ik_{\alpha,\eta} z} \mathbf{v}_\eta - \xi_{\alpha,\eta} E'_{\alpha,\eta} e^{-ik_{\alpha,\eta} z} \mathbf{v}_\eta, \tag{S260}$$

where

$$\xi_{\alpha,\eta} = -i\eta \sqrt{\frac{\varepsilon_{\alpha,\eta}}{\mu_{\alpha,\eta}}}, \qquad k_{\alpha,\eta} = k_0 \sqrt{\varepsilon_{\alpha,\eta} \mu_{\alpha,\eta}}. \tag{S261}$$

The effective permittivity and permeability seen by the circular polarization $\eta = \pm 1$ are

$$\varepsilon_{\alpha,\eta} = \varepsilon_\alpha - \eta g_\alpha, \qquad \mu_{\alpha,\eta} = \mu_\alpha - \eta f_\alpha. \tag{S262}$$

We also define

$$Y_{\alpha,\eta} = \sqrt{\frac{\varepsilon_{\alpha,\eta}}{\mu_{\alpha,\eta}}}, \qquad Z_{\alpha,\eta} = \sqrt{\frac{\mu_{\alpha,\eta}}{\varepsilon_{\alpha,\eta}}} = \frac{1}{Y_{\alpha,\eta}}. \tag{S263}$$

The circular basis vectors are defined as

$$\mathbf{v}_\eta = \frac{1}{\sqrt{2}} \begin{pmatrix} 1 \\ i\eta \\ 0 \end{pmatrix}, \qquad \mathbf{e}_z \times \mathbf{v}_\eta = -i\eta \mathbf{v}_\eta. \tag{S264}$$

We now place electric and magnetic planar currents at the local interface between regions 1 and 2. Their boundary conditions are

$$\mathbf{e}_z \times (\mathbf{H}_2 - \mathbf{H}_1) = \frac{4\pi}{c} \mathbf{J}_e, \tag{S265}$$

$$\mathbf{e}_z \times (\mathbf{E}_2 - \mathbf{E}_1) = -\frac{4\pi}{c} \mathbf{J}_m. \tag{S266}$$

Using Eq. (S264), the boundary conditions in each circular sector become

$$E_{1,\eta} + E'_{1,\eta} = E_{2,\eta} + E'_{2,\eta} + i\eta \frac{4\pi}{c} J_{m,\eta}, \tag{S267}$$

$$\xi_{1,\eta} \left( E_{1,\eta} - E'_{1,\eta} \right) = \xi_{2,\eta} \left( E_{2,\eta} - E'_{2,\eta} \right) - i\eta \frac{4\pi}{c} J_{e,\eta}. \tag{S268}$$

Here, $J_{e,\eta}$ and $J_{m,\eta}$ are the circular components of the electric and magnetic planar currents, respectively. Dividing Eq. (S268) by $\xi_{2,\eta}$ and using $\xi_{\alpha,\eta} = -i\eta Y_{\alpha,\eta}$ yields

$$E_{1,\eta} + E'_{1,\eta} = E_{2,\eta} + E'_{2,\eta} + i\eta \frac{4\pi}{c} J_{m,\eta}, \tag{S269}$$

$$\frac{Y_{1,\eta}}{Y_{2,\eta}} \left( E_{1,\eta} - E'_{1,\eta} \right) = E_{2,\eta} - E'_{2,\eta} + \frac{4\pi}{c} Z_{2,\eta} J_{e,\eta}. \tag{S270}$$

These equations can be written in matrix form as

$$\begin{pmatrix} 1 & 1 \\ \dfrac{Y_{1,\eta}}{Y_{2,\eta}} & -\dfrac{Y_{1,\eta}}{Y_{2,\eta}} \end{pmatrix} \begin{pmatrix} E_{1,\eta} \\ E'_{1,\eta} \end{pmatrix} = \begin{pmatrix} 1 & 1 \\ 1 & -1 \end{pmatrix} \begin{pmatrix} E_{2,\eta} \\ E'_{2,\eta} \end{pmatrix} + \frac{4\pi}{c} \begin{pmatrix} i\eta J_{m,\eta} \\ Z_{2,\eta} J_{e,\eta} \end{pmatrix}. \tag{S271}$$

Solving for the amplitudes on the right side of the planar current, we obtain the inhomogeneous transfer relation

$$\begin{pmatrix} E_{2,\eta} \\ E'_{2,\eta} \end{pmatrix} = M^{(\eta)}_{1\to 2} \begin{pmatrix} E_{1,\eta} \\ E'_{1,\eta} \end{pmatrix} + \mathbf{b}^{(\eta)}_{J,1\to 2}, \tag{S272}$$

where

$$M^{(\eta)}_{1\to 2} = \frac{1}{2} \begin{pmatrix} 1 + \dfrac{Y_{1,\eta}}{Y_{2,\eta}} & 1 - \dfrac{Y_{1,\eta}}{Y_{2,\eta}} \\ 1 - \dfrac{Y_{1,\eta}}{Y_{2,\eta}} & 1 + \dfrac{Y_{1,\eta}}{Y_{2,\eta}} \end{pmatrix}, \tag{S273}$$

and

$$\mathbf{b}^{(\eta)}_{J,1\to 2} = \frac{2\pi}{c} \begin{pmatrix} -Z_{2,\eta} J_{e,\eta} - i\eta J_{m,\eta} \\ Z_{2,\eta} J_{e,\eta} - i\eta J_{m,\eta} \end{pmatrix}. \tag{S274}$$

Note that the electric current contributes with opposite signs to the right- and left-going amplitudes, whereas the magnetic current contributes with the same sign.

We assume that electric and magnetic planar currents are along the $x$ axis,

$$\mathbf{J}_e = J_e\, \mathbf{e}_x, \qquad \mathbf{J}_m = J_m\, \mathbf{e}_x. \tag{S275}$$

Since

$$\mathbf{e}_x = \frac{\mathbf{v}_+ + \mathbf{v}_-}{\sqrt{2}}, \tag{S276}$$

their circular components are

$$J_{e,\eta} = \frac{J_e}{\sqrt{2}}, \qquad J_{m,\eta} = \frac{J_m}{\sqrt{2}}. \tag{S277}$$

The source vector therefore becomes

$$\mathbf{b}^{(\eta)}_{J,1\to 2} = \frac{2\pi}{c\sqrt{2}} \begin{pmatrix} -Z_{2,\eta} J_e - i\eta J_m \\ Z_{2,\eta} J_e - i\eta J_m \end{pmatrix}. \tag{S278}$$

If the source lies inside a single homogeneous gyrotropic layer, then

$$Y_{1,\eta} = Y_{2,\eta}, \qquad Z_{1,\eta} = Z_{2,\eta}, \tag{S279}$$

and Eq. (S273) reduces to the identity:

$$M^{(\eta)}_{1\to 2} = \mathbb{1}. \tag{S280}$$

In this case, the source relation becomes

$$\begin{pmatrix} E_{2,\eta} \\ E'_{2,\eta} \end{pmatrix} = \begin{pmatrix} E_{1,\eta} \\ E'_{1,\eta} \end{pmatrix} + \frac{2\pi}{c\sqrt{2}} \begin{pmatrix} -Z_{1,\eta} J_e - i\eta J_m \\ Z_{1,\eta} J_e - i\eta J_m \end{pmatrix}. \tag{S281}$$

If the source is placed exactly at an interface between two gyrotropic layers, for example with layer $A$ on the left and layer $B$ on the right, then

$$Y_{1,\eta} = Y_{A,\eta}, \qquad Y_{2,\eta} = Y_{B,\eta}, \qquad Z_{2,\eta} = Z_{B,\eta}. \tag{S282}$$

Equation (S272) then gives the complete interface-plus-source transfer step.

### B. Planar currents between two gyrotropic unit cells

We now consider a source placed at the center of a finite multilayer containing an even number of unit cells. In this geometry, the center of the slab lies at the interface between two adjacent unit cells. Since each unit cell consists of layer $A$ followed by layer $B$, the central interface is of the type

$$B|A. \tag{S283}$$

A source shifted to the left of the center therefore lies inside a $B$ layer, whereas a source shifted to the right lies inside an $A$ layer.

We place the central interface at $z = 0$. The layer immediately to its left is layer $B$, occupying

$$-\frac{a}{2} < z < 0, \tag{S284}$$

while the layer immediately to its right is layer $A$, occupying

$$0 < z < \frac{a}{2}. \tag{S285}$$

Both layers have thickness $a/2$, so that $a$ is the lattice period (size of the unit cell).

The propagation matrix in layer $\alpha = A, B$ is

$$P_{\alpha,\eta}(z) = \begin{pmatrix} e^{ik_{\alpha,\eta}z} & 0 \\ 0 & e^{-ik_{\alpha,\eta}z} \end{pmatrix}, \qquad k_{\alpha,\eta} = k_0\sqrt{\varepsilon_{\alpha,\eta}\mu_{\alpha,\eta}}. \tag{S286}$$

The interface matrix between adjacent gyrotropic layers is

$$M^{(\eta)}_{\alpha\to\beta} = \frac{1}{2}\begin{pmatrix} 1 + \dfrac{Y_{\alpha,\eta}}{Y_{\beta,\eta}} & 1 - \dfrac{Y_{\alpha,\eta}}{Y_{\beta,\eta}} \\ 1 - \dfrac{Y_{\alpha,\eta}}{Y_{\beta,\eta}} & 1 + \dfrac{Y_{\alpha,\eta}}{Y_{\beta,\eta}} \end{pmatrix}. \tag{S287}$$

For the finite multilayer, we define the propagators by one unit cell to the right and to the left directions:

$$\mathcal{U}^R_\eta = P_{B,\eta}\left(\frac{a}{2}\right) M^{(\eta)}_{A\to B} P_{A,\eta}\left(\frac{a}{2}\right) M^{(\eta)}_{B\to A}, \tag{S288}$$

and

$$\mathcal{U}^L_\eta = P_{A,\eta}\left(-\frac{a}{2}\right) M^{(\eta)}_{B\to A} P_{B,\eta}\left(-\frac{a}{2}\right) M^{(\eta)}_{A\to B}. \tag{S289}$$

The block $\mathcal{U}^R_\eta$ advances the amplitudes by one unit cell to the right, whereas $\mathcal{U}^L_\eta$ advances them by one unit cell to the left.

Let $N_R$ and $N_L$ be the numbers of complete unit cells to the right and left of the central $B|A$ interface, respectively. For a ten-unit-cell slab centered at this interface, we take

$$N_R = N_L = 5. \tag{S290}$$

#### 1. Source shifted into the left B layer

Let the planar currents be shifted to the left of the central interface, so that they lie inside the $B$ layer at

$$z = -s, \qquad 0 < s < \frac{a}{2}. \tag{S291}$$

Since the source lies inside a homogeneous $B$ layer, the local source relation is purely additive:

$$\mathbf{V}_{+,\eta} = \mathbf{V}_{-,\eta} + \mathbf{b}_{B,\eta}, \tag{S292}$$

where

$$\mathbf{b}_{B,\eta} = \frac{2\pi}{c\sqrt{2}}\begin{pmatrix} -Z_{B,\eta}J_e - i\eta J_m \\ Z_{B,\eta}J_e - i\eta J_m \end{pmatrix}. \tag{S293}$$

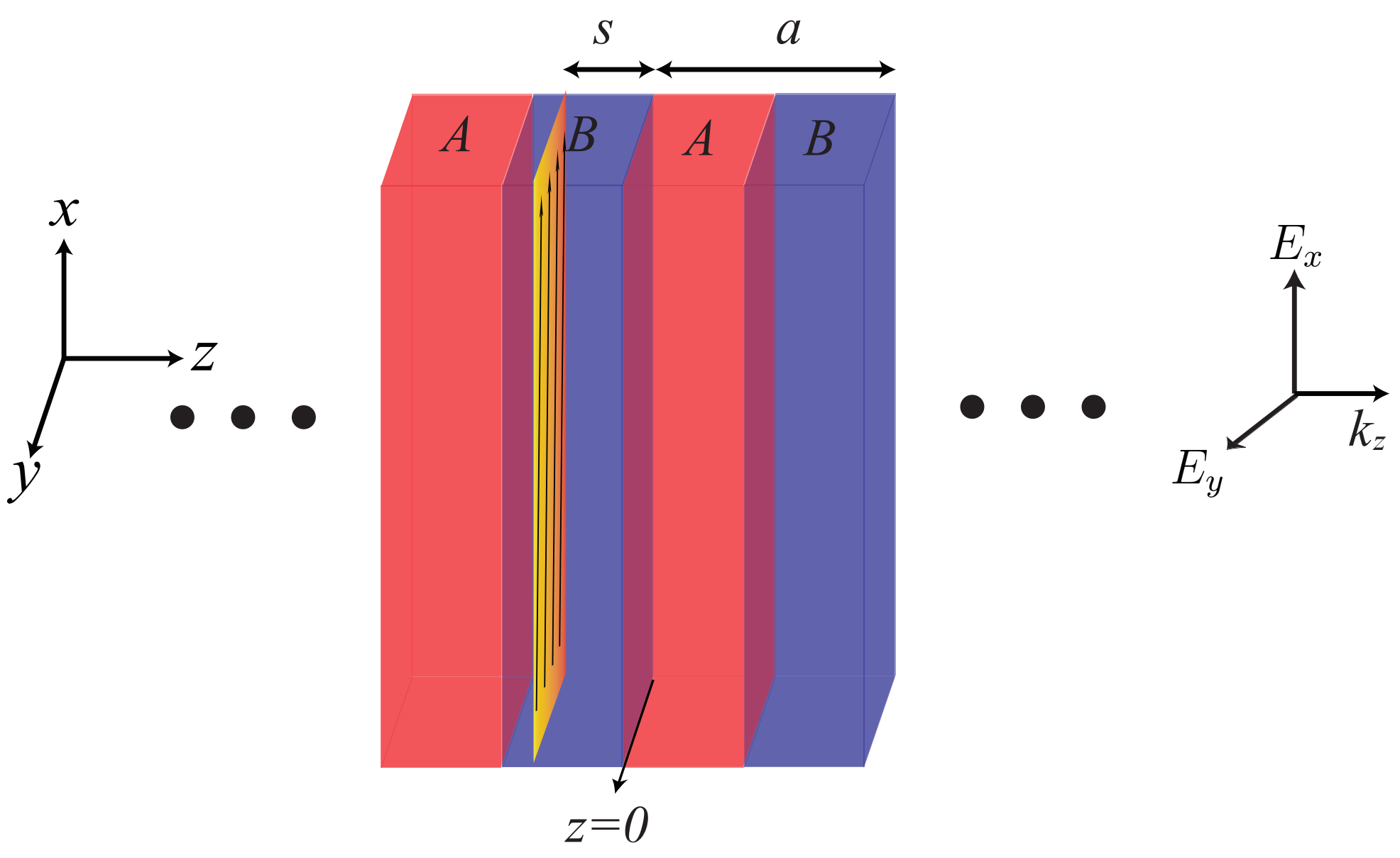


FIG. S10. Planar currents in a multilayered bi-gyrotropic slab. The source is placed inside the left $B$ layer near the central $B|A$ interface between two neighboring unit cells. Only the central part of the finite multilayer is shown explicitly.

Here, $\mathbf{V}_{+,\eta}$ and $\mathbf{V}_{-,\eta}$ denote the amplitude vectors immediately to the right and left of the source, respectively.

The right-emitted field is related to the amplitudes immediately to the right of the source by

$$\mathbf{V}_{R,\eta} = T^{(B)}_{R,\eta}(s; N_R)\mathbf{V}_{+,\eta}, \qquad \mathbf{V}_{R,\eta} = \begin{pmatrix} E^R_\eta \\ 0 \end{pmatrix}, \tag{S294}$$

where

$$T^{(B)}_{R,\eta}(s; N_R) = M^{(\eta)}_{B\to\text{vac}} \left(\mathcal{U}^R_\eta\right)^{N_R-1} P_{B,\eta}\left(\frac{a}{2}\right) M^{(\eta)}_{A\to B} P_{A,\eta}\left(\frac{a}{2}\right) M^{(\eta)}_{B\to A} P_{B,\eta}(s). \tag{S295}$$

Similarly, the left-emitted field is related to the amplitudes immediately to the left of the source by

$$\mathbf{V}_{L,\eta} = T^{(B)}_{L,\eta}(s; N_L)\mathbf{V}_{-,\eta}, \qquad \mathbf{V}_{L,\eta} = \begin{pmatrix} 0 \\ E^L_\eta \end{pmatrix}, \tag{S296}$$

with

$$T^{(B)}_{L,\eta}(s; N_L) = M^{(\eta)}_{A\to\text{vac}} \left(\mathcal{U}^L_\eta\right)^{N_L-1} P_{A,\eta}\left(-\frac{a}{2}\right) M^{(\eta)}_{B\to A} P_{B,\eta}\left[-\left(\frac{a}{2} - s\right)\right]. \tag{S297}$$

We define

$$\mathcal{R}^{(B)}_\eta(s) = \left[T^{(B)}_{R,\eta}(s; N_R)\right]^{-1}, \qquad \mathcal{L}^{(B)}_\eta(s) = \left[T^{(B)}_{L,\eta}(s; N_L)\right]^{-1}. \tag{S298}$$

Writing

$$\mathcal{R}^{(B)}_\eta(s) = \begin{pmatrix} r^{(\eta,B)}_{11}(s) & r^{(\eta,B)}_{12}(s) \\ r^{(\eta,B)}_{21}(s) & r^{(\eta,B)}_{22}(s) \end{pmatrix}, \qquad \mathcal{L}^{(B)}_\eta(s) = \begin{pmatrix} \ell^{(\eta,B)}_{11}(s) & \ell^{(\eta,B)}_{12}(s) \\ \ell^{(\eta,B)}_{21}(s) & \ell^{(\eta,B)}_{22}(s) \end{pmatrix}, \tag{S299}$$

the source equation becomes

$$\mathcal{R}^{(B)}_\eta(s) \begin{pmatrix} E^R_\eta \\ 0 \end{pmatrix} - \mathcal{L}^{(B)}_\eta(s) \begin{pmatrix} 0 \\ E^L_\eta \end{pmatrix} = \mathbf{b}_{B,\eta}. \tag{S300}$$

Solving this system yields the right-emitted circular amplitude

$$E^{R,(B)}_\eta(s) = -\frac{2\pi}{c\sqrt{2}} \frac{Z_{B,\eta} J_e \left[\ell^{(\eta,B)}_{22}(s) + \ell^{(\eta,B)}_{12}(s)\right] + i\eta J_m \left[\ell^{(\eta,B)}_{22}(s) - \ell^{(\eta,B)}_{12}(s)\right]}{r^{(\eta,B)}_{11}(s)\ell^{(\eta,B)}_{22}(s) - r^{(\eta,B)}_{21}(s)\ell^{(\eta,B)}_{12}(s)}. \tag{S301}$$

### 2. Source shifted into the right A layer

We next shift the planar currents to the right of the central interface, so that they lie inside the $A$ layer at

$$z = s, \qquad 0 < s < \frac{a}{2}. \tag{S302}$$

The local source relation is

$$\mathbf{V}_{+,\eta} = \mathbf{V}_{-,\eta} + \mathbf{b}_{A,\eta}, \tag{S303}$$

where

$$\mathbf{b}_{A,\eta} = \frac{2\pi}{c\sqrt{2}} \begin{pmatrix} -Z_{A,\eta} J_e - i\eta J_m \\ Z_{A,\eta} J_e - i\eta J_m \end{pmatrix}. \tag{S304}$$

The transfer matrices from the source to the exterior of the slab are

$$T^{(A)}_{R,\eta}(s; N_R) = M^{(\eta)}_{B\to\text{vac}} \left(\mathcal{U}^R_\eta\right)^{N_R-1} P_{B,\eta}\left(\frac{a}{2}\right) M^{(\eta)}_{A\to B} P_{A,\eta}\left(\frac{a}{2} - s\right), \tag{S305}$$

and

$$T^{(A)}_{L,\eta}(s; N_L) = M^{(\eta)}_{A\to\text{vac}} \left(\mathcal{U}^L_\eta\right)^{N_L-1} P_{A,\eta}\left(-\frac{a}{2}\right) M^{(\eta)}_{B\to A} P_{B,\eta}\left(-\frac{a}{2}\right) M^{(\eta)}_{A\to B} P_{A,\eta}(-s). \tag{S306}$$

Defining

$$\mathcal{R}^{(A)}_\eta(s) = \left[T^{(A)}_{R,\eta}(s; N_R)\right]^{-1}, \qquad \mathcal{L}^{(A)}_\eta(s) = \left[T^{(A)}_{L,\eta}(s; N_L)\right]^{-1}, \tag{S307}$$

and denoting their matrix elements by $r^{(\eta,A)}_{ij}(s)$ and $\ell^{(\eta,A)}_{ij}(s)$, respectively, we obtain

$$E^{R,(A)}_\eta(s) = -\frac{2\pi}{c\sqrt{2}} \frac{Z_{A,\eta} J_e \left[\ell^{(\eta,A)}_{22}(s) + \ell^{(\eta,A)}_{12}(s)\right] + i\eta J_m \left[\ell^{(\eta,A)}_{22}(s) - \ell^{(\eta,A)}_{12}(s)\right]}{r^{(\eta,A)}_{11}(s)\ell^{(\eta,A)}_{22}(s) - r^{(\eta,A)}_{21}(s)\ell^{(\eta,A)}_{12}(s)}. \tag{S308}$$

### 3. Source at the central B|A interface

Finally, consider the planar current placed exactly at the central interface $z = 0$. The local transfer step contains both the $B \to A$ interface matrix and the additive source term:

$$\mathbf{V}_{A,\eta} = M^{(\eta)}_{B\to A} \mathbf{V}_{B,\eta} + \mathbf{b}_{B\to A,\eta}. \tag{S309}$$

Since the transfer relation is solved for the amplitudes on the right-hand $A$ side, the relevant impedance in the electric-current contribution is $Z_{A,\eta}$. Thus,

$$\mathbf{b}_{B\to A,\eta} = \frac{2\pi}{c\sqrt{2}} \begin{pmatrix} -Z_{A,\eta} J_e - i\eta J_m \\ Z_{A,\eta} J_e - i\eta J_m \end{pmatrix}. \tag{S310}$$

The right and left transfer matrices are

$$T^{(BA)}_{R,\eta}(N_R) = M^{(\eta)}_{B\to\text{vac}} \left(\mathcal{U}^R_\eta\right)^{N_R-1} P_{B,\eta}\left(\frac{a}{2}\right) M^{(\eta)}_{A\to B} P_{A,\eta}\left(\frac{a}{2}\right), \tag{S311}$$

and

$$T^{(BA)}_{L,\eta}(N_L) = M^{(\eta)}_{A\to\text{vac}} \left(\mathcal{U}^L_\eta\right)^{N_L-1} P_{A,\eta}\left(-\frac{a}{2}\right) M^{(\eta)}_{B\to A} P_{B,\eta}\left(-\frac{a}{2}\right). \tag{S312}$$

We define

$$\mathcal{R}^{(BA)}_\eta = \left[T^{(BA)}_{R,\eta}\right]^{-1}, \qquad \mathcal{L}^{(BA)}_\eta = \left[T^{(BA)}_{L,\eta}\right]^{-1}, \tag{S313}$$

so that the source equation takes the form

$$\mathcal{R}_{\eta}^{(BA)}\begin{pmatrix}E_{\eta}^{R}\\0\end{pmatrix}-M_{B\to A}^{(\eta)}\mathcal{L}_{\eta}^{(BA)}\begin{pmatrix}0\\E_{\eta}^{L}\end{pmatrix}=\mathbf{b}_{B\to A,\eta}. \tag{S314}$$

For compactness, we introduce

$$\mathcal{K}_{\eta}^{(BA)}=M_{B\to A}^{(\eta)}\mathcal{L}_{\eta}^{(BA)}. \tag{S315}$$

Writing

$$\mathcal{R}_{\eta}^{(BA)}=\begin{pmatrix}r_{11}^{(\eta,BA)} & r_{12}^{(\eta,BA)}\\ r_{21}^{(\eta,BA)} & r_{22}^{(\eta,BA)}\end{pmatrix},\qquad \mathcal{K}_{\eta}^{(BA)}=\begin{pmatrix}k_{11}^{(\eta,BA)} & k_{12}^{(\eta,BA)}\\ k_{21}^{(\eta,BA)} & k_{22}^{(\eta,BA)}\end{pmatrix}, \tag{S316}$$

the right-emitted circular amplitude is

$$E_{\eta}^{R,(BA)}=-\frac{2\pi}{c\sqrt{2}}\frac{Z_{A,\eta}J_e\left(k_{22}^{(\eta,BA)}+k_{12}^{(\eta,BA)}\right)+i\eta J_m\left(k_{22}^{(\eta,BA)}-k_{12}^{(\eta,BA)}\right)}{r_{11}^{(\eta,BA)}k_{22}^{(\eta,BA)}-r_{21}^{(\eta,BA)}k_{12}^{(\eta,BA)}}. \tag{S317}$$

The Cartesian components of the right-emitted electric field are reconstructed from the two circular sectors according to

$$E_x^R=\frac{E_+^R+E_-^R}{\sqrt{2}},\qquad E_y^R=\frac{i\left(E_+^R-E_-^R\right)}{\sqrt{2}}. \tag{S318}$$

### C. Microscopic source-position dependence

We now use the transfer-matrix solution derived above to analyze how the right-emitted field depends on the microscopic position of the planar current inside the central unit cell. The source position is parametrized by $u=z/a$, where $a$ is the unit-cell period and $u=0$ corresponds to the central $B|A$ interface. Negative values of $u$ place the source inside the left $B$ layer, while positive values place it inside the right $A$ layer.

Fig. S11 shows the co- and cross-polarized right-emitted field amplitudes for several positions of the planar electric current. The co-polarized component remains of the same order for all source positions, but its detailed spectral dependence changes with $u$. In contrast, the cross-polarized component is highly sensitive to the microscopic position of the current. It is strongly suppressed when the source is placed exactly at the central $B|A$ interface, while it becomes non-negligible when the source is shifted into either neighboring layer. This confirms that the source probes the local gyrotropic environment rather than only the unit-cell-averaged effective response.

This position dependence has a simple microscopic origin. Under the invisibility condition used here, the circular impedances of the two gyrotropic layers are matched,

$$Y_{A,\eta}=Y_{B,\eta},$$

so that the interfaces between adjacent gyrotropic layers do not produce an additional jump or reflection in the circular basis. The polarization dynamics instead comes from the bulk propagation phases accumulated by the two circular polarization sectors. In a complete unit cell, these phases are compensated: the phase accumulated by a given circular sector in layer $A$ is exchanged with the phase accumulated by the opposite sector in layer $B$. Equivalently,

$$k_{A,+}+k_{B,+}=k_{A,-}+k_{B,-},$$

for equal layer thicknesses. Therefore, after propagation through a full unit cell, the two circular components acquire the same total phase and no net polarization rotation remains.

The central $B|A$ interface is precisely the boundary between two complete unit cells. A source placed at this interface emits waves that, on either side, propagate through complete compensated unit cells. For this reason, the cross-polarized emission is strongly suppressed. In contrast, if the source is placed inside layer $A$ or $B$, or at the internal $A|B$ interface of a unit cell, the emitted wave first propagates through an uncompensated fraction of a gyrotropic layer before the unit-cell compensation is restored. This uncompensated propagation produces a relative phase between the two circular components of the initially linearly polarized source, generating a finite cross-polarized field.

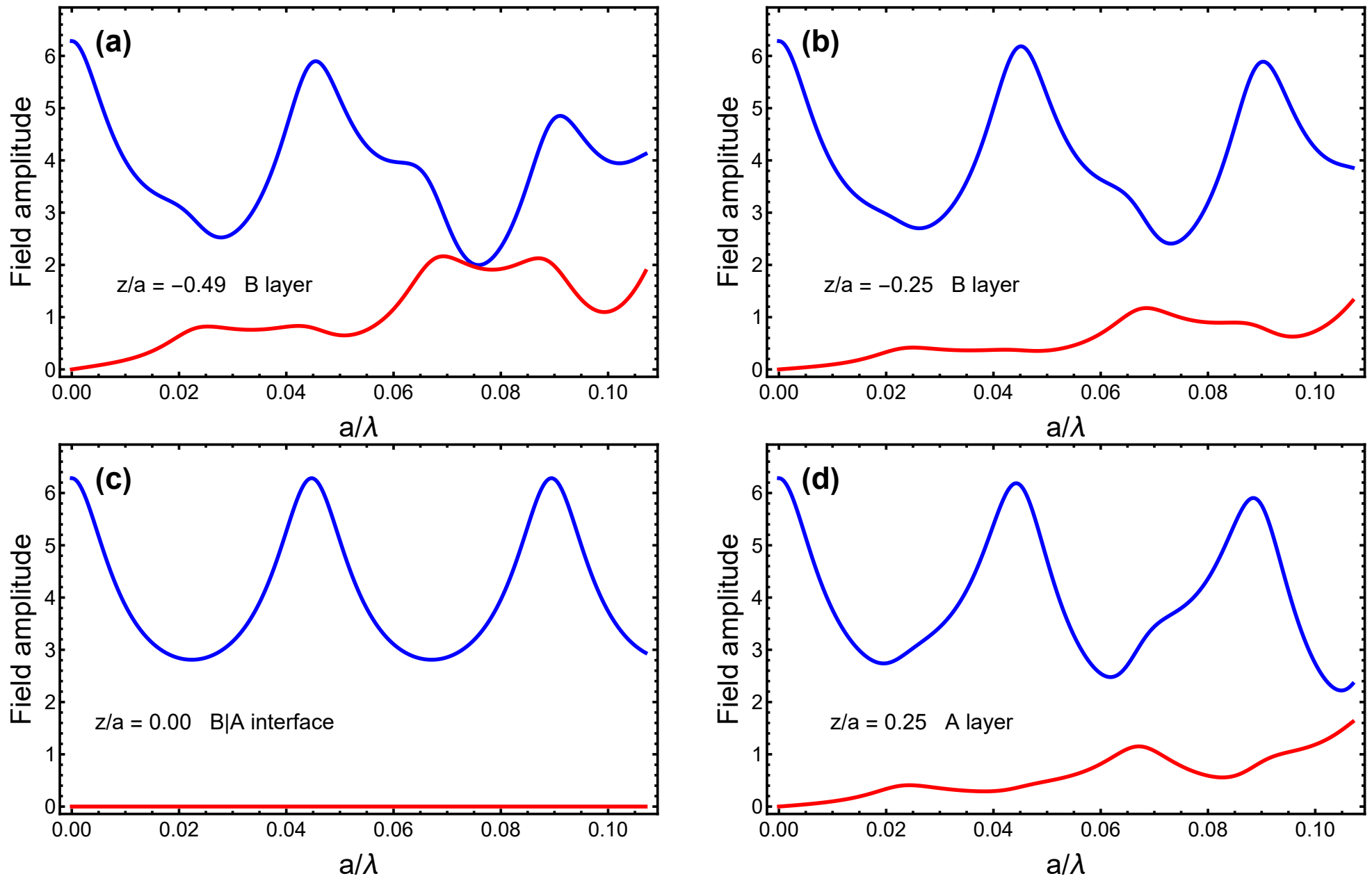


FIG. S11. Microscopic position dependence of emission from an $x$-directed electric planar current, with $J_e = 1$ and $J_m = 0$ in the adopted normalization. Right-emitted field amplitudes generated by an $x$-polarized electric planar current placed at different positions $u = z/a$ inside the central unit cell of the multilayer. The slab contains 10 unit cells, with a central $B|A$ interface. Blue and red curves show the right-emitted electric-field amplitudes $|E_x^R|$ (co-polarized, $x$ direction) and $|E_y^R|$ (cross-polarized, $y$ direction), respectively. The source position is parametrized by $u = z/a$, where $a$ is the unit-cell period. Panels correspond to (a) $u = -0.49$ and (b) $u = -0.25$, inside the left $B$ layer; (c) $u = 0$, at the central $B|A$ interface; and (d) $u = 0.25$, inside the right $A$ layer. The cross-polarized component vanishes at the central interface but becomes finite when the source is shifted into either neighboring layer, revealing the sensitivity of the emitted field to the microscopic source position.

For completeness, we also examine the microscopic position dependence for a purely magnetic planar current directed along the $x$ axis in Fig. S12. In this case, $J_e = 0$ and $J_m \neq 0$. An $x$-directed magnetic current primarily generates a $y$-polarized electric field; consequently, the electric-field components are approximately interchanged relative to the electric excitation. Nevertheless, we retain the same coordinate convention and refer to $E_x$ and $E_y$ as the co- and cross-polarized components, respectively. The magnetic-current emission exhibits the same microscopic source-position sensitivity, arising from the uncompensated gyrotropic propagation between the localized source and the nearest complete unit-cell boundary.

To probe the effective material properties of a multilayer and to reduce the dependence on an arbitrary microscopic placement, we consider a finite-width current distribution spread over one unit cell. This thick-current description provides the results that are compared to the homogenized biaxionic model in the following section.

### D. Green functions of the planar current for the even-cell geometry

To compute the response of the structure to the extended ("thick") current, it is convenient to introduce separate right-emitted Green functions for electric and magnetic planar currents. Both currents are directed along the $x$ axis, and both Green functions describe the emitted electric field. For a source placed inside layer $q = A, B$, linearity gives

$$E_\eta^{R,(q)}(s) = J_e G_{\eta,e}^{R,(q)}(s) + J_m G_{\eta,m}^{R,(q)}(s). \tag{S319}$$

Here, $G_{\eta,e}^{R,(q)}$ is the field per unit electric planar current with $J_m = 0$, whereas $G_{\eta,m}^{R,(q)}$ is the field per unit magnetic planar current with $J_e = 0$. The current amplitudes $J_e$ and $J_m$ are Cartesian amplitudes along $x$, so that their circular components are $J_{e,\eta} = J_e/\sqrt{2}$ and $J_{m,\eta} = J_m/\sqrt{2}$.

For compactness, define

$$\mathcal{D}_\eta^{(q)}(s) = r_{11}^{(\eta,q)}(s)\ell_{22}^{(\eta,q)}(s) - r_{21}^{(\eta,q)}(s)\ell_{12}^{(\eta,q)}(s). \tag{S320}$$

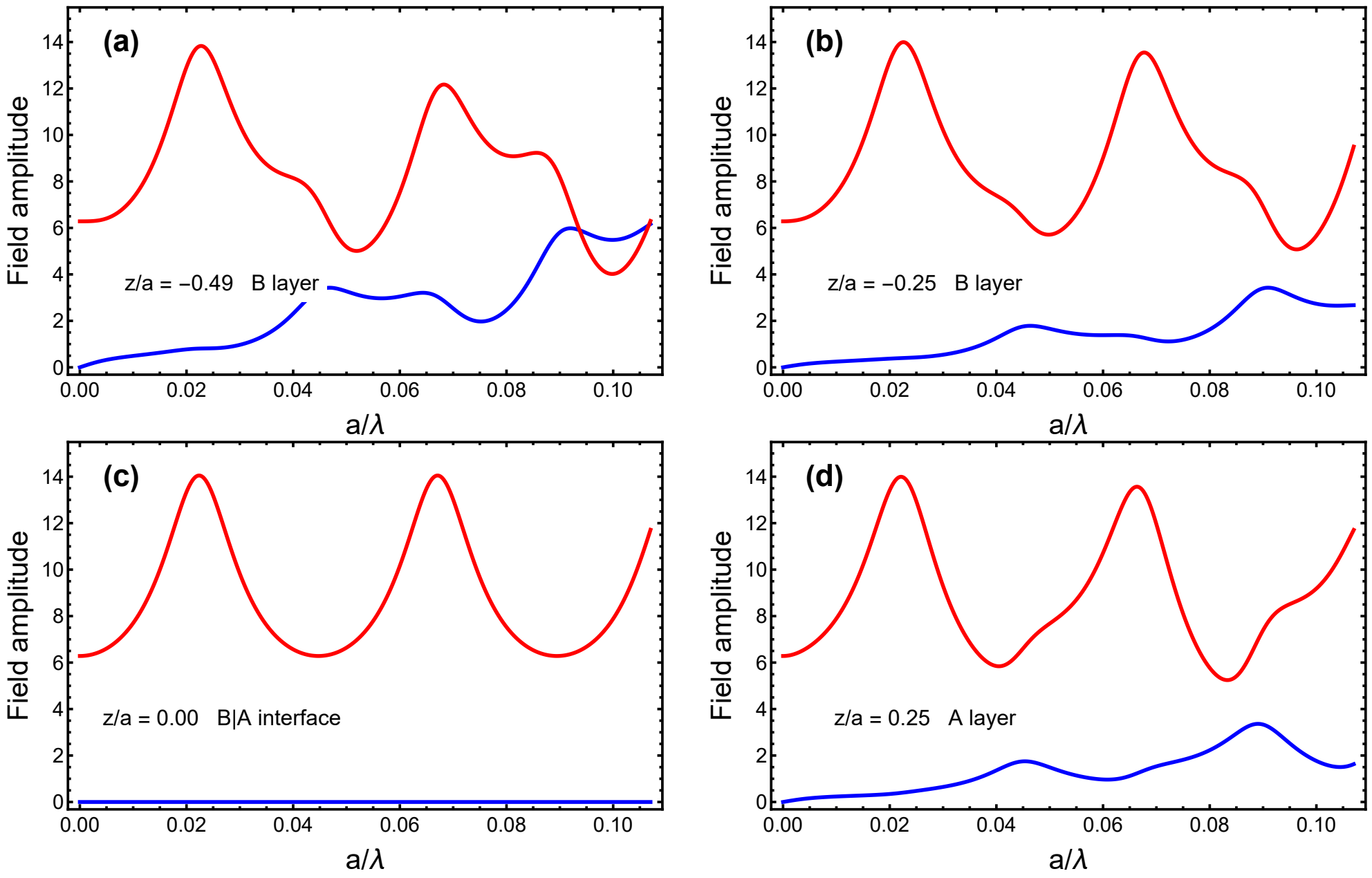


FIG. S12. Microscopic position dependence of emission from an $x$-directed magnetic planar current, with $J_e = 0$ and $J_m = 1$ in the adopted normalization. The slab contains 10 unit cells, with a central $B|A$ interface. Blue and red curves show the right-emitted electric-field amplitudes $|E_x^R|$ (co-polarized) and $|E_y^R|$ (cross-polarized), respectively. Panels correspond to (a) $u = -0.49$ and (b) $u = -0.25$, inside the left $B$ layer; (c) $u = 0$, at the central $B|A$ interface; and (d) $u = 0.25$, inside the right $A$ layer. The fixed Cartesian polarization labels are retained although an $x$-directed magnetic current generates a $y$-polarized electric field in the nongyrotropic limit.

Using the solution for the planar current derived above, we obtain

$$G_{\eta,e}^{R,(q)}(s) = -\frac{2\pi}{c\sqrt{2}} Z_{q,\eta} \frac{\ell_{22}^{(\eta,q)}(s) + \ell_{12}^{(\eta,q)}(s)}{\mathcal{D}_\eta^{(q)}(s)}, \tag{S321}$$

$$G_{\eta,m}^{R,(q)}(s) = -\frac{2\pi i\eta}{c\sqrt{2}} \frac{\ell_{22}^{(\eta,q)}(s) - \ell_{12}^{(\eta,q)}(s)}{\mathcal{D}_\eta^{(q)}(s)}, \qquad q = A, B. \tag{S322}$$

The two Green functions have the same denominator, determined by the multilayer geometry and material parameters. Their numerators differ because electric and magnetic currents impose jumps in the tangential magnetic and electric fields, respectively.

For a source exactly at the central $B|A$ interface, we similarly write

$$E_\eta^{R,(BA)} = J_e G_{\eta,e}^{R,(BA)} + J_m G_{\eta,m}^{R,(BA)}. \tag{S323}$$

Introducing

$$\mathcal{D}_\eta^{(BA)} = r_{11}^{(\eta,BA)} k_{22}^{(\eta,BA)} - r_{21}^{(\eta,BA)} k_{12}^{(\eta,BA)}, \tag{S324}$$

the interface Green functions are

$$G_{\eta,e}^{R,(BA)} = -\frac{2\pi}{c\sqrt{2}} Z_{A,\eta} \frac{k_{22}^{(\eta,BA)} + k_{12}^{(\eta,BA)}}{\mathcal{D}_\eta^{(BA)}}, \tag{S325}$$

$$G_{\eta,m}^{R,(BA)} = -\frac{2\pi i\eta}{c\sqrt{2}} \frac{k_{22}^{(\eta,BA)} - k_{12}^{(\eta,BA)}}{\mathcal{D}_\eta^{(BA)}}. \tag{S326}$$

The factor $Z_{A,\eta}$ in the electric-source Green function appears because the local interface-source relation is solved for the amplitudes on the right-hand $A$ side. No impedance factor multiplies the magnetic current in the corresponding source vector.

Here, $s$ denotes the positive displacement from the central $B|A$ interface. The physical source coordinate is $z = -s$ for $q = B$ and $z = +s$ for $q = A$, with $0 < s < a/2$. We introduce the dimensionless variables

$$\rho = \frac{a}{\lambda}, \qquad u = \frac{z}{a}. \tag{S327}$$

The left half of the central region is a $B$ layer, $-1/2 < u < 0$, while the right half is an $A$ layer, $0 < u < 1/2$. For either source type $\sigma = e, m$, the Green function entering the finite-width calculation is therefore

$$G^{R}_{\eta,\sigma}(\rho, u) = \begin{cases} G^{R,(B)}_{\eta,\sigma}(-ua), & -\dfrac{1}{2} < u < 0, \\ G^{R,(BA)}_{\eta,\sigma}, & u = 0, \\ G^{R,(A)}_{\eta,\sigma}(ua), & 0 < u < \dfrac{1}{2}. \end{cases} \tag{S328}$$

The dependence on $\rho$ on the right-hand side is implicit in the transfer matrices.

For fixed material parameters, the dependence on $\rho$ enters through the bulk propagation phases,

$$k_{\alpha,\eta} z = 2\pi\rho \frac{z}{a} \sqrt{\varepsilon_{\alpha,\eta}\mu_{\alpha,\eta}}. \tag{S329}$$

Consequently,

$$P_{\alpha,\eta}(z) = \begin{pmatrix} \exp\left[i2\pi\rho \frac{z}{a}\sqrt{\varepsilon_{\alpha,\eta}\mu_{\alpha,\eta}}\right] & 0 \\ 0 & \exp\left[-i2\pi\rho \frac{z}{a}\sqrt{\varepsilon_{\alpha,\eta}\mu_{\alpha,\eta}}\right] \end{pmatrix}. \tag{S330}$$

Changing $\rho$ thus changes the propagation phases while leaving the interface matrices unchanged, provided the material parameters are kept fixed.

### E. Finite-width electric and magnetic currents centered at the interface between two unit cells

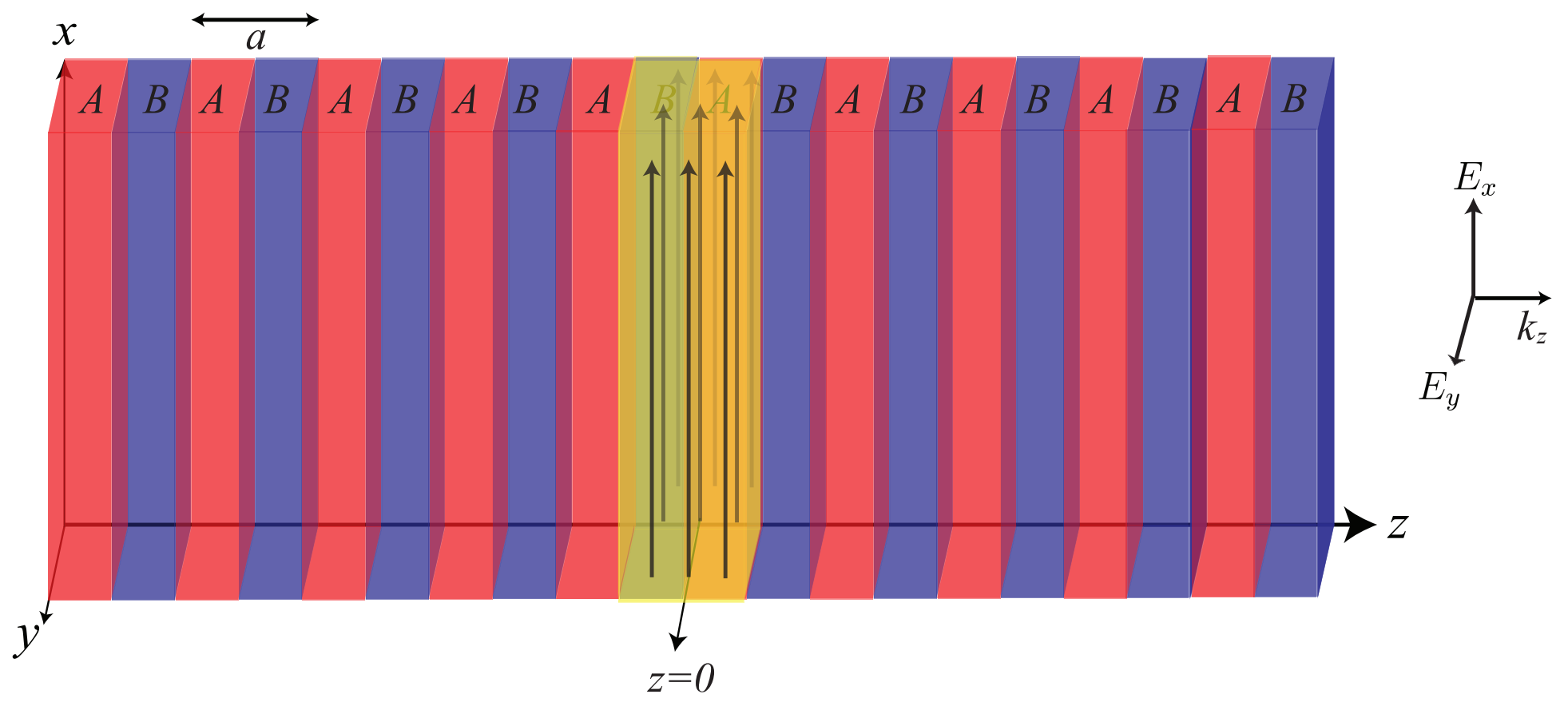


FIG. S13. Finite-width source in the even-cell multilayer geometry. The source distribution is centered at the central $B|A$ interface and is represented numerically as a coherent superposition of planar currents. The illustrated distribution spans one lattice period, occupying both the left $B$ layer and the right $A$ layer. Each elementary plane carries electric and magnetic currents directed along $x$.

We now construct the response to electric and magnetic current distributions of finite thickness. Both currents are directed along $x$ and are supported within the central region of the multilayer. Their spatial profiles may be chosen independently.

Let $j_e(z)$ and $j_m(z)$ denote the corresponding volume-current density amplitudes. An infinitesimal slice of thickness $dz = a\,du$ carries the current amplitudes

$$dJ_e(u) = a\,j_e(au)\,du, \qquad dJ_m(u) = a\,j_m(au)\,du. \tag{S331}$$

By linearity, the right-emitted circular electric field is

$$E_\eta^{R,\mathrm{thick}}(\rho) = \int_{u_-}^{u_+} \left[G_{\eta,e}^R(\rho,u)\,dJ_e(u) + G_{\eta,m}^R(\rho,u)\,dJ_m(u)\right]. \tag{S332}$$

The integration interval contains the support of both source profiles. For a distribution centered at the central interface and having thickness $h \le a$, we take

$$u_- = -\frac{h}{2a}, \qquad u_+ = \frac{h}{2a}. \tag{S333}$$

A source spanning one full period corresponds to $h = a$.

For profiles with nonzero integrated current, we write

$$dJ_\sigma(u) = J_{\sigma,\mathrm{tot}} w_\sigma(u)\,du, \qquad \sigma = e, m, \tag{S334}$$

where

$$J_{\sigma,\mathrm{tot}} = \int_{au_-}^{au_+} j_\sigma(z)\,dz, \qquad \int_{u_-}^{u_+} w_\sigma(u)\,du = 1. \tag{S335}$$

The total currents and profiles may be complex, allowing a relative phase between the electric and magnetic sources as well as spatial phase variations. A source with zero integrated current but a nonzero local distribution is treated directly using Eq. (S332).

The emitted field becomes

$$E_\eta^{R,\mathrm{thick}}(\rho) = \sum_{\sigma=e,m} J_{\sigma,\mathrm{tot}} \int_{u_-}^{u_+} w_\sigma(u) G_{\eta,\sigma}^R(\rho,u)\,du. \tag{S336}$$

To probe the effective properties, we consider spatially uniform profile of the current. Such source directly probes the effective material parameters without significant excitation of the higher-order Bloch harmonics. For electric and magnetic currents with the same uniform profile,

$$w_e(u) = w_m(u) = \frac{1}{u_+ - u_-}, \qquad u_- < u < u_+, \tag{S337}$$

this reduces to

$$E_\eta^{R,\mathrm{thick}}(\rho) = \frac{1}{u_+ - u_-} \int_{u_-}^{u_+} \left[J_{e,\mathrm{tot}} G_{\eta,e}^R(\rho,u) + J_{m,\mathrm{tot}} G_{\eta,m}^R(\rho,u)\right] du. \tag{S338}$$

With this normalization, changing the source thickness while keeping $J_{e,\mathrm{tot}}$ and $J_{m,\mathrm{tot}}$ fixed preserves the integrated planar current amplitudes.

In the numerical calculation, the integrals are approximated by a finite sum of $N_s$ planar current planes. Using the midpoint rule,

$$u_n = u_- + \left(n - \frac{1}{2}\right)\Delta u, \qquad \Delta u = \frac{u_+ - u_-}{N_s}, \qquad n = 1, \ldots, N_s. \tag{S339}$$

For each source type, the discretized amplitudes of planar currents can be normalized as

$$J_{\sigma,n} = J_{\sigma,\mathrm{tot}} \frac{w_\sigma(u_n)\Delta u}{\sum_{m=1}^{N_s} w_\sigma(u_m)\Delta u}, \qquad \sigma = e, m. \tag{S340}$$

This ensures $\sum_{n=1}^{N_s} J_{\sigma,n} = J_{\sigma,\mathrm{tot}}$, provided that the discrete normalization denominator is nonzero. For a uniform profile, the weights reduce to

$$J_{e,n} = \frac{J_{e,\mathrm{tot}}}{N_s}, \qquad J_{m,n} = \frac{J_{m,\mathrm{tot}}}{N_s}. \tag{S341}$$

The right-emitted field is then

$$E_\eta^{R,\mathrm{thick}}(\rho) \simeq \sum_{n=1}^{N_s} \left[ J_{e,n} G_{\eta,e}^R(\rho, u_n) + J_{m,n} G_{\eta,m}^R(\rho, u_n) \right]. \tag{S342}$$

Alternatively, when the volume-current profiles are prescribed directly, the midpoint weights are $J_{\sigma,n} = a\, j_\sigma(au_n)\Delta u$, without an additional normalization.

For a symmetric interval and an even number $N_s$ of current planes, the midpoint grid avoids $u = 0$. In particular, for the full central region $-1/2 < u < 1/2$, half of the sampling points lie inside the left $B$ layer and half inside the right $A$ layer. For an ordinary finite-width profile, the value of the Green function at the single interface point does not affect the continuum integral. If an additional delta-localized current plane is present at $u = 0$, its contribution must instead be included explicitly using the interface Green functions. The finite-width geometry and its discretization are illustrated in Fig. S13.

After evaluating the two circular amplitudes, the Cartesian components of the right-emitted electric field are reconstructed as

$$E_x^{R,\mathrm{thick}} = \frac{E_+^{R,\mathrm{thick}} + E_-^{R,\mathrm{thick}}}{\sqrt{2}}, \qquad E_y^{R,\mathrm{thick}} = \frac{i\left(E_+^{R,\mathrm{thick}} - E_-^{R,\mathrm{thick}}\right)}{\sqrt{2}}. \tag{S343}$$

The spatial summation and the superposition of electric and magnetic source contributions are performed over complex amplitudes. The magnitudes $|E_x^{R,\mathrm{thick}}|$ and $|E_y^{R,\mathrm{thick}}|$ are taken only afterward, retaining interference between different source positions and between the two source types. Throughout, $E_x$ and $E_y$ retain the co- and cross-polarized labels, respectively.

For a ten-unit-cell slab, five complete unit cells lie on either side of the central $B|A$ interface. This geometry is included through the finite-stack transfer matrices entering $G_{\eta,e}^R(\rho, u)$ and $G_{\eta,m}^R(\rho, u)$. Changing the number of unit cells modifies these Green functions but does not alter the superposition formula (S342). Purely electric or purely magnetic excitation is recovered by setting the magnetic or electric current distribution, respectively, to zero.

### F. Validation of effective biaxionic response with sources

We compare the rigorous multilayer calculation to the effective description based on renormalized sources. The microscopic slab contains 10 unit cells and is excited by an $x$-directed electric current uniformly distributed over the central region $-1/2 < u < 1/2$, where $u = z/a$. No magnetic current is imposed in the microscopic calculation. The emitted field is obtained by coherently summing the complex amplitudes generated by the elementary current planes. By extending over one complete period, the source samples both neighboring layers and eliminates the sensitivity to an individual microscopic source position.

The layer parameters are $\varepsilon_A = \varepsilon_B = 5$, $\mu_A = \mu_B = 1$, $(g_A, f_A) = (4, 0.8)$, and $(g_B, f_B) = (-4, -0.8)$. They satisfy the invisibility condition

$$\frac{g_0}{\varepsilon} = \frac{f_0}{\mu}, \qquad \tilde{\chi} = \frac{\mu}{\varepsilon}\chi, \tag{S344}$$

with $\chi$ and $\tilde{\chi}$ determined by the first-order homogenization of the bigyrotropic unit cell.

In turn, the effective response is evaluated under the same invisibility condition, using an ordinary dielectric slab with the same effective permittivity, permeability, and total thickness, excited by the renormalized currents. For the purely electric excitation considered here, the prescribed volume-current densities are

$$\mathbf{j}_e(z) = j_e(z)\, \mathbf{e}_x, \qquad \mathbf{j}_m(z) = 0, \tag{S345}$$

and the corresponding renormalized sources are

$$\mathbf{j}_e^{(\mathrm{ren})}(z) = j_e(z)\, \mathbf{e}_x, \qquad \mathbf{j}_m^{(\mathrm{ren})}(z) = -\tilde{\chi}\, j_e(z)\, \mathbf{e}_x. \tag{S346}$$

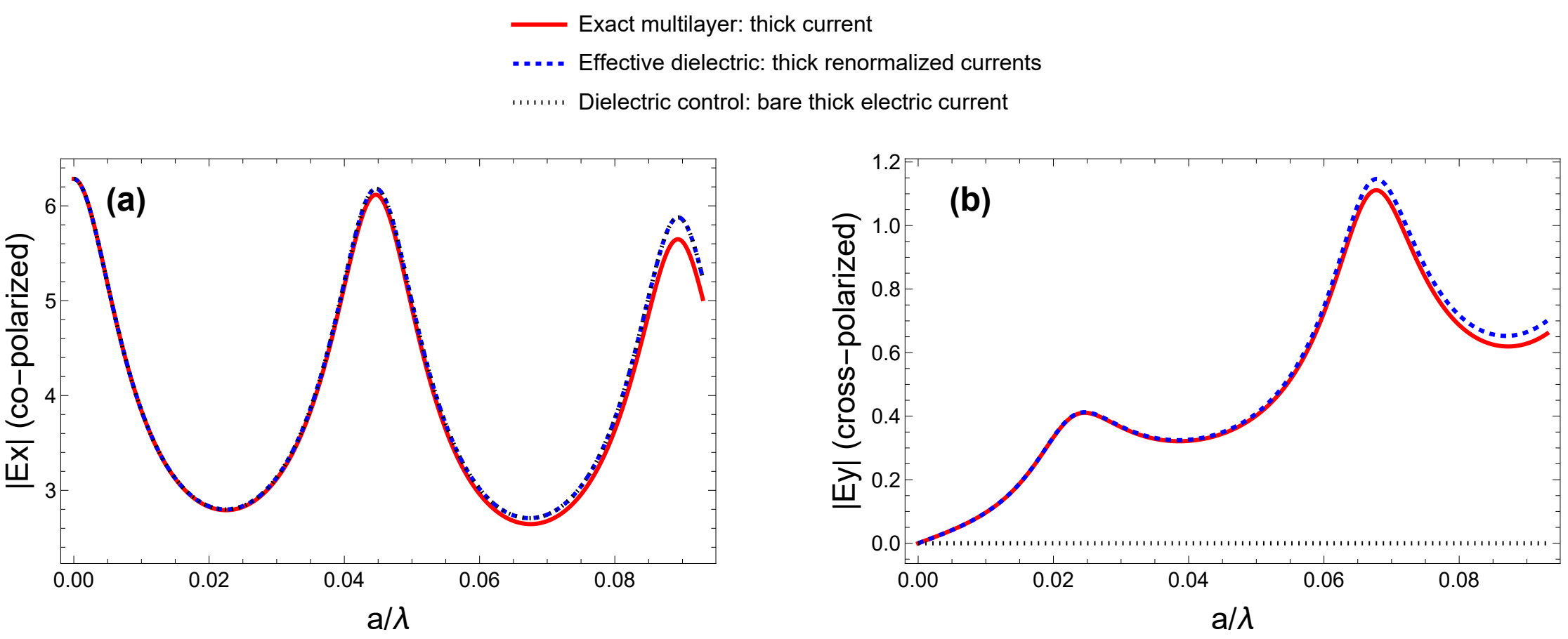


FIG. S14. Validation of the effective biaxionic response using renormalized sources. The microscopic slab contains 10 unit cells with $\varepsilon_A = \varepsilon_B = 5$, $\mu_A = \mu_B = 1$, $(g_A, f_A) = (4, 0.8)$, and $(g_B, f_B) = (-4, -0.8)$. An $x$-directed electric current with unit integrated amplitude is uniformly distributed over $-1/2 < u < 1/2$. Red solid curves show the exact multilayer calculation, blue dashed curves show the effective dielectric slab excited by the renormalized electric and magnetic currents, and black dotted curves show the dielectric control excited only by the bare electric current. All three calculations retain the finite source thickness and coherently sum the elementary contributions from planar currents. Panels show (a) the co-polarized amplitude $|E_x^R|$ and (b) the cross-polarized amplitude $|E_y^R|$.

This prescription is applied throughout the source region. Consequently, the effective electric and magnetic currents retain the same finite spatial support as the microscopic electric current. The effective field is calculated by coherently summing their contributions at the same source positions, including propagation and reflection within the dielectric slab.

As a reference, we also calculate the field emitted by the same dielectric slab excited only by the electric current. This reference calculation uses the same current profile and integrated electric-current amplitude, but its magnetic current is zero.

Figure S14 compares the right-emitted electric-field amplitudes. The effective description with renormalized sources closely reproduces the microscopic result in both polarization channels within the homogenization regime. Since $j_e^{(\mathrm{ren})} = j_e$, the effective and control calculations give identical co-polarized fields. Their distinction appears in the cross-polarized channel: the bare electric current in the ordinary dielectric produces no $E_y^R$, whereas the renormalized magnetic current generates a finite cross-polarized contribution.

This comparison tests the effective theory through radiation from an internal source. Although the biaxionic slab under invisibility condition exhibits the same scattering properties as an ordinary dielectric, its emission under electric excitation contains an additional polarization component. In the equivalent dielectric calculation, this component is accounted for by the renormalized magnetic current in Eq. (S346).

## S10. FULL-WAVE SIMULATION OF THE EXCITATION BY THE SOURCE

Next we validate the developed effective description for the source interaction with biaxionic medium using full-wave numerical simulations in Comsol Multiphysics software package, which allows us to treat other types of sources, different from the planar ones and closer to the experimental situation.

We study the metamaterial composed of bigyrotropic layers considered in the main text, with alternating electric and magnetic gyrotropies satisfying Eq. (S189). Its homogenized description, to the first order in the period-to-wavelength ratio, is characterized by

$$\chi_{\mathrm{eff}}(f) = -\frac{\pi}{2}\, g_e\, \xi(f), \qquad \tilde{\chi}_{\mathrm{eff}}(f) = -\frac{\pi}{2}\, g_m\, \xi(f), \tag{S347}$$

where

$$\xi(f) = \frac{a}{\lambda} = \frac{af}{c} = \frac{aq}{2\pi}, \qquad q = \frac{\omega}{c}. \tag{S348}$$

Here $\lambda = c/f$ is the vacuum wavelength and $a$ is the multilayer period. Up to the second order in $\xi$, effective permittivity and permeability are given by Eq. (S41):

$$\begin{aligned} \varepsilon_{\text{eff}} &= \varepsilon + \frac{\pi^2}{12} g_e \left(\mu g_e - \varepsilon g_m\right) \xi^2 + O(\xi^3), \\ \mu_{\text{eff}} &= \mu + \frac{\pi^2}{12} g_m \left(\varepsilon g_m - \mu g_e\right) \xi^2 + O(\xi^3). \end{aligned} \tag{S349}$$

For the parameters used in the simulations,

$$\varepsilon = 5, \qquad \mu = 1, \qquad g_e = 4, \qquad g_m = 0.8, \tag{S350}$$

the microscopic compensation condition

$$\frac{g_m}{\mu} = \frac{g_e}{\varepsilon} \tag{S351}$$

guarantees that

$$\tilde{\chi}_{\text{eff}}(f) = \frac{\mu}{\varepsilon} \chi_{\text{eff}}(f) \tag{S352}$$

and

$$\varepsilon_{\text{eff}} = \varepsilon, \qquad \mu_{\text{eff}} = \mu. \tag{S353}$$

at every frequency within the homogenization regime up to the second order in $\xi$.

The biaxionic response is not available as a native material response in COMSOL Multiphysics, while axion response can be simulated. Therefore, we take the equations of biaxion electrodynamics with the sources and perform a mapping transformation redefining the fields as

$$\mathbf{E}' = \mathbf{E} + \tilde{\chi}\,\mathbf{H}\,, \tag{S354}$$

$$\mathbf{B}' = \mathbf{B} - \tilde{\chi}\,\mathbf{D}\,. \tag{S355}$$

Such field redefinition does not affect the source terms and thus we recover the equations

$$\nabla \times (\mathbf{H}' - \chi'\,\mathbf{E}') = \frac{1}{c}\frac{\partial}{\partial t}\left(\mathbf{D}' + \chi'\,\mathbf{B}'\right) + \frac{4\pi}{c}\,\mathbf{j}_e^{(\text{ren})}\,, \tag{S356}$$

$$\nabla \cdot (\mathbf{D}' + \chi'\,\mathbf{B}') = 4\pi \rho_e^{(\text{ren})}\,, \tag{S357}$$

$$\nabla \times \mathbf{E}' = -\frac{1}{c}\frac{\partial \mathbf{B}'}{\partial t} - \frac{4\pi}{c}\,\mathbf{j}_m^{(\text{ren})}\,, \tag{S358}$$

$$\nabla \cdot \mathbf{B}' = 4\pi \rho_m^{(\text{ren})}\,. \tag{S359}$$

This mapping suggests that the fields produced by the seed sources $\mathbf{j}_e$ and $\mathbf{j}_m$ placed in the general biaxionic medium with the parameters $\varepsilon$, $\mu$, $\chi$, $\tilde{\chi}$ and measured outside of that medium are exactly the same as those produced by the renormalized currents $\mathbf{j}_e^{(\text{ren})} = \mathbf{j}_e + \chi\,\mathbf{j}_m$ and $\mathbf{j}_m^{(\text{ren})} = \mathbf{j}_m - \tilde{\chi}\,\mathbf{j}_e$ placed in the axion medium with the material parameters $\varepsilon'$, $\mu'$ and $\chi'$ given by the mapping Eqs. (S99),(S100). The latter scenario can be readily simulated in Comsol or other commercial software packages.

Therefore, to compute numerically the predictions of the effective medium model, we simulate an equivalent ordinary dielectric with

$$\varepsilon_{\text{eff}} = 5, \qquad \mu_{\text{eff}} = 1, \tag{S360}$$

and replace the physical electric antenna by the frequency-dependent renormalized source pair

$$\mathbf{J}_{\text{e}}^{(\text{ren})}(\mathbf{r}, f) = \mathbf{J}_{\text{e}}(\mathbf{r}), \tag{S361}$$

$$\mathbf{J}_{\text{m}}^{(\text{ren})}(\mathbf{r}, f) = -\tilde{\chi}_{\text{eff}}(f)\mathbf{J}_{\text{e}}(\mathbf{r}), \tag{S362}$$

Because both effective biaxionic parameters depend on frequency through $\xi(f)$, the amplitudes of both mapped sources are recalculated at every frequency point of the sweep. In COMSOL, Eqs. (S347) are introduced as frequency-dependent variables. The electric and magnetic currents have identical spatial support and orientation, while their relative amplitudes and phases are fixed by Eq. (S362).

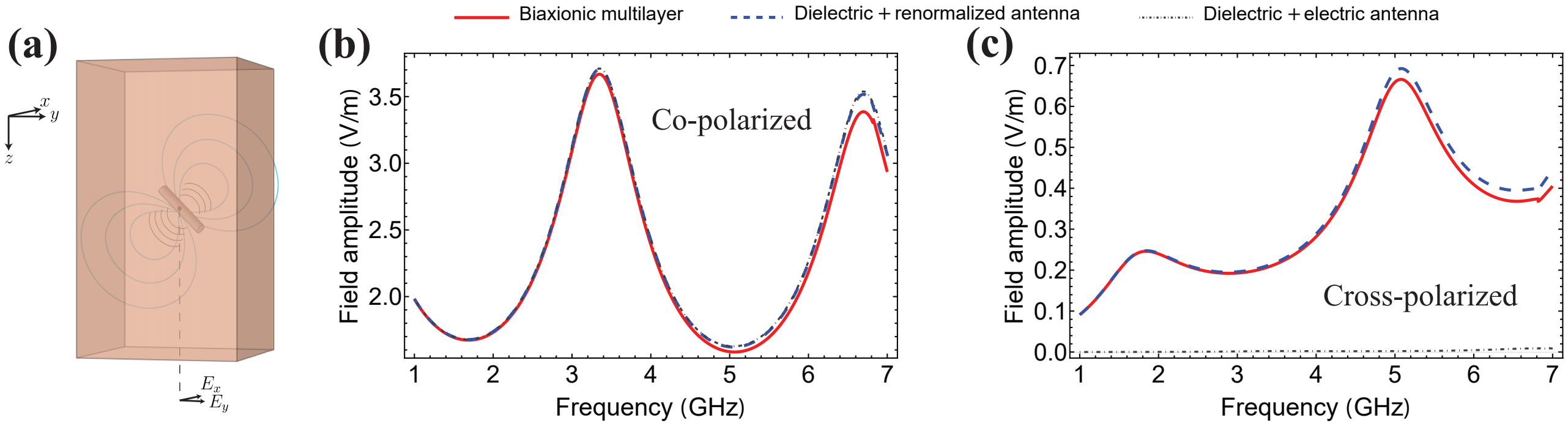


FIG. S15. Full-wave simulation of biaxionic medium excited by the source. Comparison between COMSOL simulations of the microscopic bi-gyrotropic multilayer and the effective model. The multilayer contains 10 unit cells with period $a = 4\,\mathrm{mm}$, giving a total thickness of 40 mm. The source is a tilted antenna centered inside the slab, lying in the $Oyz$ plane with length 8.94 mm and tilt angle 26.57° with respect to the $y$-axis. Red solid curves show the microscopic multilayer excited by an electric antenna [Fig. 2(d-f) in the main text]. Blue dashed curves show the effective simulation, where the dielectric slab is excited by the renormalized electric–magnetic antenna. Gray dot-dashed curves show the reference case of a dielectric slab excited only by the electric antenna. Panels show (b) the co-polarized field amplitude and (c) the cross-polarized field amplitude measured on the right side of the slab. The effective model reproduces the multilayer response, including the finite cross-polarized emission, whereas the reference case of a dielectric excited by the electric source does not.

The simulated dielectric slab contains a total thickness of 40 mm, 10 unit cells with period $a = 4\,\mathrm{mm}$. Periodic boundary conditions with period 20 mm are imposed in the transverse $x$ and $y$ directions, corresponding to an infinite array of identical in-phase sources. The antenna with current 1 mA, is centered at $z = 0$, lies in the $yz$ plane, has length 8.94 mm, and is tilted by 26.57° relative to the $y$ axis. Its projections along the $y$ and $z$ directions are therefore 8 mm and 4 mm, respectively. The latter coincides with one complete unit-cell period.

We compare the microscopic bigyrotropic multilayer excited by the physical electric antenna in the main text, with the equivalent dielectric slab excited by the mapped electric–magnetic source pair. As a reference, we also consider the same dielectric slab excited only by the original electric antenna, with the renormalized magnetic current omitted. The radiated field is evaluated on the right side of the slab at a distance of 20mm from the slab, and decomposed into the co-polarized component $E_y$ and the cross-polarized component $E_x$.

The results are presented in Fig. S15. The dielectric excited by the renormalized pair of electric and magnetic sources accurately reproduces the microscopic multilayer response for both polarization components throughout the simulated frequency range. In particular, it captures the finite cross-polarized emission generated by the embedded electric antenna. By contrast, the reference case of an ordinary dielectric excited purely by the electric source produces no appreciable cross-polarized field. These results confirm that the developed effective description is accurate and captures the interaction of macroscopic sources with biaxionic media.

## S11. INTRINSIC ELECTROMAGNETIC PROPERTIES IN FINITE SPHERICAL STRUCTURES

### A. Plane-wave scattering and control geometries

In the main text, we considered a finite spherical shell with a central void, constructed from alternating radially magnetized gyrotropic layers. The constitutive relations of the layers are

$$\mathbf{D} = \varepsilon \mathbf{E} \pm i g_e \, \mathbf{e}_r \times \mathbf{E}, \qquad \mathbf{B} = \mu \mathbf{H} \pm i g_m \, \mathbf{e}_r \times \mathbf{H}, \tag{S363}$$

where $\mathbf{e}_r$ is a unit vector in the radial direction. The structure contains 12 layers, corresponding to 6 unit cells, with layer thickness 2 mm, outer radius 30 mm and a central vacuum void of radius 6 mm. The material parameters are

$$\varepsilon = 5, \qquad \mu = 1, \qquad g_e = 4, \qquad g_m = 0.8,$$

which satisfy the same invisibility condition as in the planar multilayer. In this section, we compare the plane-wave scattering from biaxionic sphere satisfying invisibility condition with several spherical reference structures.

The first reference system is a homogeneous dielectric shell with $\varepsilon = 5$ and $\mu = 1$. As expected, this structure produces only the co-polarized electric dipole moment $p_x$, while the transverse component $p_y$ vanishes. The induced

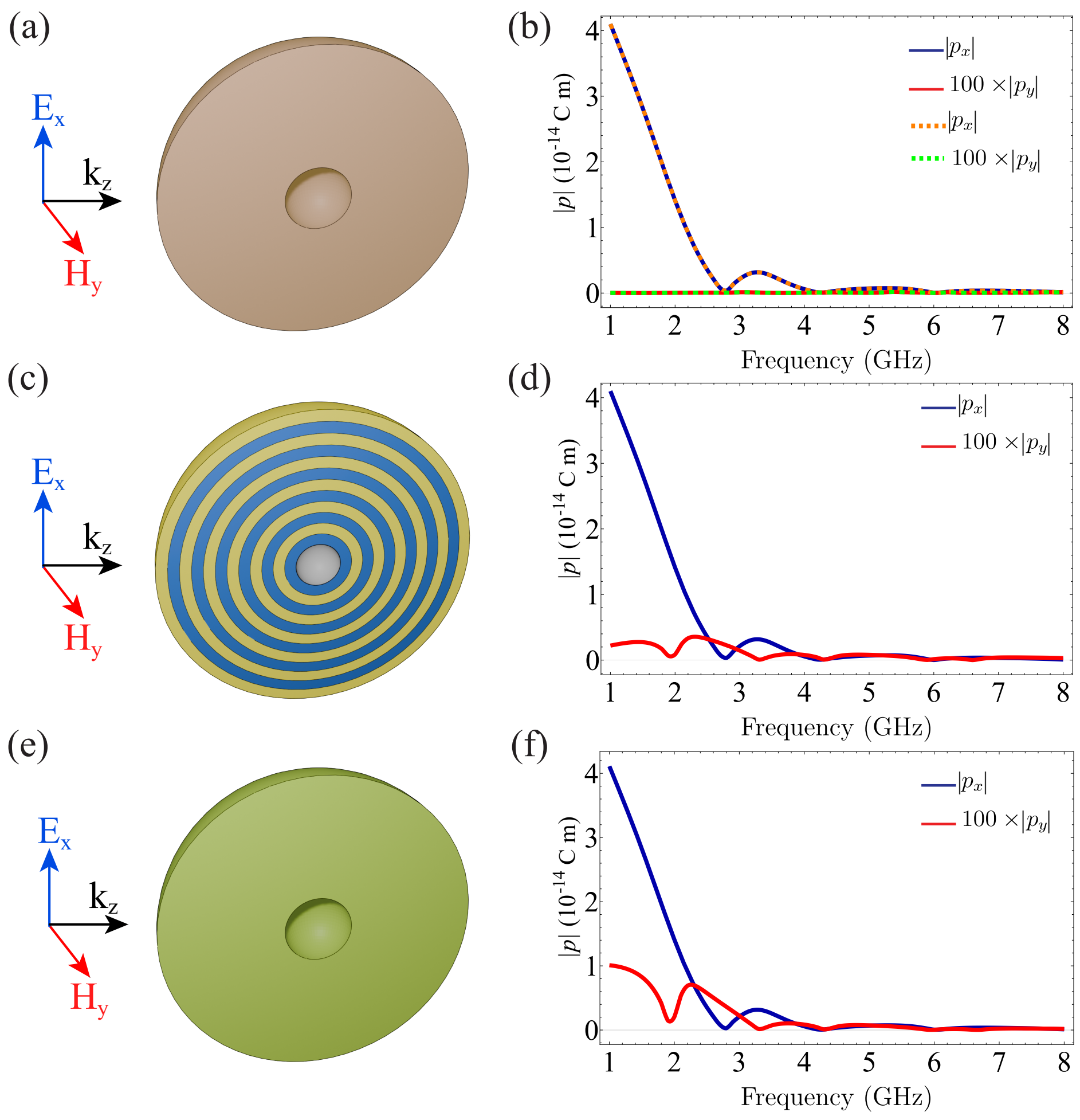


FIG. S16. Plane-wave scattering from spherical reference structures. All structures have outer radius 30 mm and a central vacuum void of radius 6 mm. The excitation is an $x$-polarized plane wave propagating along the $z$ direction. Panels (a,b) show the homogeneous dielectric shell with $\varepsilon = 5$ and $\mu = 1$. In panel (b), solid curves denote the dipole moments of the dielectric shell, while dotted curves denote the corresponding dipole moments of the hidden biaxionic multilayer sphere from the main text. The two results overlap within numerical accuracy: only the co-polarized dipole moment $p_x$ is excited, while the transverse component $p_y$ is suppressed. Panels (c,d) show a radially multilayered gyrotropic sphere with the same geometry but with $g_m = 0$, for which the magnetoelectric response is uncompensated and a finite $p_y$ appears. Panels (e,f) show a homogeneous spherical shell with constant axion coupling $\chi = 0.35$, included as a phenomenological reference for an uncompensated magnetoelectric response. The comparison demonstrates that the hidden biaxionic multilayer sphere is scattering-indistinguishable from a dielectric sphere under external plane-wave excitation, in contrast to the uncompensated axion-like reference structures.

dipole moments as a function of frequency are presented by the solid lines in Fig. S16(a,b). The dashed lines in the same figure indicate the dipole moments of the multilayered sphere extracted from the full-wave numerical simulation. We observe that the two sets of curves fully match. This confirms that under plane-wave excitation the compensated biaxionic spherical multilayer is indistinguishable from the ordinary dielectric sphere.

We then compare this behavior with the reference systems whose nonreciprocal magnetoelectric response is uncompensated. We simulate a sphere made of radially magnetized gyrotropic layers with the same geometry and arrangement but with $g_m = 0$ [Fig. S16(c)]. In this case, the compensation between the axion and dual-axion responses is removed, and a finite transverse dipole moment $p_y$ appears [Fig. S16(d)].

Next, we consider a homogeneous spherical shell with a constant axion coupling $\chi = 0.35$ [Fig. S16(e,f)]. This model is included only as a phenomenological reference for a sphere with an uncompensated magnetoelectric response. In both cases, the incident wave induces an orthogonal electric dipole moment $p_y$ revealing magnetoelectric coupling in scattering.

Therefore, the simulations for the spherical structures support the same physical conclusion as our previous calculations for the planar geometry: the intrinsic property is invisible to plane-wave scattering, but it can be distinguished

from the ordinary dielectric response once the system is probed by the sources placed inside the metamaterial, as shown in the main text.

### B. Theoretical analysis for spherical biaxionic scatterer

Next we support these numerical results by the analytical calculation for a finite spherical meta-atom made of biaxionic medium with the material parameters $\varepsilon, \mu, \chi, \tilde{\chi}$ placed in vacuum. We examine the field produced by the point electric dipole $\mathbf{d}_0$ placed in the center of this meta-atom. For conceptual simplicity, we consider the quasi-static case, which highlights the key physical features.

From the symmetry of the system we write the following ansatz for the fields inside (index $i$) and outside of the sphere:

$$\mathbf{E}_i = \mathbf{E}_{i0} + \frac{1}{r^3}\left[3(\mathbf{n}\cdot\mathbf{d}_0)\mathbf{n} - \mathbf{d}_0\right], \quad \mathbf{B}_i = \mathbf{B}_{i0}, \tag{S364}$$

$$\mathbf{E} = \mathbf{E}_0 + \frac{1}{r^3}\left[3(\mathbf{n}\cdot\mathbf{d}_{\rm ind})\mathbf{n} - \mathbf{d}_{\rm ind}\right], \quad \mathbf{B} = \mathbf{B}_0 + \frac{1}{r^3}\left[3(\mathbf{n}\cdot\mathbf{m}_{\rm ind})\mathbf{n} - \mathbf{m}_{\rm ind}\right]. \tag{S365}$$

Here, index $i$ stands for the inner volume of the sphere, whereas non-indexed fields are the fields outside, $\mathbf{d}_0$ is the source dipole and $\mathbf{d}_{\rm ind}, \mathbf{m}_{\rm ind}$ are effective electric and magnetic moments which characterize the structure of the field outside.

In writing this ansatz, we do an important step: in the vicinity of electric source only electric field features the singularity due to electric dipole, while the magnetic field stays regular. This directly corresponds to the "no-cavity" option Eqs. (S62),(S63), valid for the macroscopic sources and derived in Sec. S2 from the microscopic theory.

We now proceed writing the boundary conditions as

$$[\mathbf{B}]_n = \tilde{\chi}\,\varepsilon\, E_{i,n}, \quad [\mathbf{H}]_t = \chi \mathbf{E}_{i,t} \tag{S366}$$

$$[\mathbf{D}]_n = -\chi B_{i,n}, \quad [\mathbf{E}]_t = -\frac{\tilde{\chi}}{\mu}\mathbf{B}_{i,t}, \tag{S367}$$

where notation $[\mathbf{A}]_{\{n,t\}} = \mathbf{A}_{i,\{n,t\}} - \mathbf{A}_{\{n,t\}}$ is introduced for brevity as a difference of normal or tangential components, respectively. Resolving the obtained system of linear equations, we obtain the induced dipole moments for the arbitrary combination of $\chi$ and $\tilde{\chi}$:

$$d_{\rm ind} = \frac{3(\mu+2)(\chi\tilde{\chi}+1)}{\varepsilon\left(\mu\chi\tilde{\chi}+\mu+2\tilde{\chi}^2+2\right)+2\left(\mu\chi^2+\mu+2\chi\tilde{\chi}+2\right)} d_0\,, \tag{S368}$$

$$m_{\rm ind} = -\frac{3(\chi\tilde{\chi}+1)(\mu\chi+2\tilde{\chi})}{\varepsilon\left(\mu\chi\tilde{\chi}+\mu+2\tilde{\chi}^2+2\right)+2\left(\mu\chi^2+\mu+2\chi\tilde{\chi}+2\right)} d_0\,, \tag{S369}$$

where $d_0$ is a dipole moment of the source inside. In the limits $\chi = 0$ or $\tilde{\chi} = 0$ we recover already known [1] results for the material with either pure dual axion response or pure axion response.

Now we assess the situation of invisibility condition by requiring that $\tilde{\chi} = \frac{\mu}{\varepsilon}\chi$. Then the general solution turns into

$$d_{\rm ind} = \frac{3\,d_0}{\varepsilon+2}, \quad m_{\rm ind} = -\frac{3\,\tilde{\chi}\,d_0}{\mu+2}. \tag{S370}$$

Thus, the effective electric dipole moment is the same as in a usual dielectric sphere excited by the dipole. However, magnetic dipole moment of the structure does not vanish, even though the magnetoelectric signatures are absent in the scattering experiments.

In realistic metamaterial, both $\chi$ and $\tilde{\chi}$ responses vanish in the static limit. Therefore, these magneto-electric effects should be probed by the time-dependent sources, which motivates our analysis of biaxionic sphere excited by the oscillating point electric dipole placed in its center. The boundary conditions remain the same. The anzatz for the fields inside and outside of the sphere for a dipole oriented along the $z$-axis is expressed through vector spherical

harmonics as

$$\mathbf{E}_I = \frac{1}{\sqrt{\varepsilon}}\Big(\frac{i}{\tilde{q}}\nabla\times\Big[\Big(a_{e,\,s}\,h^{(1)}(\tilde{q}r)+a_{e,\,i}\,j(\tilde{q}r)\Big)\;\mathbf{X}_{1,0}\Big]\;-\;a_{m,\,i}\,j(\tilde{q}r)\,\mathbf{X}_{1,0}\Big), \tag{S371}$$

$$\mathbf{H}_I = \frac{1}{\sqrt{\mu}}\Big(\frac{i}{\tilde{q}}\nabla\times[a_{m,\,i}\,j(\tilde{q}r)\,\mathbf{X}_{1,0}]\;+\;\Big(a_{e,\,s}\,h^{(1)}(\tilde{q}r)+a_{e,\,i}\,j(\tilde{q}r)\Big)\;\mathbf{X}_{1,0}\Big), \tag{S372}$$

$$\mathbf{E}_O = \frac{i}{q}\nabla\times\Big[a_e\,h^{(1)}(qr)\,\mathbf{X}_{1,0}\Big]\;-\;a_m h^{(1)}(qr)\,\mathbf{X}_{1,0}, \tag{S373}$$

$$\mathbf{H}_O = \frac{i}{q}\nabla\times\Big[a_m h^{(1)}(qr)\,\mathbf{X}_{1,0}\Big]\;+\;a_e\,h^{(1)}(qr)\,\mathbf{X}_{1,0}. \tag{S374}$$

Here $q = \omega/c$ and $\tilde{q} = \sqrt{\varepsilon\mu}\,\omega/c$ are wavenumbers in vacuum and in the sphere; $j(x)$ and $h^{(1)}(x)$ are spherical Bessel function and Hankel function of the first kind, corresponding to the regular at $x = 0$ electric and magnetic eigenmodes of the sphere and irregular at $x = 0$ outgoing radiation, respectively. These functions are actually $j_1(x)$ and $h_1^{(1)}(x)$, and we omit the lower index for brevity. $\mathbf{X}_{l,m} = [l(l+1)]^{-\frac{1}{2}}\hat{\mathbf{L}}Y_{l,m}$ is a vector spherical harmonic, with $\hat{\mathbf{L}}$ being angular momentum operator and $Y_{l,m}$ being scalar spherical harmonic; $a_{A,\,B}$ are amplitude coefficients of the expansion, such that $A = \{e, m\}$ for electric and magnetic modes and $B = \{s, i, e\}$ for the source dipole, field inside the sphere, and radiation outside, respectively. One of them, $a_{e,\,s}$ is related to the source dipole as $a_{e,\,s} = -iq^3\sqrt{8\pi/3}\,d_0\varepsilon\mu\sqrt{\mu}$, while the remaining four coefficients are to be determined from the boundary conditions.

Note that the chosen ansatz satisfies the same "no cavity" option as in the static limit: electric field is singular at the origin, while magnetic field remains regular.

For monochromatic radiative solutions it is sufficient to resolve boundary conditions only for the tangential components:

$$\mathbf{H}_{It} - \chi\mathbf{E}_{It} = \mathbf{H}_{Ot}\,, \qquad \mathbf{E}_{It} + \tilde{\chi}\mathbf{H}_{It} = \mathbf{E}_{Ot}\,. \tag{S375}$$

Extracting the tangential components from the general expressions and introducing auxiliary notation

$$\alpha = \frac{j(\tilde{q}R)}{h^{(1)}(qR)}, \quad \beta = \frac{D_R j(\tilde{q}R)}{D_R h^{(1)}(qR)}, \tag{S376}$$

$$\gamma = \frac{h^{(1)}(\tilde{q}R)}{h^{(1)}(qR)}, \quad \delta = \frac{D_R h^{(1)}(\tilde{q}R)}{D_R h^{(1)}(qR)}\,, \tag{S377}$$

where $D_R(f(qR)) = \frac{d}{dR}[Rf(qR)]$, we obtain the following matrix system on the amplitude coefficients

$$\begin{bmatrix} -\frac{\chi}{\varepsilon\sqrt{\mu}}\beta & \frac{1}{\mu\sqrt{\varepsilon}}\beta & 0 & -1 \\ \frac{1}{\varepsilon\sqrt{\mu}}\beta & \frac{\tilde{\chi}}{\mu\sqrt{\varepsilon}}\beta & -1 & 0 \\ \frac{\alpha}{\sqrt{\mu}} & \frac{\chi}{\sqrt{\varepsilon}}\alpha & -1 & 0 \\ \frac{\tilde{\chi}}{\sqrt{\mu}}\alpha & -\frac{1}{\sqrt{\varepsilon}}\alpha & 0 & 1 \end{bmatrix} \begin{bmatrix} a_{e,i} \\ a_{m,i} \\ a_e \\ a_m \end{bmatrix} = a_{e,s} \begin{bmatrix} \chi\frac{1}{\varepsilon\sqrt{\mu}}\,\delta \\ -\frac{1}{\varepsilon\sqrt{\mu}}\,\delta \\ -\frac{\gamma}{\sqrt{\mu}} \\ -\tilde{\chi}\frac{\gamma}{\sqrt{\mu}} \end{bmatrix}. \tag{S378}$$

The solution of this system for the outside amplitudes is

$$a_e = -\frac{a_{e,\,s}}{\sqrt{\mu}}\frac{(\beta\gamma-\alpha\delta)}{\alpha\beta\,\frac{\mu(1+\chi^2)+\varepsilon(1+\tilde{\chi}^2)}{1+\chi\tilde{\chi}} - (\alpha^2\varepsilon\mu+\beta^2)}(\beta-\mu\alpha)\,, \tag{S379}$$

$$a_m = \frac{a_{e,\,s}}{\sqrt{\mu}}\frac{(\beta\gamma-\alpha\delta)}{\alpha\beta\,\frac{\mu(1+\chi^2)+\varepsilon(1+\tilde{\chi}^2)}{1+\chi\tilde{\chi}} - (\alpha^2\varepsilon\mu+\beta^2)}(\tilde{\chi}\beta-\mu\chi\alpha)\,. \tag{S380}$$

As can be directly checked, imposing the invisibility condition $\chi = \frac{\varepsilon}{\mu}\tilde{\chi}$ does not nullify the magnetic dipole amplitude outside of the sphere. Moreover, since the connection between dipole moments and the respective amplitudes is known, we can calculate induced external dipole moments normalized by the value of the seed source dipole moment as

$$\frac{d}{d_0} = \frac{a_e}{a_{e,\,s}}\varepsilon\mu\sqrt{\mu}\,, \qquad \frac{m}{d_0} = \frac{a_e}{a_{e,\,s}}\sqrt{\varepsilon\mu}\sqrt{\mu}\,. \tag{S381}$$

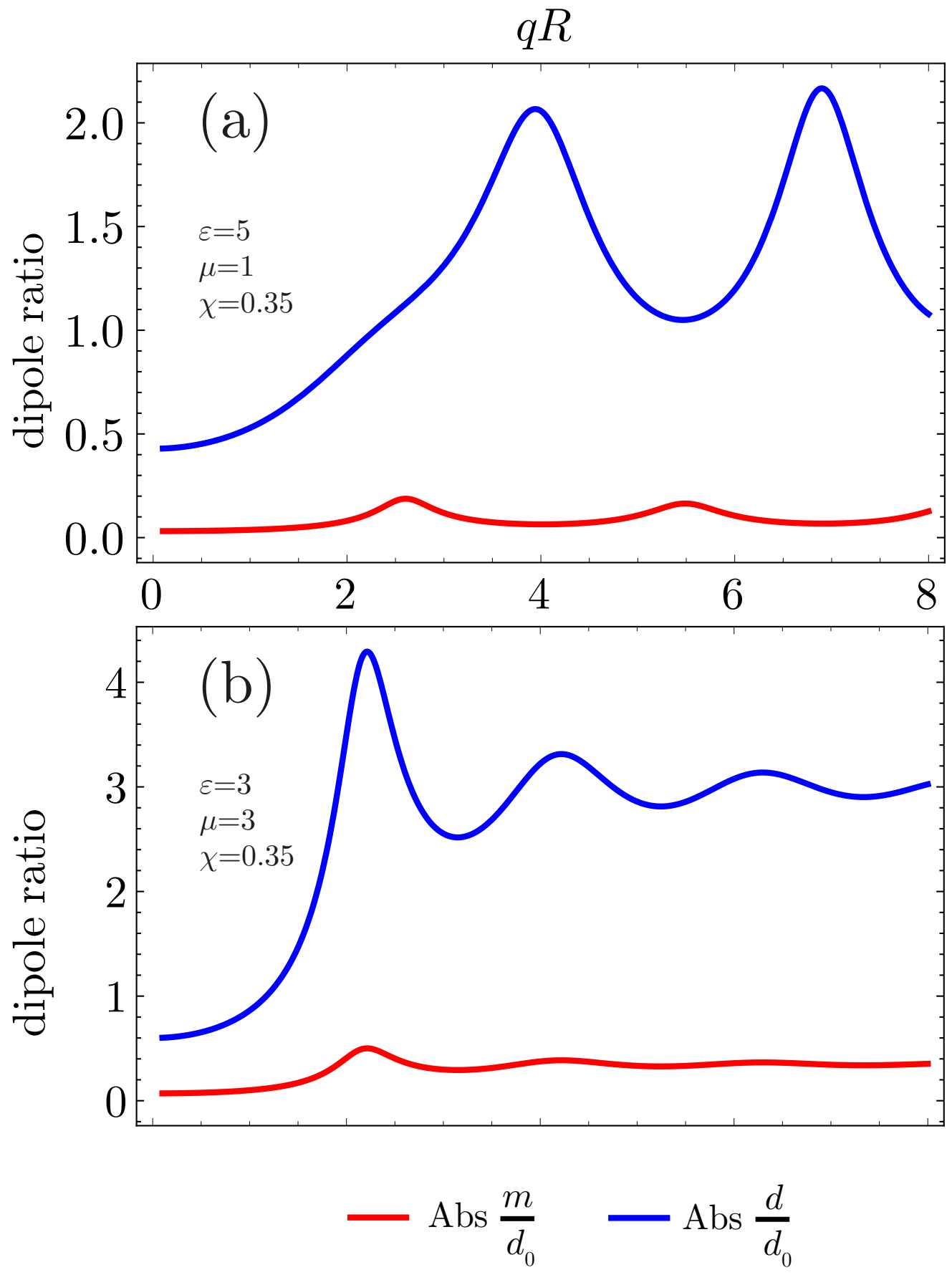


FIG. S17. Dependence of the induced external dipole moments of the biaxionic sphere with the source electric dipole $d_0$ moment inside on the dimensionless size parameter of the sphere $qR$ under invisibility condition $\chi = \frac{\varepsilon}{\mu}\tilde{\chi}$ (S135). (a) $\varepsilon = 5, \mu = 1, \chi = 0.35$. (b) $\varepsilon = 3, \mu = 3, \chi = 0.35$

.

The result is plotted on Fig. S17. As can be readily seen, induced magnetic dipole moment does not vanish under this condition. Interestingly, depending on the ratio between $\varepsilon$ and $\mu$ the maxima of the induced electric and magnetic dipoles are either matched [Fig. S17(b)] or mismatched [Fig. S17(a)]. If the medium features $\varepsilon = \mu = 1$ and simultaneously satisfies the invisibility condition, we obtain constant frequency-independent normalized dipole moments $d/d_0 = 1$ and $m/d_0 = \chi$.

### C. Eigenmodes of biaxionic sphere

In many situations, the eigenmodes of photonic structure give a key to understanding its scattering properties. In this section we derive the eigenmodes of a biaxionic sphere and examine their behavior as the parameters of the sphere approach the invisibility condition.

We start with the same system of the boundary conditions (S378), but consider a sourceless case $a_{e,s} = 0$. This

defines the eigenmodes of biaxionic sphere:

$$\begin{bmatrix} -\dfrac{\chi}{\varepsilon\sqrt{\mu}}\beta & \dfrac{1}{\mu\sqrt{\varepsilon}}\beta & 0 & -1 \\ \dfrac{1}{\varepsilon\sqrt{\mu}}\beta & \dfrac{\tilde{\chi}}{\mu\sqrt{\varepsilon}}\beta & -1 & 0 \\ \dfrac{\alpha}{\sqrt{\mu}} & \dfrac{\chi}{\sqrt{\varepsilon}}\alpha & -1 & 0 \\ \dfrac{\tilde{\chi}}{\sqrt{\mu}}\alpha & -\dfrac{1}{\sqrt{\varepsilon}}\alpha & 0 & 1 \end{bmatrix} \begin{bmatrix} a_{e,i} \\ a_{m,i} \\ a_e \\ a_m \end{bmatrix} = \begin{bmatrix} 0 \\ 0 \\ 0 \\ 0 \end{bmatrix} . \tag{S382}$$

In this notation, $\alpha = \frac{j(\tilde{q}R)}{h^{(1)}(qR)}$, $\beta = \frac{D_R j(\tilde{q}R)}{D_R h^{(1)}(qR)}$, with $D_R(f(qR)) = \frac{d}{dR}[Rf(qR)]$. We reduce the system via resolving lines 3 and 4 of (S382) with respect to $a_{e,i}$ and $a_{m,i}$:

$$\frac{\alpha}{\sqrt{\mu}}a_{e,i} + \frac{\chi\alpha}{\sqrt{\varepsilon}}a_{m,i} = a_e, \qquad \frac{\tilde{\chi}\alpha}{\sqrt{\mu}}a_{e,i} - \frac{\alpha}{\sqrt{\varepsilon}}a_{m,i} = -a_m\,. \tag{S383}$$

Equivalently, in matrix form, that is

$$\begin{pmatrix} \frac{1}{\sqrt{\mu}} & \frac{\chi}{\sqrt{\varepsilon}} \\ \frac{\tilde{\chi}}{\sqrt{\mu}} & -\frac{1}{\sqrt{\varepsilon}} \end{pmatrix} \begin{pmatrix} a_{e,i} \\ a_{m,i} \end{pmatrix} = \begin{pmatrix} \frac{a_e}{\alpha} \\ -\frac{a_m}{\alpha} \end{pmatrix} . \tag{S384}$$

The solution of this system is expressed as follows

$$a_{e,i} = \frac{\sqrt{\mu}}{\alpha(1+\chi\tilde{\chi})}\,(a_e - \chi a_m), \qquad a_{m,i} = \frac{\sqrt{\varepsilon}}{\alpha(1+\chi\tilde{\chi})}\,(a_m + \tilde{\chi}a_e)\,. \tag{S385}$$

We substitute this back into lines 1 and 2 of the system (S382) to obtain

$$\begin{pmatrix} \beta\Big(\frac{1}{\varepsilon} + \frac{\tilde{\chi}^2}{\mu}\Big) - \alpha(1+\chi\tilde{\chi}) & \beta\Big(\frac{\tilde{\chi}}{\mu} - \frac{\chi}{\varepsilon}\Big) \\ \beta\Big(\frac{\tilde{\chi}}{\mu} - \frac{\chi}{\varepsilon}\Big) & \beta\Big(\frac{\chi^2}{\varepsilon} + \frac{1}{\mu}\Big) - \alpha(1+\chi\tilde{\chi}) \end{pmatrix} \begin{pmatrix} a_e \\ a_m \end{pmatrix} = 0 \tag{S386}$$

Note that the off-diagonal elements contain a quantity proportional to the *extrinsic property* $\theta$ (S87) $\frac{\tilde{\chi}}{\mu} - \frac{\chi}{\varepsilon} = -\theta\frac{1+\chi\tilde{\chi}}{\varepsilon\mu}$. Further, using the other two invariants of the mapping group (S88),(S89) given as $\mu_T = \frac{\mu+\varepsilon\tilde{\chi}^2}{1+\chi\tilde{\chi}}$ and $\varepsilon_T = \frac{\varepsilon+\mu\chi^2}{1+\chi\tilde{\chi}}$, as well as $n^2 = \varepsilon\mu$, we rewrite the system as

$$\begin{pmatrix} \beta\mu_T - \alpha\, n^2 & -\beta\theta \\ -\beta\theta & \beta\varepsilon_T - \alpha\, n^2 \end{pmatrix} \begin{pmatrix} a_e \\ a_m \end{pmatrix} = 0\,. \tag{S387}$$

The determinant of the system yields the equation for the eigenfrequencies:

$$(\beta\mu_T - \alpha\, n^2)(\beta\varepsilon_T - \alpha\, n^2) - \beta^2\theta^2 = 0\,. \tag{S388}$$

Rearranging the terms using $\varepsilon_T\mu_T - \theta^2 = n^2$, we obtain the following form

$$(\beta - \mu_T\alpha)(\beta - \varepsilon_T\alpha) - \alpha^2\theta^2 = 0\,. \tag{S389}$$

Here, $\alpha$ and $\beta$ are the functions of sphere size and frequency. We therefore observe that the hybridization of electric and magnetic multipole modes is caused exclusively by the extrinsic property $\theta$, while intrinsic property $\psi$ affects neither the frequencies, nor the structure ($a_e/a_m$ ratio) of the eigenmodes, and its only imprint is in the values of $a_{e,i}$ and $a_{m,i}$ coefficients characterizing the field inside.

Since $\alpha$ and $\beta$ contain transcendental functions of the dimensionless size parameter of the sphere $qR$, the obtained equation can only be resolved numerically. As expected, the roots of this equation correspond exactly to the poles of the multipole coefficients when compared to the results of the previous section, Eqs. (S379) and (S380). If the extrinsic property is set to zero $\theta = 0$, the eigenmodes of the sphere have the same structure in the far field as those for a usual dielectric with effective permittivity $\varepsilon_T$ and permeability $\mu_T$. If there are no axionic effects at all $\chi = \tilde{\chi} = 0$, the extrinsic property is absent $\theta = 0$ and effective dielectric parameters coincide with the actual ones $\varepsilon_T = \varepsilon, \mu_T = \mu$. In such case, we recover the eigenmodes of a dielectric sphere.

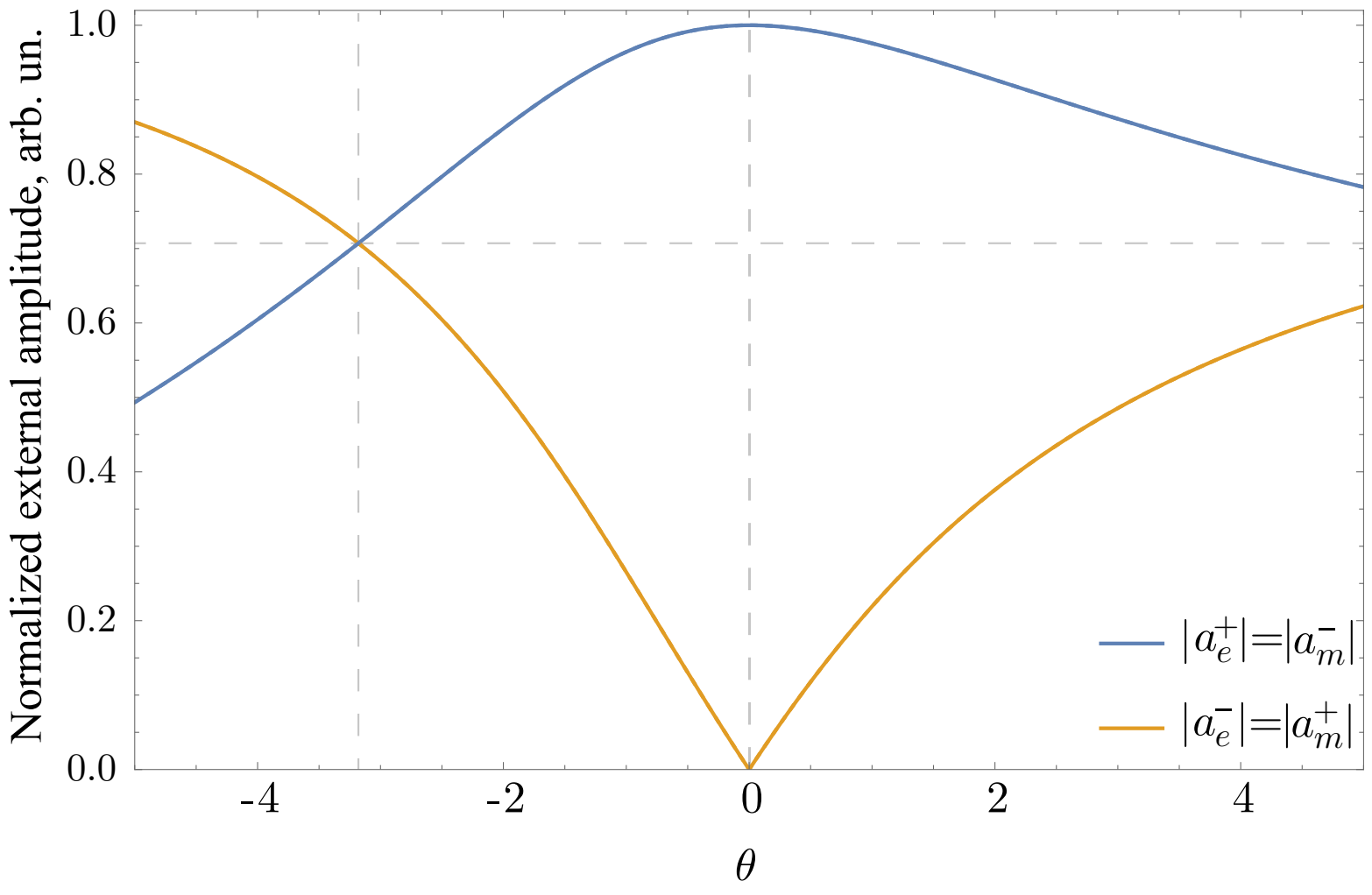


FIG. S18. Dependence of the normalized dipole moments of the dipole eigenmodes. At $\theta = 0$ the eigenmodes become either purely electric or purely magnetic, contrasting with the situation when the axion response hybridizes eigenmodes. $\varepsilon = 5, \mu = 1, \chi = 0.35$.

.

One may also solve the above equation with respect to $x = \frac{\beta}{\alpha}$ as a function of extrinsic property:

$$x(\theta)_\pm = \frac{(\varepsilon_T + \mu_T) \pm \sqrt{(\varepsilon_T - \mu_T)^2 + 4\theta^2}}{2} \tag{S390}$$

After the implementation of the constraint (S390), we may resolve (S386) with respect to the multipole amplitudes $a_e$ and $a_m$. The solution in this case reads

$$r_\pm = \left(\frac{a_m}{a_e}\right)_\pm = -\left(\frac{a_e}{a_m}\right)_\mp = \frac{\mu_T\, x_\pm(\theta) - n^2}{\theta\, x_\pm(\theta)} = \frac{\theta\, x_\pm(\theta)}{\varepsilon_T\, x_\pm(\theta) - n^2}, \tag{S391}$$

Note that the amplitude ratio is purely real for both plus and minus modes, which is typical for axionic hybridization and contrasts with other sorts of hybridization, e.g., in chiral meta-atoms. Also, the external amplitudes depend only on the invariants of the mapping $\varepsilon_T, \mu_T$ and $\theta$.

Figure S18 depicts the absolute values of the electric and magnetic dipole amplitudes outside the sphere with respect to the extrinsic property $\theta$. As can be seen, at $\theta = 0$, which corresponds to the axion invisibility condition, the hybridization of the eigenmodes disappears, resulting in the distinct electric and magnetic eigenmodes.

From (S385) we may also obtain the ratio of internal amplitudes $a_{e,i}$, $a_{m,i}$ as

$$\rho_\pm = \left(\frac{a_{m,i}}{a_{e,i}}\right)_\pm = \sqrt{\frac{\varepsilon}{\mu}}\frac{r_\pm + \tilde{\chi}}{1 - \chi r_\pm}\,. \tag{S392}$$

As we have shown above, at the axion invisibility condition $\theta = 0$ the minus branch can be seen by an outside observer as fully magnetic $a_{e-} = 0$, whereas the plus branch is fully electric $a_{m+} = 0$. Therefore

$$\rho_+ = \sqrt{\frac{\varepsilon}{\mu}}\tilde{\chi}\,, \qquad \rho_- = -\sqrt{\frac{\varepsilon}{\mu}}\frac{1}{\chi}\,. \tag{S393}$$

Expressing both of these ratios through the intrinsic property $\psi = \frac{\varepsilon\tilde{\chi} + \mu\chi}{1 - \chi\tilde{\chi}}$ and the refractive index $n = \sqrt{\varepsilon\mu}$ at the axion invisibility condition $\tilde{\chi} = \frac{\varepsilon}{\mu}\chi$, we obtain

$$\rho_\pm = \frac{-n \pm \sqrt{\psi^2 + n^2}}{\psi}\,. \tag{S394}$$

This demonstrates that even if the external field amplitudes of the mode appear non-hybridized, the internal field structure is richer, and the degree of internal hybridization is related to the intrinsic property $\psi$.

Hence, at the invisibility condition, the eigenfrequencies of biaxionic spherical meta-atom are identical to those of an ordinary dielectric sphere with the same permittivity and permeability. Moreover, the exterior field of each eigenmode has a definite multipolar character: the modes are represented either by pure electric or by pure magnetic multipoles, with no magneto-electric hybridization outside the sphere. Nevertheless, the fields inside the sphere are hybridized. The relative weights of the internal electric and magnetic multipoles are controlled by the intrinsic property $\psi$, which shapes the internal modal structure even when nonreciprocal magneto-electric response is visible neither in the exterior field, nor in the eigenfrequencies.

## S12. REFLECTION AND TRANSMISSION IN A TIME-VARYING BIAXIONIC MEDIUM

In this section, we derive analytically reflection and transmission matrices for electromagnetic waves at temporal interfaces in biaxionic media, characterized both by axion ($\chi$) and dual-axion ($\tilde{\chi}$) responses. We first consider a single temporal boundary, where the material parameters experience an abrupt change at a fixed moment of time, and then generalize our analysis to a finite temporal slab, i.e. two consecutive stepwise changes of material parameters back and forth at different moments of time. As we show below, the invisibility condition remains valid also in the time-varying case. In particular, the complete temporal biaxionic slab does not produce cross-polarized time transmission, while the cross-polarized time-reflected component vanishes when the invisibility condition is satisfied.

### A. Reflection and transmission at a single time boundary

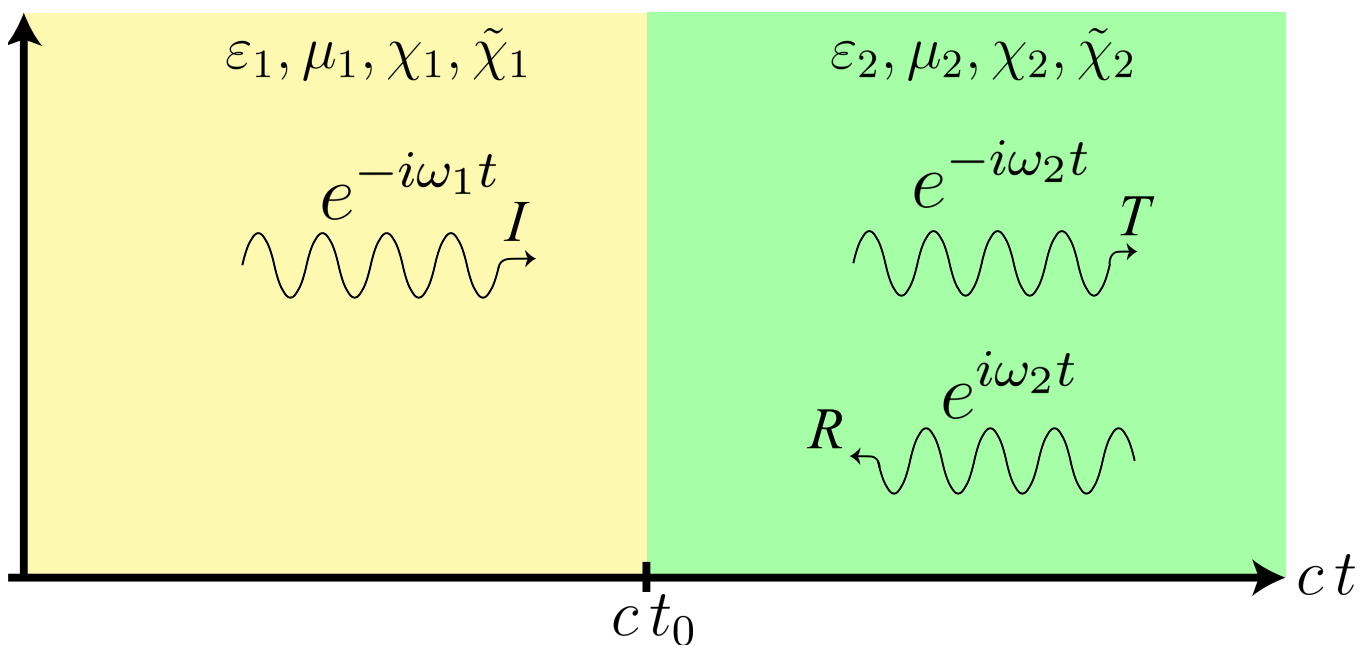


FIG. S19. Schematic of scattering at a single temporal boundary between two homogeneous biaxionic media. At $t = t_0$, the material parameters are abruptly changed from $(\varepsilon_1, \mu_1, \chi_1, \tilde{\chi}_1)$ to $(\varepsilon_2, \mu_2, \chi_2, \tilde{\chi}_2)$. The incident monochromatic wave generates time-transmitted ($T$) and time-reflected ($R$) components with frequencies $+\omega_2$ and $-\omega_2$, respectively, while the wave vector is conserved across the temporal boundary.

We begin with a temporal boundary between two homogeneous biaxionic media, schematically shown in Fig. S19. The medium parameters are $\varepsilon_1, \mu_1, \chi_1, \tilde{\chi}_1$ for $t < t_0$ and $\varepsilon_2, \mu_2, \chi_2, \tilde{\chi}_2$ for $t > t_0$. The abrupt change of the material parameters changes the frequency content of the incident monochromatic wave and generates time-transmitted and time-reflected components. Since the system remains homogeneous in space, the wave vector is conserved across the temporal boundary.

The conditions at the temporal boundary are recovered from Maxwell's equations by integrating them across the discontinuity at $t = t_0$:

$$\mathbf{B}^{(1)} - \mathbf{B}^{(2)} = \left(\tilde{\chi}_1 \mathbf{D}^{(1)} - \tilde{\chi}_2 \mathbf{D}^{(2)}\right), \tag{S395}$$

$$\mathbf{D}^{(1)} - \mathbf{D}^{(2)} = -\left(\chi_1 \mathbf{B}^{(1)} - \chi_2 \mathbf{B}^{(2)}\right). \tag{S396}$$

with $\mathbf{D}^{(i)} = \varepsilon_i \mathbf{E}^{(i)}$. We consider plane waves propagating along the $z$ axis in an infinite medium.

$$\mathbf{B}^{(1)} = \mathbf{B}_i(z,t) = \mathbf{B}_i(k_i, \omega_i) e^{ik_i z} e^{-i\omega_i t}, \tag{S397}$$

$$\mathbf{B}^{(2)} = \mathbf{B}_t(z,t) + \mathbf{B}_r(z,t) = \mathbf{B}_t(k_t, \omega_t) e^{ik_t z} e^{-i\omega_t t} + \mathbf{B}_r(k_r, \omega_r) e^{ik_r z} e^{i\omega_r t}, \tag{S398}$$

where the subscripts $i$, $r$, and $t$ label the incident, time-reflected, and time-transmitted components, respectively. According to the time-boundary conditions in Eqs. (S395) and (S396), it is evident that $k_i = k_t = k_r = k$, $\omega_i = \omega_1 = ck/\sqrt{\mu_1\varepsilon_1}$, and $\omega_r = \omega_t = \omega_2 = ck/\sqrt{\mu_2\varepsilon_2}$.

The temporal boundary conditions, Eqs. (S395) and (S396), can be recast in the matrix form as

$$\begin{pmatrix}\mathbf{B}^{(2)}\\ \mathbf{D}^{(2)}\end{pmatrix} = \frac{1}{1+\chi_2\tilde{\chi}_2}\begin{pmatrix}(1+\chi_1\tilde{\chi}_2)\,\hat{I} & (\tilde{\chi}_2-\tilde{\chi}_1)\,\hat{I}\\ (\chi_1-\chi_2)\,\hat{I} & (1+\tilde{\chi}_1\chi_2)\,\hat{I}\end{pmatrix}\begin{pmatrix}\mathbf{B}^{(1)}\\ \mathbf{D}^{(1)}\end{pmatrix}, \tag{S399}$$

where $\hat{I}$ is the $2\times2$ identity matrix. For the scattering calculation, it is convenient to express the electric displacement fields in terms of the magnetic field amplitudes. For plane waves propagating along $\mathbf{n} = \mathbf{e}_z$,

$$\mathbf{n}\times\mathbf{D}_i = \sqrt{\frac{\varepsilon_1}{\mu_1}}\mathbf{B}_i, \qquad \mathbf{n}\times\mathbf{D}_t = \sqrt{\frac{\varepsilon_2}{\mu_2}}\mathbf{B}_t, \qquad \mathbf{n}\times\mathbf{D}_r = -\sqrt{\frac{\varepsilon_2}{\mu_2}}\mathbf{B}_r, \tag{S400}$$

Therefore, immediately before and after the temporal boundary,

$$\begin{pmatrix}\mathbf{B}^{(1)}\\ \hat{\mathbf{n}}\times\mathbf{D}^{(1)}\end{pmatrix} = \begin{pmatrix}\mathbf{B}_i\\ \sqrt{\frac{\varepsilon_1}{\mu_1}}\mathbf{B}_i\end{pmatrix}, \qquad \begin{pmatrix}\mathbf{B}^{(2)}\\ \hat{\mathbf{n}}\times\mathbf{D}^{(2)}\end{pmatrix} = \begin{pmatrix}\mathbf{B}_t+\mathbf{B}_r\\ \sqrt{\frac{\varepsilon_2}{\mu_2}}\left(\mathbf{B}_t-\mathbf{B}_r\right)\end{pmatrix}. \tag{S401}$$

Substituting these relations into Eq. (S399), the temporal boundary conditions can be written entirely in terms of the magnetic-field amplitudes as

$$\begin{pmatrix}\mathbf{B}_t+\mathbf{B}_r\\ \sqrt{\frac{\varepsilon_2}{\mu_2}}\Big(\mathbf{B}_t-\mathbf{B}_r\Big)\end{pmatrix} = \begin{pmatrix}M_{11} & M_{12}\\ M_{21} & M_{22}\end{pmatrix}\begin{pmatrix}\mathbf{B}_i\\ \sqrt{\frac{\varepsilon_1}{\mu_1}}\mathbf{B}_i\end{pmatrix}, \tag{S402}$$

with

$$M_{11} = \frac{1+\chi_1\tilde{\chi}_2}{1+\chi_2\tilde{\chi}_2}\begin{pmatrix}1 & 0\\ 0 & 1\end{pmatrix}, \qquad M_{22} = \frac{1+\tilde{\chi}_1\chi_2}{1+\chi_2\tilde{\chi}_2}\begin{pmatrix}1 & 0\\ 0 & 1\end{pmatrix}. \tag{S403}$$

$$M_{12} = \frac{\tilde{\chi}_1-\tilde{\chi}_2}{1+\chi_2\tilde{\chi}_2}\begin{pmatrix}0 & -1\\ 1 & 0\end{pmatrix}, \qquad M_{21} = \frac{\chi_1-\chi_2}{1+\chi_2\tilde{\chi}_2}\begin{pmatrix}0 & -1\\ 1 & 0\end{pmatrix}. \tag{S404}$$

From the transfer matrix, the transmission and reflection matrices are obtained as

$$\hat{t} = \frac{1}{2\sqrt{\frac{\varepsilon_2}{\mu_2}}}\left[\sqrt{\frac{\varepsilon_2}{\mu_2}}M_{11} + \sqrt{\frac{\varepsilon_1\varepsilon_2}{\mu_1\mu_2}}M_{12} + M_{21} + \sqrt{\frac{\varepsilon_1}{\mu_1}}M_{22}\right], \tag{S405}$$

$$\hat{r} = \left[M_{11} + \sqrt{\frac{\varepsilon_1}{\mu_1}}M_{12} - \hat{t}\right], \tag{S406}$$

that describe the amplitudes $\mathbf{B}_t = \hat{t}\,\mathbf{B}_i$ and $\mathbf{B}_r = \hat{r}\,\mathbf{B}_i$. To simplify the transmission and reflection coefficients, we introduce the wave admittance of each medium,

$$Y_j = \sqrt{\frac{\varepsilon_j}{\mu_j}}, \qquad j = 1, 2. \tag{S407}$$

Their explicit components are

$$t_{xx} = t_{yy} = \frac{1}{2Y_2(1+\chi_2\tilde{\chi}_2)}\left[Y_2(1+\chi_1\tilde{\chi}_2) + Y_1(1+\tilde{\chi}_1\chi_2)\right], \tag{S408}$$

$$t_{xy} = -t_{yx} = -\frac{1}{2Y_2(1+\chi_2\tilde{\chi}_2)}\left[\Delta\chi + Y_1Y_2\Delta\tilde{\chi}\right], \tag{S409}$$

$$r_{xx} = r_{yy} = \frac{1}{2Y_2(1+\chi_2\tilde{\chi}_2)}\left[Y_2(1+\chi_1\tilde{\chi}_2) - Y_1(1+\tilde{\chi}_1\chi_2)\right], \tag{S410}$$

$$r_{xy} = -r_{yx} = \frac{1}{2Y_2(1+\chi_2\tilde{\chi}_2)}\left[\Delta\chi - Y_1Y_2\Delta\tilde{\chi}\right], \tag{S411}$$

with $\Delta\tilde{\chi} = \tilde{\chi}_1 - \tilde{\chi}_2$ and $\Delta\chi = \chi_1 - \chi_2$. These expressions show that the temporal boundary generically produces both co-polarized and cross-polarized temporal scattering. In particular, off-diagonal components responsible for the cross-polarized scattering are controlled by the differences $\Delta\chi$, and $\Delta\tilde{\chi}$ and therefore vanish only under specific conditions.

### B. Reflection and transmission in biaxionic time slab

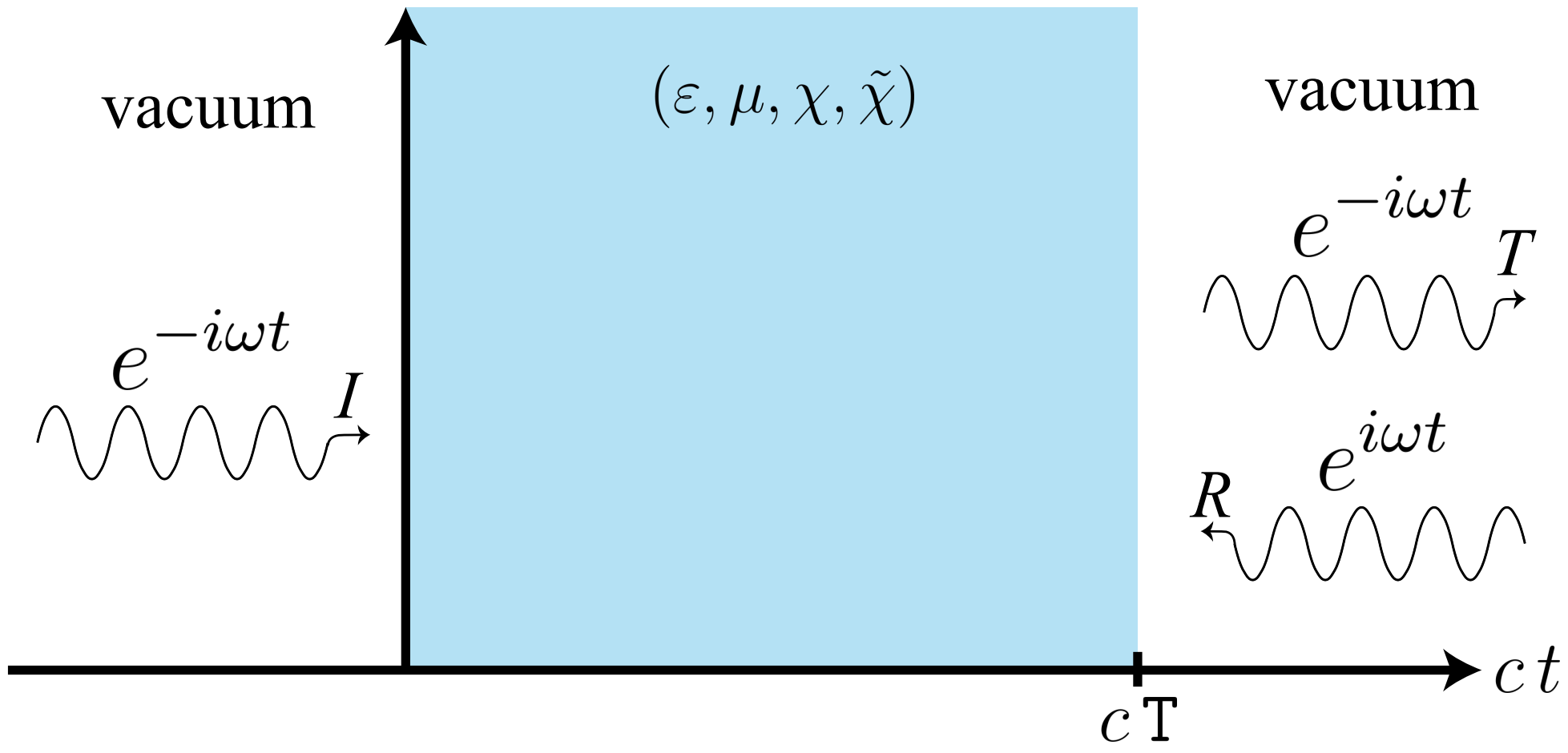


FIG. S20. Schematic of a biaxionic temporal slab in vacuum. At $t = 0$, the medium is abruptly switched from vacuum to the biaxionic state $(\varepsilon, \mu, \chi, \tilde{\chi})$, which persists for a time interval $\mathtt{T}$. At $t = \mathtt{T}$, the original vacuum parameters are restored. The two temporal boundaries and the intermediate bulk evolution generate the final time-transmitted ($T$) and time-reflected ($R$) components.

We now consider the finite temporal slab schematically shown in Fig. S20. The medium is initially vacuum, is abruptly switched at $t = 0$ to a biaxionic state characterized by $(\varepsilon, \mu, \chi, \tilde{\chi})$, and is returned to vacuum at $t = \mathtt{T}$. The problem therefore consists of two temporal boundaries separated by a finite interval of bulk evolution in the intermediate biaxionic phase. The total temporal transfer matrix is obtained by combining the transfer matrices of the two temporal interfaces with the propagation matrix describing the field evolution between $t = 0$ and $t = \mathtt{T}$. This construction is the direct temporal analog of the standard transfer-matrix formalism for a spatial slab in Sec. S4.

The bulk evolution inside the intermediate homogeneous medium is first written in terms of the magnetic induction and electric displacement fields as

$$\begin{pmatrix} \mathbf{B}(T) \\ \mathbf{D}(T) \end{pmatrix} = \hat{\mathcal{P}}_{BD}(T) \begin{pmatrix} \mathbf{B}(0) \\ \mathbf{D}(0) \end{pmatrix}, \tag{S412}$$

where

$$\hat{\mathcal{P}}_{BD}(T) = \begin{pmatrix} \cos(\omega' T)\hat{I} & -i\sin(\omega' T)\sqrt{\frac{\mu'}{\varepsilon'}}\,\hat{J} \\ i\sin(\omega' T)\sqrt{\frac{\varepsilon'}{\mu'}}\,\hat{J} & \cos(\omega' T)\hat{I} \end{pmatrix}, \tag{S413}$$

with

$$\hat{J} = \begin{pmatrix} 0 & -1 \\ 1 & 0 \end{pmatrix}. \tag{S414}$$

Here, $\varepsilon'$ and $\mu'$ are the permittivity and permeability of the intermediate medium, and

$$\omega' = \frac{ck}{\sqrt{\varepsilon'\mu'}}. \tag{S415}$$

For the subsequent calculation, it is convenient to use $\mathbf{n}\times\mathbf{D}$ instead of $\mathbf{D}$. In this basis, the bulk evolution reads

$$\begin{pmatrix}\mathbf{B}(T)\\ \hat{\mathbf{n}}\times\mathbf{D}(T)\end{pmatrix}=\begin{pmatrix}\cos(\omega' T) & -i\sin(\omega' T)\sqrt{\dfrac{\mu'}{\varepsilon'}}\\ -i\sin(\omega' T)\sqrt{\dfrac{\varepsilon'}{\mu'}} & \cos(\omega' T)\end{pmatrix}\begin{pmatrix}\mathbf{B}(0)\\ \hat{\mathbf{n}}\times\mathbf{D}(0)\end{pmatrix}. \tag{S416}$$

Inside the intermediate medium, the total fields can be decomposed as

$$\mathbf{B}(t)=\mathbf{B}_t(t)+\mathbf{B}_r(t),\qquad \mathbf{n}\times\mathbf{D}(t)=\sqrt{\frac{\varepsilon'}{\mu'}}\left[\mathbf{B}_t(t)-\mathbf{B}_r(t)\right]. \tag{S417}$$

Therefore, the bulk propagation inside the a pure-time axion slab of thickness $\mathtt{T}$ can be written in the form

$$\begin{pmatrix}\mathbf{B}_t(\mathtt{T})+\mathbf{B}_r(\mathtt{T})\\ \sqrt{\dfrac{\varepsilon}{\mu}}\Big(\mathbf{B}_t(\mathtt{T})-\mathbf{B}_r(\mathtt{T})\Big)\end{pmatrix}=\begin{pmatrix}\cos(\omega'\mathtt{T}) & -i\sin(\omega'\mathtt{T})\sqrt{\dfrac{\mu'}{\varepsilon'}}\\ -i\sin(\omega'\mathtt{T})\sqrt{\dfrac{\varepsilon'}{\mu'}} & \cos(\omega'\mathtt{T})\end{pmatrix}\begin{pmatrix}\mathbf{B}_i(0)\\ \sqrt{\dfrac{\varepsilon'}{\mu'}}\mathbf{B}_i(0)\end{pmatrix}, \tag{S418}$$

where the propagation matrix accounts for the phase accumulation $e^{-i\omega' t}$ of the forward temporal component and $e^{+i\omega' t}$ of the backward temporal component inside the slab. Using the temporal-boundary transfer matrix derived in Eq. (S402), the full transfer matrix of the temporal slab is

$$\hat{M}=\hat{M}_{2\to1}\hat{M}(\mathtt{T})\hat{M}_{1\to2} \tag{S419}$$

where $\hat{M}_{1\to2}$ and $\hat{M}_{2\to1}$ describe the temporal boundaries at $t=0$ and $t=\mathtt{T}$, respectively, and $\hat{M}(\mathtt{T})$ describes the propagation between the two switchings of the material parameters. The corresponding transmission and reflection matrices are obtained from $\hat{M}$ as

$$\hat{t}=\frac{1}{2}\Big(M_{11}+M_{12}+M_{21}+M_{22}\Big), \tag{S420}$$

$$\hat{r}=M_{11}+M_{12}-\hat{t}, \tag{S421}$$

so that $\mathbf{B}_t=\hat{t}\,\mathbf{B}_i$ and $\mathbf{B}_r=\hat{r}\,\mathbf{B}_i$.

To study the temporal biaxionic slab embedded in vacuum, we set $\chi_1=\tilde{\chi}_1=0$, $\chi_2=\chi$ and $\tilde{\chi}_2=\tilde{\chi}$. The components of the transmission and reflection matrices then take the form

$$t_{xx}=t_{yy}=\cos(\omega\,\mathtt{T})-\frac{i}{2Y(1+\chi\tilde{\chi})}\left(Y^2+Y^2\tilde{\chi}^2+1+\chi^2\right)\sin(\omega\,\mathtt{T}), \tag{S422}$$

$$t_{xy}=-t_{yx}=0, \tag{S423}$$

$$r_{xx}=r_{yy}=\frac{i}{2Y(1+\chi\tilde{\chi})}\left(-Y^2+Y^2\tilde{\chi}^2+1-\chi^2\right)\sin(\omega\,\mathtt{T}), \tag{S424}$$

$$r_{xy}=-r_{yx}=\frac{i}{Y(1+\chi\tilde{\chi})}\left(\tilde{\chi}Y^2-\chi\right)\sin(\omega\,\mathtt{T}). \tag{S425}$$

Here, $\varepsilon,\mu$ are the permittivity and permeability of biaxionic medium. These expressions show that a temporal biaxionic slab generally produces cross-polarized temporal reflection. Importantly, the off-diagonal time-reflected component vanishes under invisibility condition

$$\tilde{\chi}=\frac{1}{Y^2}\chi=\frac{\mu}{\varepsilon}\chi. \tag{S426}$$

Therefore, the same condition that suppresses polarization rotation in the scattering from the time-independent structures remains valid for the metamaterial modulated in time. As long as invisibility condition Eq. (S426) is fulfilled, the traces of time-varying $\chi$ and $\tilde{\chi}$ are not observable in the scattering experiments.

## S13. WAVE PROPAGATION IN TIME-VARYING BIAXIONIC MEDIUM

In this section, we investigate the propagation of electromagnetic waves in a homogeneous medium whose axion and dual axion responses are modulated in time, while keeping $\varepsilon$ and $\mu$ constant. Our goal is to understand the impact of such time-varying $\chi$ and $\tilde{\chi}$ on the field dynamics inside and outside of the structure.

### A. Field equations for the time-varying biaxionic medium

We consider a general biaxionic medium with the time-dependent and spatially homogeneous axion and dual-axion responses, $\chi = \chi(t)$ and $\tilde{\chi} = \tilde{\chi}(t)$, while $\varepsilon$ and $\mu$ are constant. In the absence of the sources, the Maxwell equations read

$$\nabla \times \Big(\mu^{-1}\mathbf{B} - \chi\mathbf{E}\Big) - \frac{1}{c}\frac{\partial}{\partial t}\Big(\varepsilon\mathbf{E} + \chi\mathbf{B}\Big) = \mathbf{0}, \tag{S427}$$

$$\nabla \cdot \Big(\varepsilon\mathbf{E} + \chi\mathbf{B}\Big) = 0, \tag{S428}$$

$$\nabla \times \Big(\mathbf{E} + \tilde{\chi}\,\mu^{-1}\mathbf{B}\Big) + \frac{1}{c}\frac{\partial}{\partial t}\Big(\mathbf{B} - \tilde{\chi}\,\varepsilon\mathbf{E}\Big) = \mathbf{0}, \tag{S429}$$

$$\nabla \cdot \Big(\mathbf{B} - \tilde{\chi}\,\varepsilon\mathbf{E}\Big) = 0. \tag{S430}$$

Expanding Eqs. (S427) and (S429), we obtain

$$\nabla \times \big(\mu^{-1}\mathbf{B}\big) - \frac{1}{c}\frac{\partial}{\partial t}\big(\varepsilon\mathbf{E}\big) = \chi\Big(\nabla \times \mathbf{E} + \frac{1}{c}\frac{\partial \mathbf{B}}{\partial t}\Big) + \frac{\dot{\chi}}{c}\mathbf{B}\,, \tag{S431}$$

$$\nabla \times \mathbf{E} + \frac{1}{c}\frac{\partial \mathbf{B}}{\partial t} = -\tilde{\chi}\Big(\nabla \times \big(\mu^{-1}\mathbf{B}\big) - \frac{1}{c}\frac{\partial}{\partial t}\big(\varepsilon\mathbf{E}\big)\Big) + \frac{\dot{\tilde{\chi}}}{c}\varepsilon\mathbf{E}\,, \tag{S432}$$

where we assume that $\chi$ and $\tilde{\chi}$ do not depend on coordinates, i.e. propagation in the bulk is considered.

By combining Eqs. (S431)–(S432), we recover two equations

$$\nabla \times \big(\mu^{-1}\mathbf{B}\big) - \frac{1}{c}\frac{\partial}{\partial t}\big(\varepsilon\mathbf{E}\big) = \frac{1}{\Delta}\left[\frac{\dot{\chi}}{c}\mathbf{B} + \frac{\chi\dot{\tilde{\chi}}}{c}\varepsilon\mathbf{E}\right], \tag{S433}$$

and

$$\nabla \times \mathbf{E} + \frac{1}{c}\frac{\partial \mathbf{B}}{\partial t} = \frac{1}{\Delta}\left[\frac{\dot{\tilde{\chi}}}{c}\varepsilon\mathbf{E} - \frac{\dot{\chi}\tilde{\chi}}{c}\mathbf{B}\right], \tag{S434}$$

where we introduce

$$\Delta(t) = 1 + \chi(t)\tilde{\chi}(t). \tag{S435}$$

These equations are exact and already reveal the essential physical mechanism of the time-varying biaxionic medium. Even in the absence of external charges and currents, temporal modulation of the axion and dual-axion responses acts as an effective source driving the electromagnetic field dynamics. Indeed, Eqs. (S433) and (S434) can be interpreted as Maxwell equations with effective electric and magnetic currents,

$$\frac{4\pi}{c}\mathbf{J}_{\mathrm{e}}^{\mathrm{eff}} = \frac{1}{\Delta}\left[\frac{\dot{\chi}}{c}\mathbf{B} + \frac{\chi\dot{\tilde{\chi}}}{c}\varepsilon\mathbf{E}\right], \qquad -\frac{4\pi}{c}\mathbf{J}_{\mathrm{m}}^{\mathrm{eff}} = \frac{1}{\Delta}\left[\frac{\dot{\tilde{\chi}}}{c}\varepsilon\mathbf{E} - \frac{\dot{\chi}\tilde{\chi}}{c}\mathbf{B}\right]. \tag{S436}$$

Importantly, the invisibility condition $\tilde{\chi} = (\mu/\varepsilon)\chi$ does not trivialize these equations. Even when the extrinsic electromagnetic property vanishes, the bulk dynamics is still governed by the time-dependent magnetoelectric couplings. The divergence equations (S428) and (S430) then imply

$$\varepsilon\nabla \cdot \mathbf{E} + \chi\nabla \cdot \mathbf{B} = 0, \qquad \nabla \cdot \mathbf{B} - \varepsilon\tilde{\chi}\,\nabla \cdot \mathbf{E} = 0, \tag{S437}$$

so that for $\Delta \neq 0$ one has

$$\nabla \cdot \mathbf{E} = 0, \qquad \nabla \cdot \mathbf{B} = 0. \tag{S438}$$

Therefore, throughout the bulk-propagation analysis we still can use

$$\nabla \times \nabla \times \mathbf{E} = -\nabla^2\mathbf{E}, \qquad \nabla \times \nabla \times \mathbf{B} = -\nabla^2\mathbf{B}. \tag{S439}$$

By taking the curl of Eq. (S433) and the curl of Eq. (S434), one can derive second-order wave equations. For the magnetic and electric fields, this yields

$$\nabla^2\mathbf{B} - \frac{\varepsilon\mu}{c^2}\frac{\partial^2\mathbf{B}}{\partial t^2} + \frac{\mu\dot{\chi} + \varepsilon\dot{\tilde{\chi}}}{c\Delta}\nabla\times\mathbf{B} - \frac{\varepsilon\mu\dot{\Delta}}{c^2\Delta}\frac{\partial\mathbf{B}}{\partial t} - \frac{\varepsilon\mu}{c^2}\left[\frac{d}{dt}\left(\frac{\tilde{\chi}\dot{\chi}}{\Delta}\right) + \frac{\dot{\chi}\dot{\tilde{\chi}}}{\Delta}\right]\mathbf{B} + \frac{\mu\varepsilon}{c^2}\frac{d}{dt}\left(\frac{\dot{\tilde{\chi}}}{\Delta}\right)\varepsilon\mathbf{E} = 0. \tag{S440}$$

$$\nabla^2\mathbf{E} - \frac{\varepsilon\mu}{c^2}\frac{\partial^2\mathbf{E}}{\partial t^2} + \frac{\mu\dot{\chi} + \varepsilon\dot{\tilde{\chi}}}{c\Delta}\nabla\times\mathbf{E} - \frac{\varepsilon\mu\dot{\Delta}}{c^2\Delta}\frac{\partial\mathbf{E}}{\partial t} - \frac{\varepsilon\mu}{c^2}\left[\frac{d}{dt}\left(\frac{\chi\dot{\tilde{\chi}}}{\Delta}\right) + \frac{\dot{\chi}\dot{\tilde{\chi}}}{\Delta}\right]\mathbf{E} - \frac{\mu}{c^2}\frac{d}{dt}\left(\frac{\dot{\chi}}{\Delta}\right)\mathbf{B} = 0. \tag{S441}$$

Eqs. (S440) and (S441) make the nontrivial structure of the problem explicit. The terms proportional to $\nabla\times\mathbf{E}$ and $\nabla\times\mathbf{B}$ are helicity-sensitive and therefore distinguish opposite circular polarizations. The terms proportional to $\partial_t\mathbf{E}$ and $\partial_t\mathbf{B}$ act as temporal driving contributions analogous to those appearing in photonic time crystals and time-varying dielectrics. Finally, the explicit coupling between $\mathbf{E}$ and $\mathbf{B}$ demonstrates that the electric and magnetic sectors remain dynamically mixed even in the absence of the external sources.

Therefore, even under the invisibility condition it is evident that the propagation inside the bulk remains nontrivial despite the absence of observable magnetoelectric scattering signatures outside the medium.

The second-order wave equations above make the polarization dependence and temporally driven structure of the biaxionic medium explicit. However, for the Floquet analysis and numerical treatment below, it is more convenient to work directly with the equivalent first-order system Eqs. (S433)–(S434).

### B. Bulk propagation in a time-varying biaxionic medium

To analyze the bulk propagation in a periodically modulated biaxionic medium, we rewrite Eqs. (S433)–(S434) in explicit first-order form,

$$\frac{\partial\mathbf{B}}{\partial t} = -c\,\nabla\times\mathbf{E} + \varepsilon\frac{\dot{\tilde{\chi}}}{\Delta}\mathbf{E} - \frac{\tilde{\chi}\dot{\chi}}{\Delta}\mathbf{B}\,, \tag{S442}$$

$$\frac{\partial\mathbf{E}}{\partial t} = \frac{c}{\mu\varepsilon}\nabla\times\mathbf{B} - \frac{1}{\varepsilon}\frac{\dot{\chi}}{\Delta}\mathbf{B} - \frac{\chi\dot{\tilde{\chi}}}{\Delta}\mathbf{E}\,, \tag{S443}$$

with

$$\Delta(t) = 1 + \chi(t)\tilde{\chi}(t), \qquad \dot{\Delta}(t) = \chi(t)\dot{\tilde{\chi}}(t) + \tilde{\chi}(t)\dot{\chi}(t). \tag{S444}$$

For a homogeneous medium it is natural to consider plane-wave solutions propagating along the $z$ direction,

$$\mathbf{E}(\mathbf{r},t) = \mathbf{E}(t)e^{ikz}, \qquad \mathbf{B}(\mathbf{r},t) = \mathbf{B}(t)e^{ikz}. \tag{S445}$$

Because time-translation symmetry is broken, whereas spatial homogeneity is preserved, the wavevector $\mathbf{k}$ remains a conserved quantity, while the frequency is no longer conserved. For propagation along the $z$ axis, it is convenient to decompose the fields into circular components,

$$\mathbf{B}_\eta(t) = B_\eta(t)\mathbf{e}_\eta, \qquad \mathbf{E}_\eta(t) = E_\eta(t)\mathbf{e}_\eta, \tag{S446}$$

Here, $\mathbf{e}_\eta = \mathbf{e}_\pm$ is given by

$$\mathbf{e}_\eta = \frac{1}{\sqrt{2}}\begin{pmatrix}1\\ i\eta\\ 0\end{pmatrix}. \tag{S447}$$

The curl operator in Eqs. (S442,S443) becomes algebraic,$\nabla\times\;\rightarrow\; ik\mathbf{e}_z^\times$. The bulk problem then reduces to two independent first-order systems, one for each polarization $\eta = \pm 1$. For the symmetric periodic modulation considered below, the two helicity sectors yield identical Floquet spectra, so only one branch is shown in the numerically calculated plots. Equations (S442)–(S443) become

$$\frac{dB_\eta}{dt} = -\eta ck\,E_\eta + \varepsilon\frac{\dot{\tilde{\chi}}}{\Delta}E_\eta - \frac{\tilde{\chi}\dot{\chi}}{\Delta}B_\eta, \tag{S448}$$

$$\frac{dE_\eta}{dt} = \eta\frac{ck}{\varepsilon\mu}B_\eta - \frac{1}{\varepsilon}\frac{\dot{\chi}}{\Delta}B_\eta - \frac{\chi\dot{\tilde{\chi}}}{\Delta}E_\eta. \tag{S449}$$

These equations constitute a convenient starting point for the numerical analysis. For arbitrary $\chi(t)$ and $\tilde{\chi}(t)$ they define a nonautonomous linear system, while for periodic modulation they define a temporal Floquet problem. In the periodically modulated case, the biaxionic medium may therefore be regarded as a biaxionic photonic time crystal. Its bulk evolution can display nontrivial polarization-dependent temporal dynamics, including sideband generation, polarization reshaping, and modified phase accumulation inside the slab.

We define field amplitudes in the vector form as:

$$\Psi_\eta(t) = \begin{pmatrix} B_\eta(t) \\ E_\eta(t) \end{pmatrix} \tag{S450}$$

Then Eqs. (S448,S449) can be written as:

$$\frac{d\Psi_\eta(t)}{dt} = M_\eta(t;k)\,\Psi_\eta(t) \tag{S451}$$

with

$$M_\eta(t;k) = \begin{pmatrix} -\dfrac{\tilde{\chi}\dot{\chi}}{\Delta} & -\eta ck + \varepsilon\dfrac{\dot{\tilde{\chi}}}{\Delta} \\ \eta\dfrac{ck}{\varepsilon\mu} - \dfrac{1}{\varepsilon}\dfrac{\dot{\chi}}{\Delta} & -\dfrac{\chi\dot{\tilde{\chi}}}{\Delta} \end{pmatrix}. \tag{S452}$$

The bulk evolution of the fields in the time-modulated biaxionic medium can be written in terms of the propagator $U_\eta$ as

$$\Psi_\eta(T) = U_\eta(T,0;k)\Psi_\eta(0)\,, \tag{S453}$$

where

$$U_\eta(T,0;k) = \mathcal{T}\exp\left[\int_0^T dt'\,M_\eta(t';k)\right] \tag{S454}$$

and $\mathcal{T}$ means time ordering.

### C. Biaxionic photonic band structure

Next we analyze the Floquet spectrum of a periodically modulated biaxionic medium. We consider a biaxionic medium whose axion and dual-axion responses are periodically modulated in time. The two responses are assumed to oscillate in phase according to

$$\chi(t) = \chi_0\cos(\Omega t + \varphi), \qquad \tilde{\chi}(t) = \tilde{\chi}_0\cos(\Omega t + \varphi). \tag{S455}$$

The corresponding temporal modulation is illustrated schematically in Fig. S21. The phase $\varphi$ fixes the temporal origin of the modulation; in the figure we use the representative value $\varphi = \pi/3$.

Since a common phase $\varphi$ corresponds only to a shift of the time origin, it does not affect the bulk Floquet frequencies. In the numerical calculations below we set $\varphi = 0$. Since both responses oscillate with the same temporal dependence, their deviation from the invisibility condition is fully determined by the relation between the modulation amplitudes $\chi_0$ and $\tilde{\chi}_0$. We therefore introduce the parameter $X$ to quantify the deviation from the invisibility condition as

$$\tilde{\chi}_0 = \left(\frac{\mu}{\varepsilon} + X\right)\chi_0\,, \tag{S456}$$

so that $X = 0$ corresponds exactly to the invisibility condition $\tilde{\chi} = (\mu/\varepsilon)\chi$, whereas $X \neq 0$ introduces an extrinsic magnetoelectric property manifested in the scattering. Substituting Eq. (S456) into the matrix Eq. (S452) yields

$$M_\eta(t;k,X) = \begin{pmatrix} -\left(\dfrac{\mu}{\varepsilon} + X\right)\dfrac{\chi\dot{\chi}}{\Delta} & -\eta ck + \varepsilon\left(\dfrac{\mu}{\varepsilon} + X\right)\dfrac{\dot{\chi}}{\Delta} \\ \eta\dfrac{ck}{\varepsilon\mu} - \dfrac{1}{\varepsilon}\dfrac{\dot{\chi}}{\Delta} & -\left(\dfrac{\mu}{\varepsilon} + X\right)\dfrac{\chi\dot{\chi}}{\Delta} \end{pmatrix} \tag{S457}$$

with

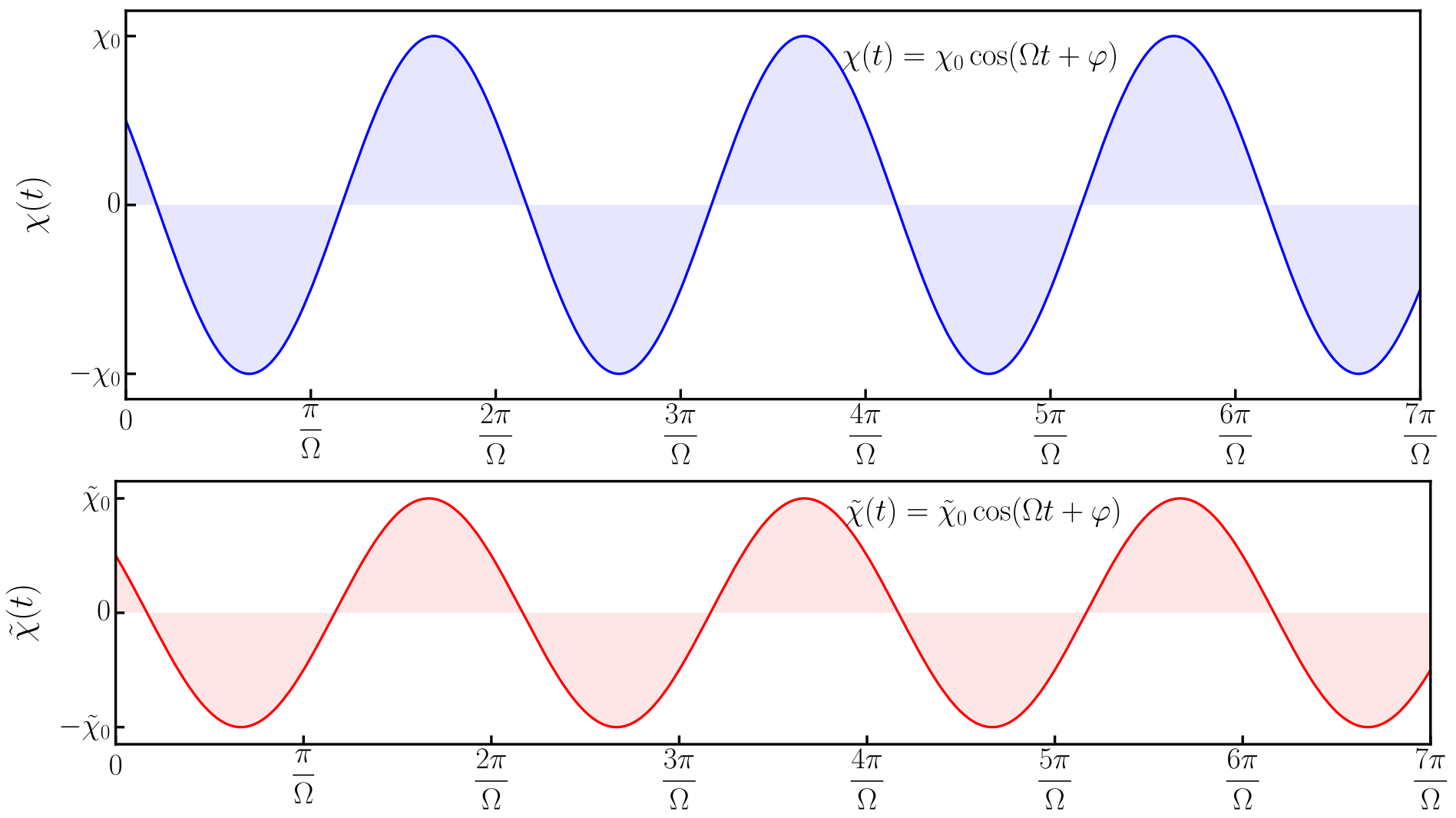


FIG. S21. Schematic of the time-periodic biaxionic response considered in this section. The axion and dual-axion parameters oscillate harmonically with the same frequency $\Omega$ and phase $\varphi$, while their modulation amplitudes $\chi_0$ and $\tilde{\chi}_0$ can be different. A representative phase $\varphi = \pi/3$ is shown.

$$\Delta(t) = 1 + \left(\frac{\mu}{\varepsilon} + X\right)\chi^2(t). \tag{S458}$$

Because the medium is homogeneous in space but periodic in time, the wavevector $k$ is conserved, whereas the frequency is not. For fixed $k$, the matrix $M_\eta(t; k, X)$ is periodic,

$$M_\eta(t + T_\Omega; k, X) = M_\eta(t; k, X), \qquad T_\Omega = \frac{2\pi}{\Omega}. \tag{S459}$$

The problem is therefore a temporal Floquet problem. According to Floquet theory, the solutions can be written as

$$\Psi_\eta(t) = e^{-i\omega_F t} p_\eta(t), \qquad p_\eta(t + T_\Omega) = p_\eta(t), \tag{S460}$$

where $\omega_F$ is the Floquet quasifrequency and $p_\eta(t)$ is a periodic vector containing the temporal sideband structure. To calculate Floquet eigenfrequencies $\omega_F$, we solve the bulk dynamic propagation over one modulation period $0 \le t \le T_\Omega$, and we define the fundamental matrix $F_\eta(t)$ by

$$\frac{d}{dt} F_\eta(t, X) = M_\eta(t; k, X) F_\eta(t, X), \tag{S461}$$

with $F_\eta(t = 0) = I$. Then, the matrix $U_\eta(T_\Omega, 0; k, X) = F_\eta(T_\Omega, X)$ is the one-period propagator describing the evolution of the field over a single modulation cycle. Since the state vector $\Psi_\eta$ has two components, the propagator $U_\eta$ is a $2 \times 2$ matrix and therefore possesses two Floquet branches labeled by the index $j = 1, 2$ for each helicity sector $\eta$.

The Floquet quasifrequencies are obtained from the eigenvalue problem

$$U_\eta(T_\Omega, 0; k, X)\,\mathbf{v}_{\eta,j} = \lambda_{\eta,j}\,\mathbf{v}_{\eta,j}, \tag{S462}$$

where $\lambda_{\eta,j}$ are the Floquet multipliers and $\mathbf{v}_{\eta,j}$ are the corresponding Floquet eigenvectors. The quasifrequencies are then defined through

$$\lambda_{\eta,j}(k) = \exp\left(-i\,\omega_{F\eta,j}(k) T_\Omega\right), \qquad \omega_{F\eta,j}(k) = \frac{i}{T_\Omega} \ln\left(\lambda_{\eta,j}(k)\right). \tag{S463}$$

The functions $\omega_{F\eta,j}(k)$ define the temporal photonic band structure of the biaxionic medium. The real part of $\omega_F$ determines the quasifrequency dispersion, while the imaginary part identifies unstable regions associated with parametric amplification or attenuation. In the following, the Floquet spectrum is plotted in the reduced Floquet zone.

The resulting biaxionic photonic band structures are shown in Fig. S22. We consider $\varepsilon = 5$, $\mu = 1$ and $\chi_0 = 0.3$. For finite $X$, the temporal modulation opens momentum bandgaps near the crossings of the folded quasifrequency branches. These gaps are identified by the appearance of a nonzero imaginary part of $\omega_F$, corresponding to temporal amplification or attenuation of the Floquet modes. As $X$ is reduced, the gaps narrow progressively. At $X = 0$, corresponding to the invisibility condition, the imaginary part vanishes and the Floquet branches become gapless over the plotted range. Thus, the same compensation condition that suppresses cross-polarized scattering also eliminates the observable Floquet instability gaps in the periodically modulated biaxionic medium.

This gap closing should not be interpreted as the absence of the internal dynamics. Even at $X = 0$, the instantaneous bulk equations remain time-dependent and can generate nontrivial sideband and polarization dynamics for suitably prepared internal fields, as discussed below. Figure S22 shows that the outside signatures associated with the time-varying biaxionic response remain hidden when the invisibility condition is satisfied.

## D. Reflection and transmission in time-varying biaxionic slab

The propagation of the wave through the temporal slab involves two temporal interfaces and bulk evolution between them. The wave propagates initially ($t < 0$) in a medium with parameters $\varepsilon_1, \mu_1$. At $t = 0$, the first temporal boundary converts the incident wave into internal slab modes,

$$\Psi_\eta(0^+) = S^{1\to 2}\Psi^{\text{in}}. \tag{S464}$$

The fields then evolve inside the time-varying biaxionic medium as

$$\Psi_\eta(T^-) = U_\eta(T, 0; k)\Psi_\eta(0^+). \tag{S465}$$

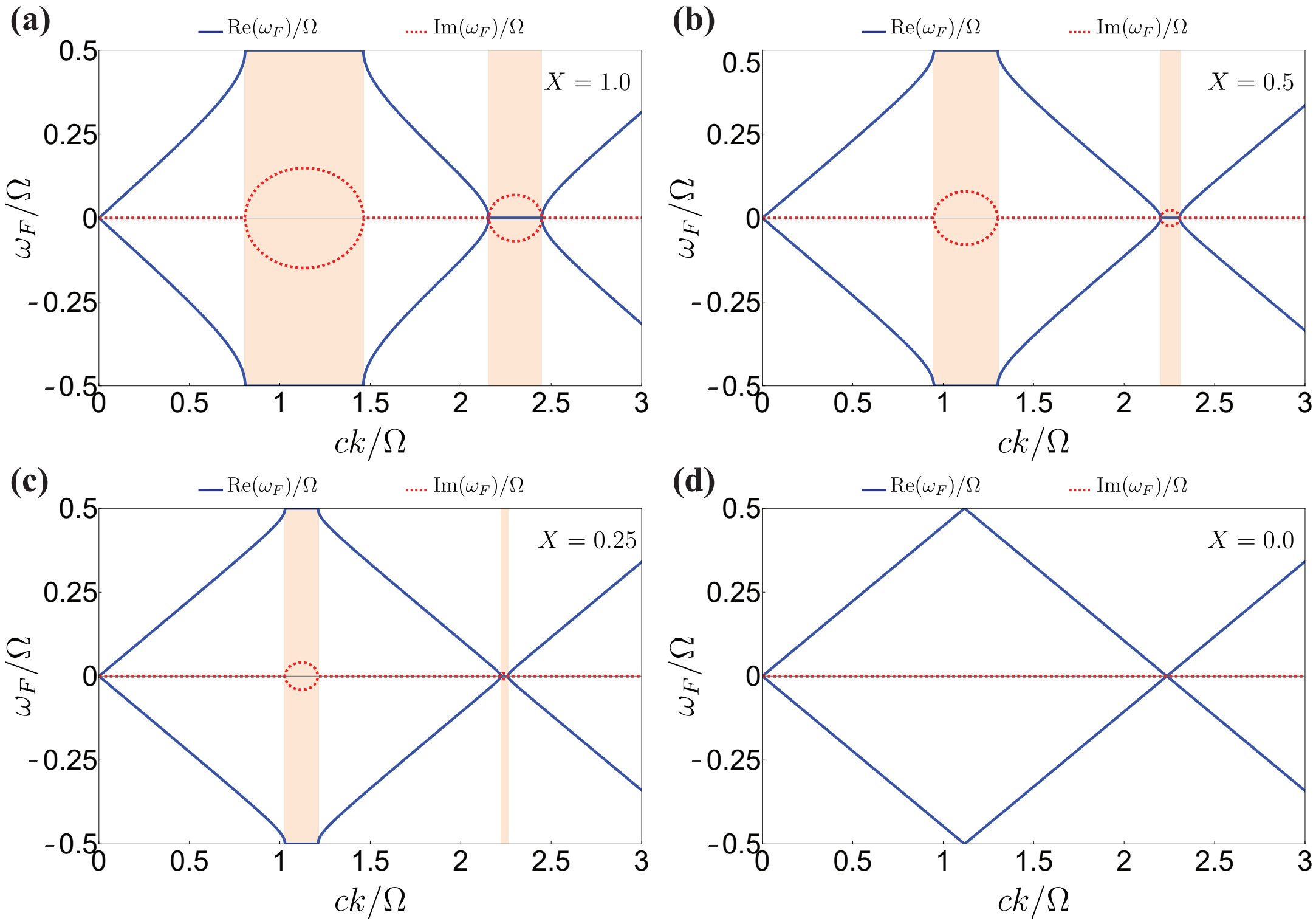


FIG. S22. Photonic band structure of a periodically modulated biaxionic medium for the different deviations from the invisibility condition. The dielectric background is $\varepsilon = 5$, $\mu = 1$ with $\chi_0 = 0.3$. The axion and dual-axion responses are modulated in time as $\chi(t) = \chi_0 \cos\Omega t$ and $\tilde{\chi}(t) = [(\mu/\varepsilon) + X]\,\chi_0 \cos\Omega t$. Blue solid curves show the real part of the Floquet quasifrequency $\omega_F/\Omega$, while red dashed curves show its imaginary part. Shaded regions indicate intervals where $\operatorname{Im}\omega_F \neq 0$, corresponding to Floquet instability gaps. As the detuning parameter $X$ is reduced from (a) $X = 1$ to (b) $X = 0.5$ and (c) $X = 0.25$, the gaps progressively shrink. At invisibility condition (d) $X = 0$, the gaps collapse and the quasifrequency remains purely real in the entire plotted range.

.

At $t = T$, the second temporal boundary converts the internal fields into the outgoing state,

$$\Psi^{\text{out}} = S^{2\to 1}\Psi_\eta(T^-). \tag{S466}$$

The total transfer matrix therefore reads

$$\Psi^{\text{out}} = S^{2\to 1}\, U_\eta(T,0;k)\, S^{1\to 2}\, \Psi^{\text{in}}. \tag{S467}$$

The boundary matrices $S^{1\to 2}$ are constructed from the temporal boundary conditions in Eqs. (S395),(S396).

$$\begin{pmatrix} B_{2,\eta} \\ E_{2,\eta} \end{pmatrix} = S^{1\to 2}\begin{pmatrix} B_{1,\eta} \\ E_{1,\eta} \end{pmatrix}, \qquad S^{1\to 2} = \frac{1}{1+\chi_2\tilde{\chi}_2}\begin{pmatrix} 1+\chi_1\tilde{\chi}_2 & -\varepsilon_1\left(\tilde{\chi}_1-\tilde{\chi}_2\right) \\ \frac{1}{\varepsilon_2}\left(\chi_1-\chi_2\right) & \frac{\varepsilon_1}{\varepsilon_2}\left(1+\tilde{\chi}_1\chi_2\right) \end{pmatrix}. \tag{S468}$$

In the numerical calculations, the temporal slab is implemented as a finite-duration modulation consisting of 10 oscillation periods of the biaxionic response, corresponding to a finite photonic time crystal.

We consider an incident monochromatic plane wave propagating along $z$ axis and linearly polarized along $x$ axis. In the circular polarization basis, Eq. (S447), the incident magnetic field can be decomposed as

$$\mathbf{B}(t=0) = B_0\,\mathbf{e}_x = \frac{B_0}{\sqrt{2}}\left(\mathbf{e}_+ + \mathbf{e}_-\right), \tag{S469}$$

The corresponding incident state in the field basis is therefore

$$\Psi^{\text{in}}_\eta = \frac{B_0}{\sqrt{2}}\begin{pmatrix} 1 \\ \dfrac{i\eta}{\sqrt{\mu_1\varepsilon_1}} \end{pmatrix}, \tag{S470}$$

where the electric-field component follows directly from the Maxwell's equations in the homogeneous exterior medium.

For each polarization $\eta = \pm 1$, the bulk equations are solved numerically over the full modulation interval to obtain the propagated state

$$\Psi^{\text{out}}_\eta = \Psi_\eta(T^+). \tag{S471}$$

To extract the temporal transmission and reflection amplitudes, the outgoing state is decomposed into the positive- and negative-frequency temporal modes of the exterior medium. For a given $\eta$, these modes are

$$\mathbf{u}^{(1)}_{\eta,+} = \begin{pmatrix} 1 \\ \dfrac{i\eta}{\sqrt{\varepsilon_1\mu_1}} \end{pmatrix}, \qquad \mathbf{u}^{(1)}_{\eta,-} = \begin{pmatrix} 1 \\ -\dfrac{i\eta}{\sqrt{\varepsilon_1\mu_1}} \end{pmatrix}. \tag{S472}$$

The positive-frequency mode is proportional to $e^{-i\omega_1 t}$, whereas the negative-frequency mode is proportional to $e^{+i\omega_1 t}$, with

$$\omega_1 = \frac{ck}{\sqrt{\varepsilon_1\mu_1}}. \tag{S473}$$

The vectors $\mathbf{u}^{(1)}_{\eta,\pm}$ are normalized to unit magnetic-field amplitude. We collect the two temporal modes into the matrix

$$C^{(1)}_\eta = \begin{pmatrix} \mathbf{u}^{(1)}_{\eta,+} & \mathbf{u}^{(1)}_{\eta,-} \end{pmatrix} = \begin{pmatrix} 1 & 1 \\ \dfrac{i\eta}{\sqrt{\varepsilon_1\mu_1}} & -\dfrac{i\eta}{\sqrt{\varepsilon_1\mu_1}} \end{pmatrix}. \tag{S474}$$

The outgoing state is then decomposed as

$$\Psi^{\text{out}}_\eta = \frac{B_0}{\sqrt{2}}C^{(1)}_\eta\begin{pmatrix} t_\eta \\ r_\eta \end{pmatrix}, \tag{S475}$$

where $t_\eta$ and $r_\eta$ are the dimensionless temporal transmission and reflection amplitudes for $(\eta)$. Equivalently,

$$\begin{pmatrix} t_\eta \\ r_\eta \end{pmatrix} = \frac{\sqrt{2}}{B_0} \left(C_\eta^{(1)}\right)^{-1} \Psi_\eta^{\mathrm{out}}. \tag{S476}$$

The Cartesian amplitudes are reconstructed using

$$\mathbf{B}^{\mathrm{t}} = \frac{B_0}{\sqrt{2}} \left(t_+ \mathbf{e}_+ + t_- \mathbf{e}_-\right), \qquad \mathbf{B}^{\mathrm{r}} = \frac{B_0}{\sqrt{2}} \left(r_+ \mathbf{e}_+ + r_- \mathbf{e}_-\right). \tag{S477}$$

Since

$$\mathbf{e}_\eta = \frac{1}{\sqrt{2}} \left(\mathbf{e}_x + i\eta \mathbf{e}_y\right), \tag{S478}$$

the co- and cross-polarized scattering amplitudes for the (x)-polarized incident field are

$$t_{xx} = \frac{t_+ + t_-}{2}, \qquad t_{xy} = \frac{i}{2}(t_+ - t_-), \tag{S479}$$

$$r_{xx} = \frac{r_+ + r_-}{2}, \qquad r_{xy} = \frac{i}{2}(r_+ - r_-). \tag{S480}$$

Figure S23 shows the resulting scattering coefficients for the different values of the detuning $X$, which controls the deviation from the invisibility condition.

For finite detunings ($X = 1, 0.5, 0.25$), the temporal slab exhibits pronounced resonant features near $ck/\Omega \approx 1$. These resonances originate from Floquet instability momentum bandgaps of time-modulated medium and are accompanied by strong amplification and activation of cross-polarized channels. As a result, both co- and cross-polarized transmission and reflection coefficients become significant.

The quantity $|t|^2 + |r|^2$ strongly deviates from unity, indicating the absence of energy conservation in the time-modulated system due to the parametric exchange with the modulation. In contrast, $|t|^2 - |r|^2$ remains constant, consistent with the conservation of momentum associated with spatial translational invariance.

As the detuning parameter is reduced, the resonant features progressively weaken. In the limit $X \to 0$, corresponding to the invisibility condition, both amplification and polarization conversion are strongly suppressed. The scattering response approaches that of the conventional dielectric, despite the presence of nontrivial time modulation in the bulk. This demonstrates that the temporal scattering problem features the same key property: intrinsic electromagnetic property cannot be identified in any kind of the scattering experiments.

### E. Polarization dynamics and Fourier spectrum of the field in the bulk

To further verify that the invisibility condition does not trivialize the internal dynamics of the time-modulated biaxionic medium, we analyze the temporal evolution of the field inside the slab. Unlike the scattering calculation above, here we focus on the field in the bulk of the medium during the period when the axion and dual axion responses are modulated.

We consider a homogeneous medium with the same periodic axion and dual-axion responses, $\chi(t)$ and $\tilde{\chi}(t)$, as before in Eq. (S455), and use as initial condition a linearly polarized plane wave propagating along the $z$ axis and polarized in the $x$ direction. This incident state is first mapped into the time-modulated medium using the temporal boundary matrix $S^{1\to 2}$, and then evolved according to the bulk equations

$$\frac{d\Psi_\eta(t)}{dt} = M_\eta(t; k, X)\Psi_\eta(t), \qquad 0 < t < T. \tag{S481}$$

After solving for each polarization independently, the Cartesian magnetic-field components are reconstructed as

$$B_x(t) = \frac{B_+(t) + B_-(t)}{\sqrt{2}}, \tag{S482}$$

$$B_y(t) = \frac{i\left[B_+(t) - B_-(t)\right]}{\sqrt{2}}. \tag{S483}$$

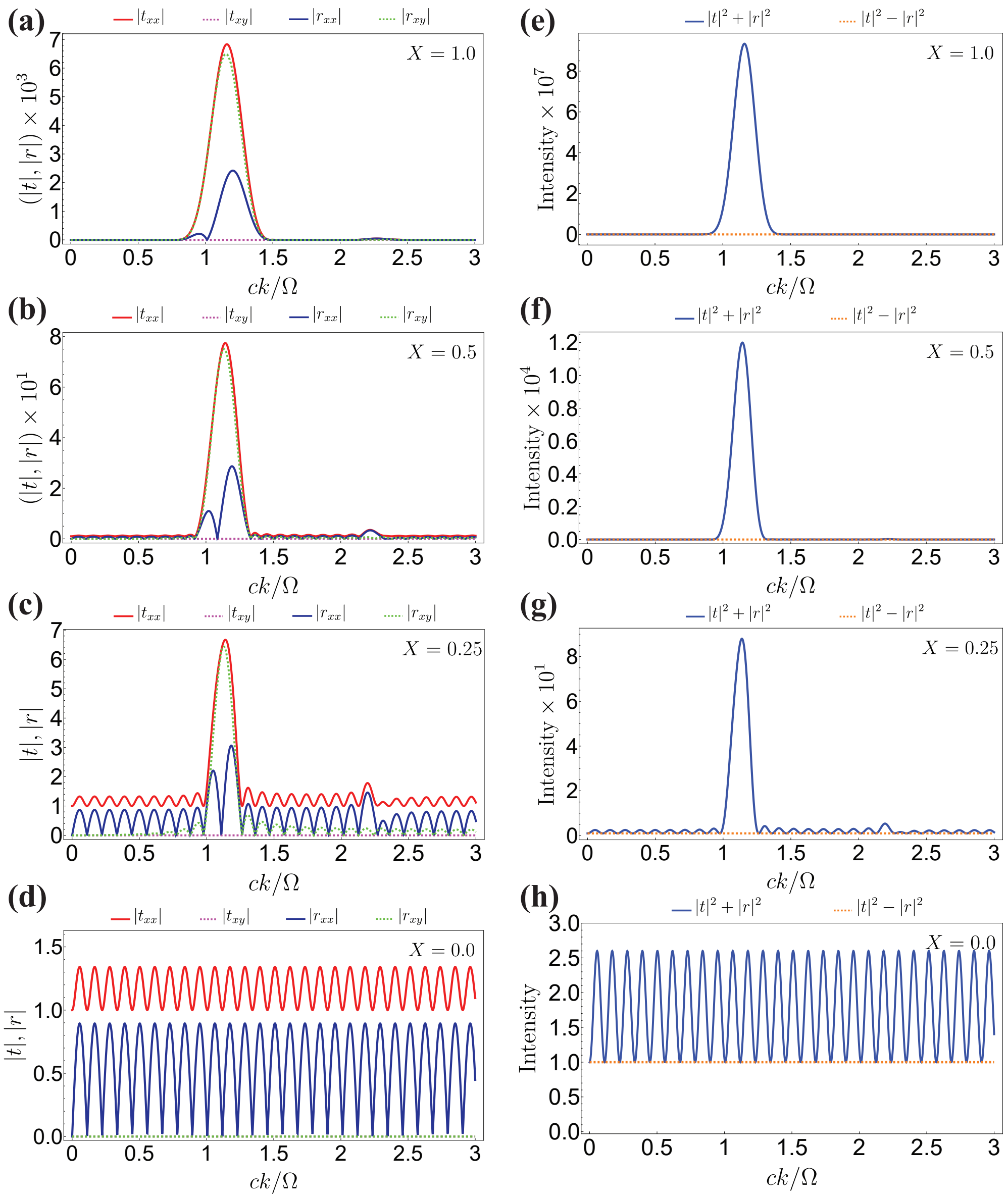


FIG. S23. Temporal scattering from a finite biaxionic slab composed of 10 modulation periods, for different values of the detuning parameter $X$. The incident wave is linearly polarized along $x$ axis and propagates in the $z$ direction. (a–d) Amplitudes of the co- and cross-polarized transmission and reflection coefficients, $t_{ij}$ and $r_{ij}$, as functions of normalized wavenumber $ck/\Omega$. (e–h) Corresponding intensity combinations $|t|^2 + |r|^2$ and $|t|^2 - |r|^2$ which quantify energy non-conservation and momentum conservation.

We then compute the Fourier spectra $B_x(\omega)$ and $B_y(\omega)$ from the time-dependent field $\mathbf{B}$ during the modulation time. For visualization purposes, the spectra are normalized by the maximum total magnetic field amplitude,

$$B_{\max} = \max_{\omega} \sqrt{|B_x(\omega)|^2 + |B_y(\omega)|^2}, \tag{S484}$$

so that the plotted quantities correspond to $|B_x(\omega)|/B_{\max}$ and $|B_y(\omega)|/B_{\max}$.

To obtain well-resolved Fourier peaks, the bulk evolution is computed over $N = 1000$ modulation periods. This long interval improves the spectral resolution of the Fourier transform and should not be confused with the finite temporal slab of 10 periods considered in the scattering calculations above.

The incident frequency is chosen as $\omega_{\text{in}}/\Omega = 0.5$. Since the wavevector is conserved at a temporal boundary, the

frequency associated with the conserved wavevector inside the dielectric background becomes

$$\frac{\omega_2}{\Omega} = \frac{\omega_{\rm in}}{\Omega\sqrt{\varepsilon_2\mu_2}} = \frac{0.5}{\sqrt{5}} \simeq 0.223. \tag{S485}$$

Accordingly, the dominant spectral peaks appear near $\omega/\Omega \simeq \pm 0.223$, together with the sidebands shifted by integer multiples of the modulation frequency.

Figure S24 shows the resulting internal Fourier spectra for the different values of $X$. In all cases, the initially monochromatic excitation develops additional Fourier components, demonstrating that the time-modulated biaxionic medium produces nontrivial bulk dynamics. Importantly, this remains true at $X = 0$, where the invisibility condition is exactly satisfied. Although the radiation which comes out of the medium remains monochromatic and does not feature any cross-polarized components, the bulk field still develops sidebands and a finite cross-polarized component $B_y$. Thus, the invisibility condition suppresses observable scattering signatures but does not eliminate the intrinsic temporal evolution of the electromagnetic field inside the medium.

This mechanism can in principle be used for information transfer protected against remote sensing. While the time-varying axion and dual axion responses can carry some information, this information remains fundamentally inacessible provided the outside observer has no physical access to the medium and is restricted to do only the scattering measurements.

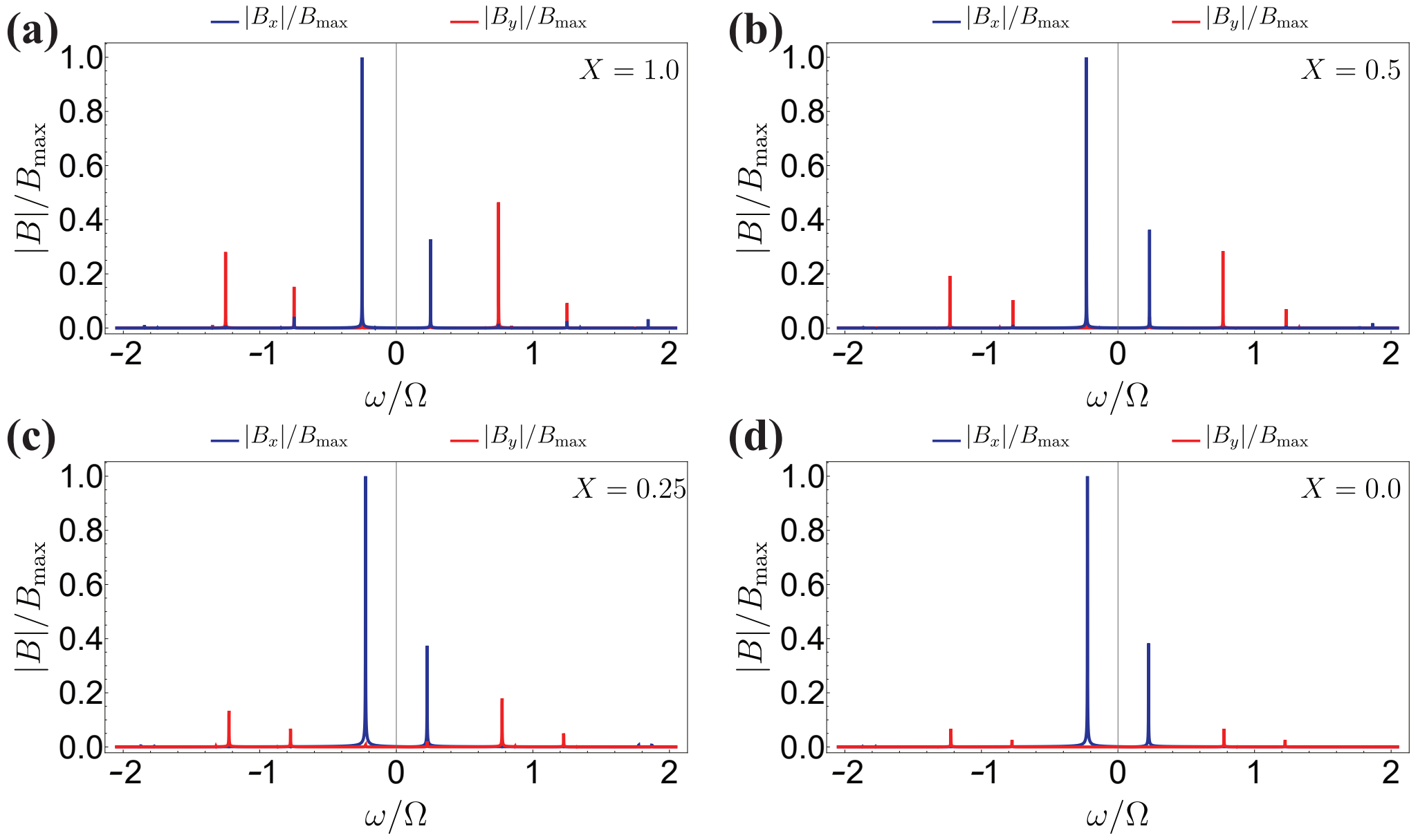


FIG. S24. Fourier spectrum of the magnetic field inside time-modulated biaxionic medium. An initially $x$-polarized monochromatic wave with $\omega_{\rm in}/\Omega = 0.5$ is injected into the medium and evolves during the modulation time. The spectra of $B_x(\omega)$ and $B_y(\omega)$ are shown for the different values of the detuning parameter: (a) $X = 1$, (b) $X = 0.5$, (c) $X = 0.25$, and (d) $X = 0$. The spectra are normalized by the maximal total magnetic-field amplitude $B_{\rm max} = \max_\omega \sqrt{|B_x(\omega)|^2 + |B_y(\omega)|^2}$. Because the background medium has $\varepsilon_2 = 5$ and $\mu_2 = 1$, the dominant frequency is shifted to $\omega/\Omega \simeq 0.5/\sqrt{5} \simeq 0.223$, with additional sidebands separated by the modulation frequency $\Omega$. Even at the invisibility condition $X = 0$, the field develops sidebands and a finite cross-polarized component $B_y$, confirming that the invisibility condition does not trivialize the internal bulk dynamics. The Fourier transform is computed from the bulk evolution over $N = 1000$ modulation periods.



\end{document}